\documentclass[fleqn,10pt]{wlscirep}
\usepackage[utf8]{inputenc}
\usepackage[T1,T2A]{fontenc}
\usepackage[backend=biber,style=nature,]{biblatex}
\usepackage{csquotes}
\usepackage{amsmath} 
\usepackage{longtable}

\newcommand\tess{{\it TESS}}

\newcommand\kepler{{\it Kepler}}

\newcommand\harps{{\it HARPS}}

\newcommand\gaia{{\it Gaia}}
\newcommand\soar{{\it SOAR}}
\newcommand\lcogt{{\it LCOGT}}

\newcommand{\var}[1]{{\operatorname{#1}}}

\def\arxivprefixesep{:}

\newcommand{\eprint}[2][]{%
{\tt\if!#1!#2\else#1\arxivprefixesep\ignorespaces#2\fi}%
}
\newcommand{\kms}{\,km\,s$^{-1}$} 
\newcommand{\ms}{\,m\,s$^{-1}$} 

\newcommand*\degr{\ensuremath{^{\circ}}}
\newcommand*\arcmin{\ensuremath{^\prime}}
\newcommand*\arcsec{\ensuremath{^{\prime\prime}}}
\def\utw{\ensuremath{\smash{\rlap{\lower5pt\hbox{$\sim$}}}}}
\def\udtw{\ensuremath{\smash{\rlap{\lower6pt\hbox{$\approx$}}}}}

\newcommand*\farcm{\ensuremath{\overset{\prime}{.}}}
\newcommand*\farcs{\ensuremath{\overset{\prime\prime}{.}}}

\DeclareUnicodeCharacter{0301}{HEREHEREHERE}

\newcommand{\Tzerobbis}[1][days]   {$1468.3905 \pm {0.0017}$~#1} 
\newcommand{\Pbbis}[1][days]   {$0.548177 \pm 0.000019$~#1} 
\newcommand{\esinbbis}[1][ ]   {$-0.07 \pm 0.24 $~#1} 
\newcommand{\ecosbbis}[1][ ]   {$0.02 \pm 0.17$~#1} 
\newcommand{\bbbis}[1][ ]   {$0.53 _{ - 0.18 } ^ { + 0.12 }$~#1} 
 
\newcommand{\rrbbis}[1][ ]   {$0.01577 _{ - 0.00070 } ^ { + 0.00072 }$~#1} 
\newcommand{\kbbis}[1][${\rm m\,s^{-1}}$]   {$1.35 \pm 0.17 $~#1} 
\newcommand{\mpbbis}[1][$M_{\oplus}$]   {$1.42 \pm 0.18$ ~#1} 
\newcommand{\rpbbis}[1][$R_{\oplus}$]   {$1.166 _{ - 0.058 } ^ { + 0.061 }$~#1} 
 
\newcommand{\ebbis}[1][ ]   {$0.063 _{ - 0.044 } ^ { + 0.073 }$~#1} 
\newcommand{\wbbis}[1][deg]   {$228.5 _{ - 160.6 } ^ { + 72.9 }$~#1} 
\newcommand{\ibbis}[1][deg]   {$82.09 _{ - 1.31 } ^ { + 2.53 }$~#1} 
\newcommand{\abbis}[1][AU]   {$0.01189 _{ - 0.00040 } ^ { + 0.00041 }$~#1} 
\newcommand{\insolationbbis}[1][${\rm F_{\oplus}}$]   {$1138 _{ - 112 } ^ { + 123 }$~#1} 
\newcommand{\denstrbbis}[1][${\rm g\,cm^{-3}}$]   {$3.38 _{ - 0.24 } ^ { + 0.26 }$~#1} 
 
\newcommand{\denpbbis}[1][${\rm g\,cm^{-3}}$]   {$4.89 _{ - 0.88 } ^ { + 1.03 }$~#1}

\newcommand{\Teqbbis}[1][K]   {$1617 \pm 41$~#1} 
\newcommand{\ttotbbis}[1][hours]   {$0.998 _{ - 0.044 } ^ { + 0.050 }$~#1} 
 
\newcommand{\Tdepthbis}[1][ppm] {$248.7 _ {22.0} ^ {22.6}$~#1}
\newcommand{\Tzerocbis}[1][days]   {$1561.96 \pm 0.20$~#1} 
\newcommand{\Pcbis}[1][days]   {$6.6356 \pm 0.0040$~#1} 
\newcommand{\esincbis}[1][ ]   {$-0.04 \pm 0.20 $~#1} 
\newcommand{\ecoscbis}[1][ ]   {$0.11 _{ - 0.22 } ^ { + 0.19 }$~#1} 
\newcommand{\kcbis}[1][${\rm m\,s^{-1}}$]   {$2.10 \pm 0.17 $~#1} 
\newcommand{\mpcbis}[1][$M_{\oplus}$]   {$5.03 \pm 0.41$~#1} 
 
\newcommand{\ecbis}[1][ ]   {$0.072 _{ - 0.050 } ^ { + 0.074 }$~#1} 
\newcommand{\wcbis}[1][deg]   {$228.0 _{ - 197.0 } ^ { + 100.0 }$~#1} 
\newcommand{\Tzerodbis}[1][days]   {$1587.22 _{ - 0.20 } ^ { + 0.22 }$~#1} 
\newcommand{\Pdbis}[1][days]   {$26.233 \pm 0.020$~#1} 
\newcommand{\esindbis}[1][ ]   {$-0.011 _{ - 0.099 } ^ { + 0.101 }$~#1} 
\newcommand{\ecosdbis}[1][ ]   {$-0.045 _{ - 0.091 } ^ { + 0.104 }$~#1} 
\newcommand{\kdbis}[1][${\rm m\,s^{-1}}$]   {$8.72 \pm 0.19$~#1} 
\newcommand{\mpdbis}[1][$M_{\oplus}$]   {$33.12 \pm 0.88 $~#1} 
 
\newcommand{\edbis}[1][ ]   {$0.016 _{ - 0.011 } ^ { + 0.017 }$~#1} 
\newcommand{\wdbis}[1][deg]   {$189.6 _{ - 93.3 } ^ { + 92.3 }$~#1} 
\newcommand{\Tzeroebis}[1][days]   {$1865.82 _{ - 1.32 } ^ { + 1.16 }$~#1} 
\newcommand{\Pebis}[1][days]   {$61.30 \pm 0.28$~#1} 
\newcommand{\esinebis}[1][ ]   {$0.04 _{ - 0.19 } ^ { + 0.18 }$~#1} 
\newcommand{\ecosebis}[1][ ]   {$0.18 _{ - 0.20 } ^ { + 0.14 }$~#1} 
\newcommand{\kebis}[1][${\rm m\,s^{-1}}$]   {$3.00 \pm 0.22$~ #1} 
\newcommand{\mpebis}[1][$M_{\oplus}$]   {$15.05 _{ - 1.11 } ^ { + 1.12 }$~#1} 
 
\newcommand{\eebis}[1][ ]   {$0.073 _{ - 0.051 } ^ { + 0.068 }$~#1} 
\newcommand{\webis}[1][deg]   {$103.7 _{ - 80.7 } ^ { + 228.2 }$~#1} 
\newcommand{\Tzerofbis}[1][days]   {$1583.92 _{ - 1.11 } ^ { + 1.10 }$~#1} 
\newcommand{\Pfbis}[1][days]   {$43.73 _{ - 0.20 } ^ { + 0.21 }$~#1}

\newcommand{\kfbis}[1][${\rm m\,s^{-1}}$]   {$1.94 \pm 0.20$~#1}

\newcommand{\qonebis}[1][]   {$0.483 _{ - 0.096 } ^ { + 0.098 }$~#1} 
\newcommand{\qtwobis}[1][]   {$0.404 \pm 0.098 $~#1} 
\newcommand{\uonebis}[1][]   {$0.55 _{ - 0.14 } ^ { + 0.15 }$~#1} 
\newcommand{\utwobis}[1][]   {$0.13 _{ - 0.13 } ^ { + 0.14 }$~#1} 
 
\newcommand{\jHARPSbis}[1][${\rm m\,s^{-1}}$]   {$1.16 \pm 0.12$~#1}

\newcommand{\smass}[1][$M_{\odot}$]{ $ 0.740 \pm 0.017$ #1} 
\newcommand{\sradius}[1][$R_{\odot}$]{ $0.678 \pm 0.016$ #1}

\newcommand{\Tzerob}[1][days]   {$1468.3909 \pm 0.0017$~#1} 
\newcommand{\Pb}[1][days]   {$0.548172 \pm 0.000019$~#1} 
\newcommand{\esinb}[1][ ]   {$-0.03 _{ - 0.23 } ^ { + 0.21 }$~#1} 
\newcommand{\ecosb}[1][ ]   {$-0.09 _{ - 0.16 } ^ { + 0.19 }$~#1} 
\newcommand{\bb}[1][ ]   {$0.51 _{ - 0.17 } ^ { + 0.12 }$~#1} 
 
\newcommand{\rrb}[1][ ]   {$0.01568 \pm 0.00068$~#1} 
\newcommand{\kb}[1][${\rm m\,s^{-1}}$]   {$1.56 \pm 0.20$~#1} 
\newcommand{\mpb}[1][$M_{\oplus}$]   {$1.64 \pm 0.21$~#1} 
\newcommand{\rpb}[1][$R_{\oplus}$]   {$1.159 \pm 0.058$~#1} 
 
\newcommand{\eb}[1][ ]   {$0.063 _{ - 0.044 } ^ { + 0.068 }$~#1} 
\newcommand{\wbs}[1][deg]   {$316.2 _{-95.9}^{+181.7}$~#1} 
\newcommand{\ib}[1][deg]   {$82.31 _{ - 1.41 } ^ { + 2.35 }$~#1} 
\newcommand{\arb}[1][ ]   {$3.769 \pm 0.090$~#1} 
\newcommand{\ab}[1][AU]   {$0.01188 \pm 0.00040$~#1} 
\newcommand{\depthb}[1][ppm]   {$245.9 _{ - 20.8 } ^ { + 21.8 }$~#1} 
 
\newcommand{\insolationb}[1][${\rm F_{\oplus}}$]   {$1140 _{ - 112 } ^ { + 121 }$~#1} 
 
\newcommand{\denstrb}[1][${\rm g\,cm^{-3}}$]   {$3.37 _{ - 0.23 } ^ { + 0.25 }$~#1} 
 
\newcommand{\Teqb}[1][K]   {$1617 \pm 41$~#1} 
\newcommand{\ttotb}[1][hours]   {$1.002 _{ - 0.045 } ^ { + 0.051 }$~#1}

\newcommand{\denpb}[1][${\rm g\,cm^{-3}}$]   {$5.78 _{ - 1.04 } ^ { + 1.23 }$~#1}

\newcommand{\Tzeroc}[1][days]   {$1562.09 \pm 0.20$~#1} 
\newcommand{\Pc}[1][days]   {$6.6299 \pm 0.0051$~#1} 
\newcommand{\esinc}[1][ ]   {$-0.03 \pm 0.23$~#1} 
\newcommand{\ecosc}[1][ ]   {$0.22 _{ - 0.27 } ^ { + 0.19 }$~#1} 
\newcommand{\kc}[1][${\rm m\,s^{-1}}$]   {$1.95 \pm 0.21$~#1} 
\newcommand{\mpc}[1][$M_{\oplus}$]   {$4.64 \pm 0.50$~#1} 
 
\newcommand{\ec}[1][ ]   {$0.111 _{ - 0.076 } ^ { + 0.102 }$~#1} 
\newcommand{\wc}[1][deg]   {$352.4 _{ - 76.2 } ^ { + 54.4 }$~#1} 
\newcommand{\Tzerod}[1][days]   {$1587.20 \pm 0.24$~#1} 
\newcommand{\Pd}[1][days]   {$26.235 \pm 0.024$~#1} 
\newcommand{\esind}[1][ ]   {$0.04 \pm 0.11$~#1} 
\newcommand{\ecosd}[1][ ]   {$0.04 \pm 0.11$~#1} 
\newcommand{\kd}[1][${\rm m\,s^{-1}}$]   {$8.83 _{ - 0.23 } ^ { + 0.25 }$~#1} 
\newcommand{\mpd}[1][$M_{\oplus}$]   {$33.54 _{ - 1.02 } ^ { + 1.07 }$~#1} 
 
\newcommand{\ed}[1][ ]   {$0.019 _{ - 0.013 } ^ { + 0.020 }$~#1} 
\newcommand{\wds}[1][deg]   {$27.2 _{ - 104.8 } ^ { + 81.9 }$~#1} 
\newcommand{\Tzeroe}[1][days]   {$1865.22 _{ - 1.03 } ^ { + 0.91 }$~#1} 
\newcommand{\Pe}[1][days]   {$60.33 _{ - 0.33 } ^ { + 0.32 }$~#1} 
\newcommand{\esine}[1][ ]   {$0.12 _{ - 0.22 } ^ { + 0.19 }$~#1} 
\newcommand{\ecose}[1][ ]   {$0.21 _{ - 0.20 } ^ { + 0.15 }$~#1} 
\newcommand{\ke}[1][${\rm m\,s^{-1}}$]   {$2.40 \pm 0.27$~#1} 
\newcommand{\mpe}[1][$M_{\oplus}$]   {$11.94 _{ - 1.34 } ^ { + 1.36 }$~#1} 
 
\newcommand{\ee}[1][ ]   {$0.106 _{ - 0.075 } ^ { + 0.092 }$~#1} 
\newcommand{\we}[1][deg]   {$28.1 _{ - 60.9 } ^ { + 43.6 }$~#1} 
\newcommand{\qone}[1][]   {$0.484 _{ - 0.093 } ^ { + 0.100 }$~#1} 
\newcommand{\qtwo}[1][]   {$0.399 _{ - 0.099 } ^ { + 0.103 }$~#1} 
\newcommand{\uone}[1][]   {$0.55 _{ - 0.15 } ^ { + 0.16 }$~#1} 
\newcommand{\utwo}[1][]   {$0.14 _{ - 0.14 } ^ { + 0.14 }$~#1}

\newcommand{\jRV}[1][${\rm m\,s^{-1}}$]   {$1.50 \pm 0.21$~#1}

\newcommand{\jAzero}[1][]   {$5.6_{-4.1}^{+30.1}$~#1} 
\newcommand{\jAuno}[1][]   {$3_{-19}^{+32}$~#1} 
\newcommand{\jAdue}[1][]   {$0.21 _{ - 0.15 } ^ { + 1.03 }$~#1} 
 
\newcommand{\jAquattro}[1][]   {$0.057 _{ - 0.041 } ^ { + 0.284 }$~#1} 
 
\newcommand{\jlambdae}[1][]   {$105.0 _{ - 50.2 } ^ { + 93.8 }$~#1} 
\newcommand{\jlambdap}[1][]   {$2.70 _{ - 1.52 } ^ { + 4.33 }$~#1} 
\newcommand{\jPGP}[1][]   {$44.57 _{ - 2.94 } ^ { + 2.05 }$~#1} 

\title{A low-eccentricity migration pathway for a 13-h-period Earth analogue in a four-planet system} 

\author[1,*]{Luisa Maria Serrano}
\author[1]{Davide Gandolfi}
\author[2]{Alexander J.~Mustill}
\author[3]{Oscar Barrag\'an}
\author[4]{Judith Korth}
\author[5]{Fei Dai}
\author[6]{Seth Redfield}
\author[7,8]{Malcolm Fridlund}
\author[9]{Kristine W.\,F. Lam}
\author[10, 11]{Matías R. Díaz}
\author[12]{Sascha Grziwa} 
\author[13]{Karen A.\,Collins}
\author[14]{John H. Livingston}
\author[15]{William D. Cochran}
\author[16]{Coel Hellier}
\author[1]{Salvatore E.\,Bellomo}
\author[17]{Trifon Trifonov}
\author[18]{Florian Rodler}
\author[18]{Javier Alarcon}
\author[19]{Jon M. Jenkins}
\author[20]{David W. Latham}
\author[21]{George Ricker}
\author[21,22,23]{Sara Seager}
\author[21]{Roland Vanderspeck} 
\author[24]{Joshua N. Winn}
\author[25]{Simon Albrecht}
\author[26]{Kevin I. Collins} 
\author[27]{Szil\'ard Csizmadia}
\author[21, 28]{Tansu Daylan} 
\author[29, 30]{Hans J. Deeg}
\author[31]{Massimiliano Esposito}
\author[21]{Michael Fausnaugh}
\author[7]{Iskra Georgieva}
\author[1, 31]{Elisa Goffo}
\author[31]{Eike Guenther}
\author[31]{Artie P. Hatzes}
\author[19]{Steve B. Howell}
\author[32]{Eric L.\ N.\ Jensen}
\author[29,30]{Rafael Luque}
\author[33]{Andrew W. Mann}
\author[29,30]{Felipe Murgas}
\author[34]{Hannah L. M. Osborne}
\author[29,30]{Enric Palle}
\author[7]{Carina M. Persson}
\author[35]{Pam Rowden}
\author[21]{Alexander Rudat}
\author[36]{Alexis M. S. Smith}
\author[37,19]{Joseph D. Twicken}
\author[34]{Vincent Van Eylen}
\author[38]{Carl Ziegler}

\affil[1]{Dipartimento di Fisica, Universit\`a degli Studi di Torino,
via Pietro Giuria 1, 10125, Torino, Italy}
\affil[2]{Lund Observatory, Department of Astronomy and Theoretical Physics, Lund University, Box 43, SE-221 00 Lund, Sweden}
\affil[3]{Sub-department of Astrophysics, Department of Physics, University of Oxford, Oxford, OX1 3RH, UK}
\affil[4]{Department of Space, Earth and Environment, Astronomy and Plasma Physics, Chalmers University of Technology, SE-412 96 Gothenburg, Sweden}
\affil[5]{Division of Geological and Planetary Sciences 1200 E California Blvd, Pasadena, CA 91125}
\affil[6]{Astronomy Department and Van Vleck Observatory, Wesleyan University, Middletown, CT 06459, USA}
\affil[7]{Leiden Observatory, Leiden University, 2333CA Leiden, The Netherlands}
\affil[8]{Department of Space, Earth and Environment, Chalmers University of Technology, Onsala Space Observatory, SE-439 92 Onsala, Sweden}
\affil[9]{Center for Astronomy and Astrophysics, TU Berlin, Hardenbergstr. 36, 10623 Berlin, Germany}
\affil[10]{Departamento de Astronomía, Universidad de Chile, Camino El Observatorio 1515, Las Condes, Santiago, Chile.}
\affil[11]{Las Campanas Observatory, Carnegie Institution of Washington, Colina El Pino, Casilla 601, La Serena, Chile.}
\affil[12]{Rheinisches Institut f\"ur Umweltforschung an der Universit\"at zu K\"oln, Aachener Strasse 209, 50931 K\"oln, Germany}
\affil[13]{Center for Astrophysics \textbar \ Harvard \& Smithsonian, 60 Garden Street, Cambridge, MA 02138, USA}
\affil[14]{Department of Astronomy, University of Tokyo, 7-3-1 Hongo, Bunkyo-ku, Tokyo 113-0033, Japan}
\affil[15]{Department of Astronomy and McDonald Observatory, University of Texas at Austin, 2515 Speedway, Austin, TX 78712, USA}
\affil[16]{Keele University, Astrophysics Group, Keele University, Staffordshire ST5 5BG, U.K.}
\affil[17]{Max Planck Institut f\"ur Astronomie, K\"onigstuhl 17, 69117 Heidelberg, Germany}
\affil[18]{European Southern Observatory, Alonso de Cordova 3107, Vitacura, Santiago de Chile, Chile}
\affil[19]{NASA Ames Research Center, Mail Stop 269-3, Bldg. T35A, Rm. 102, P.O. Box 1, Moffett Field, CA 94035-0001, USA}
\affil[20]{Harvard-Smithsonian Center for Astrophysics, 60 Garden Street, Office: P-333, Cambridge, MA 02138, MS-16, USA}
\affil[21]{Department of Physics and Kavli Institute for Astrophysics and Space Research, Massachusetts Institute of Technology, Cambridge, MA 02139, USA}
\affil[22]{Department of Earth, Atmospheric and Planetary Sciences, Massachusetts Institute of Technology, Cambridge, MA 02139, USA}
\affil[23]{Department of Aeronautics and Astronautics, MIT, 77 Massachusetts Avenue, Cambridge, MA 02139, USA}
\affil[24]{Department of Astrophysical Sciences, Princeton University, 4 Ivy Lane, Princeton, NJ 08544, USA}
\affil[25]{Stellar Astrophysics Centre, Department of Physics and Astronomy, Aarhus University, Ny Munkegade 120, DK-8000 Aarhus C, Denmark}
\affil[26]{George Mason University, 4400 University Drive, Fairfax, VA, 22030 USA}
\affil[27]{Institute of Planetary Research, German Aerospace Center (DLR), Rutherfordstrasse 2, D-12489 Berlin, Germany}
\affil[28]{KAVLI fellow}
\affil[29]{Instituto de Astrof\'\i sica de Canarias (IAC), 38205 La Laguna, Tenerife, Spain}
\affil[30]{Departamento de Astrof\'\i sica, Universidad de La Laguna (ULL), 38206 La Laguna, Tenerife, Spain}
\affil[31]{Th\"uringer Landessternwarte Tautenburg,  D-07778 Tautenburg, Germany}
\affil[32]{Department of Physics \& Astronomy, Swarthmore College, Swarthmore PA 19081, USA}
\affil[33]{Department of Physics and Astronomy, The University of North Carolina at Chapel Hill, Chapel Hill, NC 27599-3255, USA}
\affil[34]{Mullard Space Science Laboratory, University College London, Holmbury St Mary, Dorking, Surrey, RH5 6NT, UK}
\affil[35]{Royal Astronomical Society, Burlington House, Piccadilly, London W1J 0BQ, U.K.}
\affil[36]{Institute for Planetary Research, German Aerospace Center (DLR), Rutherfordstr. 2, 12489 Berlin, Germany}
\affil[37]{SETI Institute, 189 Bernardo Ave, Suite 200 Mountain View, CA 94043, USA}
\affil[38]{Department of Physics, Engineering and Astronomy, Stephen F. Austin State University, 1936 North St, Nacogdoches, TX 75962, USA}

\affil[*]{Corresponding author: Luisa Maria Serrano (luisamaria.serrano@unito.it)}

\begin{abstract}
It is commonly accepted that exoplanets with orbital periods shorter than 1\,day, also known as ultra-short period (USP) planets, formed further out within their natal protoplanetary disk, before migrating to their current-day orbits via dynamical interactions. One of the most accepted theories suggests a violent scenario involving high-eccentricity migration followed by tidal circularization. Here, we present the discovery of a four planet system orbiting the bright (V=10.5) K6 dwarf star TOI-500. The innermost planet is a transiting, Earth-sized USP planet with an orbital period of $\sim$13 hours, a mass of \mpbbis, a radius of \rpbbis, and a mean density of \denpbbis. Via Doppler spectroscopy, we discovered that the system hosts three outer planets on nearly circular orbits with periods of $6.6$, $26.2$, and $61.3$\,d and minimum masses of \mpcbis, \mpdbis, and \mpebis, respectively. The presence of both a USP planet and a low-mass object on a 6.6-day orbit indicates that the architecture of this system can be explained via a scenario in which the planets started on low-eccentricity orbits, then moved inwards through a quasi-static secular migration. Our numerical simulations show that this migration channel can bring TOI-500\,b to its current location in 2\,Gyrs, starting from an initial orbit of 0.02\,au. TOI-500 is the first four planet system known to host a USP Earth analog whose current architecture can be explained via a non-violent migration scenario. 

\end{abstract}
\begin{document}
\begingroup
\fontencoding{T1}\selectfont

\flushbottom
\maketitle

\section*{Main Text}
TOI-500 (also known as HIP\,34269, TIC\,134200185, CD-47 2804) is a high proper motion star (Table~\ref{stellar_values}) with a radial velocity of $55.6$~km\,s$^{-1}$ \cite{GaiaDR2}, a V-band magnitude of $10.54$ \cite{Mermilliod1987}, located at a distance of $47.39$~pc from the Sun \cite{GaiaDR2}. NASA's Transiting Exoplanet Survey Satellite (\tess, \cite{Ricker15}) observed TOI-500 for the first time in Sectors $6$, $7$, and $8$ between 11 December 2018 and 28 February 2019. The \tess\ Science Processing Operations Center (SPOC, \cite{SPOC}) identified the signature of a possible Earth-sized transiting planet with an orbital period of nearly $13$\,hrs. The candidate was subsequently designated as TOI-500.01 by the \tess\ Science Office and announced on 8 March 2019. We performed an independent analysis of the \tess\ light curves with the codes D\'etection Sp\'ecialis\'ee de Transits (\texttt{DST}, \cite{Cabrera2012}) and Transit Least Square (\texttt{TLS},  \cite{Hippke19b}), which confirmed the presence of the candidate (Figure~\ref{Tess_data}) and excluded additional significant transit signals. We used the Las Cumbres Observatory Global Telescope (\lcogt, \cite{Brown:2013}) to perform photometric observations of the $78$ neighboring stars up to about $\Delta$mag\,$\approx$\,10 at angular separation between $12$\arcsec\ and $2.5$\arcmin\ from TOI-500 (Figure~\ref{field}). The analysis of the retrieved light curves allowed us to exclude that those sources are contaminating eclipsing binaries mimicking the transit signal detected in the \tess\ light curves. Speckle images acquired with the $4.1$\,m Southern Astrophysical Research (\soar) telescope (Cerro Tololo Inter-American Observatory, Chile) and the $8.1$\,m \textit{Gemini} South telescope (Cerro Pach\'on, Chile) excluded the presence of nearby stars up to about $\Delta\mathrm{mag} \approx 7$, as close as $0.02\arcsec$ (\textit{Zorro}@\textit{Gemini} speckle inner working angle, Figure~\ref{fig:zorro}) and out to $3\arcsec$ (\soar\ outer limit, Figure~\ref{fig:soar}). Finally, we confirmed the planetary nature of the transit signal with an intensive radial velocity (RV) follow-up campaign carried out with the High Accuracy Radial velocity Planet Searcher (\harps, \cite{Mayor2003}) spectrograph mounted at the $3.6$\,m telescope of the European Southern Observatory (ESO, La Silla, Chile). We collected nearly $200$ \harps\ spectra of TOI-500 between 22 March 2019 and 23 March 2020. Our RV measurements also unveiled the presence of three additional Doppler signals that have no counterpart in any of the stellar activity indicators, providing strong evidence that they are induced by three additional planets (Figures~\ref{RV_data}, \ref{periodogram_planet}, and \ref{activity indexes}). TOI-500 is thus orbited by (at least) 4 planets, 3 of which are not seen to transit their host star. In order to determine the planetary parameters, we simultaneously modelled the transit photometry and radial velocity measurements using the software \texttt{pyaneti} \cite{Barragan19} (Figures~\ref{planetb_phase_folded} and \ref{Sinusoidal}). We also derived the fundamental parameters of the host star by analyzing the co-added \harps\ spectrum with the code Spectroscopy Made Easy (\texttt{SME}, \cite{Valenti96,Valenti17}). We inferred the stellar mass, radius, and age using the Bayesian web-tool \texttt{PARAM\,1.3}, \cite{daSilva06}. In order to measure the rotational period of the star, we performed a frequency analysis of the existing Wide Angle Search for Planets (\textit{WASP-South}, \cite{2006PASP..118.1407P}) ground-based photometry (Figure~\ref{fig:wasp}), and frequency (Figure~\ref{activity indexes}) and Bayesian analyses of the \harps\ activity indicators (Figure~\ref{fig:gpstimeseries}).

TOI-500\,b has an orbital period of $\mathrm{P}_\mathrm{b}=$\,\Pbbis. Its mass of $\mathrm{M}_\mathrm{b}=$\,\mpbbis\ and radius of $\mathrm{R}_\mathrm{b}=$\,\rpbbis\ imply a mean density of $\rho_\mathrm{b}=$\,\denpbbis. For the other three planets, we could only measure their minimum masses because, in the absence of transit detection, the inclinations of their orbits remain unknown. TOI-500\,c, d, and e have periods of $\mathrm{P}_\mathrm{c}=$\,\Pcbis, $\mathrm{P}_\mathrm{d}=$\,\Pdbis, and $\mathrm{P}_\mathrm{e}=$\,\Pebis, and minimum masses of $\mathrm{M}_\mathrm{c} \sin i_\mathrm{c}=$\,\mpcbis, $\mathrm{M}_\mathrm{d} \sin i_\mathrm{d}=$\,\mpdbis, and $\mathrm{M}_\mathrm{e} \sin i_\mathrm{e}=$\,\mpebis, respectively (Table~\ref{table_prior}). The four planets have nearly circular orbits, with eccentricities compatible with zero within 1.5$\sigma$. The host star TOI-500 is a K6 dwarf with a mass of $\mathrm{M}_{\star}=$\,\smass, a radius of  $\mathrm{R}_{\star}=$\,\sradius, and iron and calcium abundances of $\mathrm{[Fe/H]} = 0.12\pm0.08$\, and $\mathrm{[Ca/H]} = -0.01\pm0.10$, respectively. The interstellar extinction along the line of sight is consistent with zero, being A$_\mathrm{v} = 0.02 \pm 0.02$. The rotational period of the star is $\mathrm{P}_\mathrm{rot}=$\,\Pfbis\ in agreement with previous results by \cite{Oelkers18}, while the isochronal and gyrochronological ages are $4.7\,\pm\,4.0$\,Gyr and $5.0\,\pm\,0.2$\,Gyr, respectively. The fundamental parameters of TOI-500 are listed in Table~\ref{stellar_values}.

According to its physical properties, TOI-500\,b is a new member of a very small sample of Earth analogs with well known masses and radii. More importantly, it is the USP planet with the lowest mass and smallest radius known to date within the sample of ultra-short period planets belonging to multi-planet systems. Figure~\ref{mass_radius} shows the mass-radius diagram for USP planets with radii between $1$ and $2~\mathrm {R}_{\oplus}$ and masses $<\,10~\mathrm{M}_{\oplus}$. Most of the USP small planets (e.g., Kepler-78\,b \cite{Pepe13} and CoRoT-7\,b \cite{Leger09}), have a bulk composition comprising $\sim$30\% iron and $\sim$70\% silicates \cite{Winn18}, similar to Earth. Although, TOI-500\,b falls in this same interval of compositions, it stands out of the crowd because, after GJ-367\,b \cite{Lam21}, it is the leftmost planet in the plot, one of the smallest and lightest of its type.

Like most of the small USP planets, TOI-500\,b is expected to be a lava ocean planet \cite{Leger09}, because the close vicinity to its host star renders the surface extremely hot. Assuming the planet is a black body with zero albedo, we estimated an equilibrium temperature of $\mathrm{T}_\mathrm{eq}$ = \Teqbbis. As such, it is likely that TOI-500\,b does not have a primary atmosphere inherited from the formation process and has undergone complete photo-evaporation during the formation and evolution processes \cite{Winn18}. The extreme vicinity to the host star could also have caused the formation of a metallic (secondary) atmosphere, possibly via volcanic out-gassing, as it might have happened on 55 Cancri\,e (see \cite{2020MNRAS.498.4222T} for the estimated upper limit of atmospheric composition, and also \cite{Madhusudhan15}) and for planet Earth. A statistical analysis of the atmospheric predicted signal-to-noise ($\mathrm{S}/\mathrm{N}$) ratio of Earth-sized planets thus far discovered showed that TOI-500\,b stands among the top 10 objects, meaning it is a promising target to perform atmospheric studies with current and future instruments (see Methods and Figure~\ref{fig:snrtoi}).

Based on their minimum masses, TOI-500\,c is probably a super-Earth, while TOI-500\,d and e are more likely to be Neptunian planets \cite{Otegi20}. 
Although the three outer planets do not transit their host star, a co-planar geometry of the system is not excluded as this would imply an impact parameter $>$ 2 for planets c, d, and e,  accounting for the null detection of their transits. Dynamical simulations carried out with \texttt{rebound} \cite{Rein2012} and reboundx \cite{reboundx2020} and covering $10^8$ orbits of the USP planet, ruled out orbits with high mutual inclination. We constrained the orbital inclination for the outer two planets TOI-500\,d and e to lie between $40^{\circ}$ and $90^{\circ}$, and ruled out an inclination $\mathrm{i}_\mathrm{c}<\,30^{\circ}$ for planet c. 

Ultra-short period planets most likely did not form at their current locations, as these are often in close-in orbits within the dust sublimation radius of the proto-stellar disk \cite{Isella06}. Dynamical interactions could lead close-in super-Earths to reach high eccentricities and circularize to orbits with P < 1 day \cite{Schlaufman10}. The inward migration of USP planets is also evident through their larger orbital period ratios and larger mutual inclination, compared to longer period planets belonging to the same system \cite{Steffen13,Dai18}. How this migration occurred is still debated \cite{Lee,Petrovich19,PuLai,Millholland20}. The secular formation scenario \cite{Petrovich19,PuLai,Millholland20} suggests that the presence of several close-in planetary companions provides the dynamical interaction necessary for the migration to occur. Moreover, it requires that the outer planets have enough angular momentum deficit (see Equation~4 in \cite{Laskar97}) to launch the USP planet into an eccentric inward-migrating orbit. TOI-500 provides us with a unique opportunity to compare the predictions of different secular formation scenarios.

The bright and not very active host star allowed us to map out the inner planetary architecture of the system with hundreds of RV data points over a baseline of one year. We found that the orbital period ratio between planets b and c is about $12$, while within most \kepler\ planetary systems the two close-in planets have a period ratio lower than 4 \cite{Fabrycky12}. This higher period ratio is the hallmark of the inward migration commonly seen in other systems known to host USP planets (Figure~\ref{USP_known}). Furthermore, since the ratios of the orbital periods of the non-transiting planets are not in a first or second order but in a $1$:$4$ ($\mathrm{P}_\mathrm{c}$ and $\mathrm{P}_\mathrm{d}$) and a 3:7 ($\mathrm{P}_\mathrm{d}$ and $\mathrm{P}_\mathrm{e}$) commensurabilities, secular forces dominate the dynamical interaction between the planets.

One of the migration models for USP planets, described in \cite{PuLai} and mentioned in \cite{Petrovich19}, involves the excitation of the eccentricities to high values, leading to the formation of USP planets in highly mutually inclined orbits with respect to the outer planetary companions ($>30^{\circ}$, see \cite{PuLai} for a deep description of the model). While eccentricities are quickly damped out by tidal dissipation inside the planet on kyr to Myr timescales \cite{Winn18}, orbital inclinations are only damped by dissipation inside the host star and persist over Gyrs \cite{Winn18}.
Indeed, many USPs are observed on orbits with high mutual inclinations \cite{Dai18,Kamiaka} and our system easily accommodates itself in this model. Assuming that the USP planet TOI-500\,b emerged from the protoplanetary disc, prior to tidal migration, at an initial orbit of about 3 days (safely beyond the dust sublimation radius and Hill radius instability from planet c), we estimated that the minimum angular momentum deficit for it to undergo inward migration can be achieved as long as the outer planets currently have either eccentricities of $\sim$0.05 or mutual inclinations of $\sim$4$^{\circ}$ \cite{Petrovich19}, conditions that are well fitted by TOI-500. 

The sample of USP planets is so small that whenever we discover a new one it is important to explore which other existing migration processes might explain the final architecture. \cite{Petrovich19} described several scenarios that can justify some rare planetary system configurations involving a USP planet. One of them is the low eccentricity channel, according to which the USP planet and its companions emerge from the protoplanetary disc with relatively low orbital eccentricities, small semi-major axes, and low orbital inclinations. The eccentricities, damped by tidal forces, slowly decay towards zero in a quasi-equilibrium state, while the semi-major axes decay much more slowly than in the high-eccentricity case. In addition, since the secular behaviour remains linear, the inclination fluctuations remain small, without the large chaotic variations expected in the high-eccentricity channel. After billions of years, the system stabilizes itself with almost zero eccentricities and co-planar orbits, and it will have one small USP planet and at least one additional companion, usually a super-Earth, with an orbital period shorter than $\sim$10 days. The current architecture of TOI-500 fits well within these conditions, implying that the four planets might have undergone the low eccentricity scenario prior to fully stabilizing. With a view to provide evidence for (or against) this scenario, we performed simulations in which we tested the evolution of TOI-500 over the course of 5\,Gyrs, starting from a set of initial conditions as listed: semi-major axis $\mathrm{a}_\mathrm{b}$ between 0.02 and 0.03\,au, $e_\mathrm{b}$ and $e_\mathrm{c} = 0.05$, $e_\mathrm{d}$ and $e_\mathrm{e} \in \{0.05, 0.1,0.15,0.2,0.25\}$. When the system reached the observed configuration, the integration would stop. We found that the low eccentricity migration channel can bring TOI-500\,b to its current location in 2\,Gyrs, from an initial orbit of just beyond 0.02\,au, as long as the outer planets' eccentricities are not too low (see Methods and Figure~\ref{fig:secular}). 

Thanks to the extensive RV follow-up with \harps, TOI-500 is the first multi-planetary system with precise mass measurements comprising a small USP planet and more than one non-Jupiter-type companion for which the low eccentricity scenario has been demonstrated to predict the planetary final architecture. We emphasize that this has been done without the use of any additional assumptions. Prior to our work, \cite{PuLai} could apply a low eccentricity scenario on Kepler-10 and Kepler-290, but in the first case they needed to add another Earth-like object between planets b and c, while in the second case they assumed values for the planetary masses. TOI-500 is not the only existing system for which the low eccentricity channel may work. For instance, TOI-561 \cite{Lacedelli20, Lacedelli21} and CoRoT-7 \cite{Leger09, Queloz09} could also have migrated with the same scenario. CoRoT-7 is a two-planet system with no analogs thus far: an active star with a USP planet and a 3\,d-period Neptune. Testing the model on this system might be interesting, because it could allow predictions about the debated third planet, CoRoT-7\,d \cite{Hatzes2010, Haywood, Faria}. TOI-561 is, on the contrary, very similar to TOI-500, although planet c has a period of 11.77\,d. It could therefore be a good laboratory to test whether the theory can work, given that the orbital period of the second planet is longer than the 10\,d requirement mentioned in \cite{Petrovich19}. 

The presence of the smallest and lightest USP planet known to belong to a multi-planetary system, the close commensurability of the three outer planets, and the additional discovery of a planet with period shorter than 10 days makes TOI-500 extremely compelling and rare. The compatibility of the system with a less common secular scenario of migration discloses a new path of future exploration aiming at identifying other cases in which the low eccentricity scenario can account for the current location of USP planets. The possibility that TOI-500\,b might have a secondary atmosphere makes this system an important laboratory for future atmospheric analysis with, e.g., \texttt{ESPRESSO}, \texttt{EXPRES}, and \texttt{JWST}.

\section*{Methods}

\subsection*{Transit search}
\subsubsection*{\textit{\textbf{TESS}} observations}

\tess\ observed TOI-500 in Cycle 1 during three consecutive sectors (namely, sectors $6$, $7$, and $8$), between 11 December 2018 and 28 February 2019. The star was photometrically monitored every 2\,minutes by \tess\ camera \#3 using CCD \#4 for the first two sectors, and CCD \#3 for Sector $8$. The data from each sector has a gap of about $0.98$\,d, due to the data downlink at the satellite perigee passage. Sector 8 light curve shows an additional 1-day gap due to an instrument failure that occurred on 14 February 2019 (see Fig.~\ref{Tess_data}). 

We retrieved the \tess\ light curves of TOI-500 from the Mikulski Archive for Space Telescope (MAST) (\url{https://mast.stsci.edu}). The data products were extracted by SPOC \cite{SPOC} at NASA Ames Research Center, and include simple aperture photometry (SAP) and the so-called PDC-SAP, a systematics-corrected photometry obtained by applying to the SAP time series the Presearch Data Conditioning algorithm (PDC) developed for \kepler\ light curves (\cite{Smith2012, Stumpe2012, Stumpe2014}). SPOC conducted with its pipeline multiple transiting planet searches, which stop when the significance of a transit signal is below a given detection threshold. They produced the Data Validation Report for the combined Sector 6-8 data sets \cite{Twicken18, Li19} and published it on 4 May 2019.

The vetting team at MIT reviewed the Threshold Crossing Events (TCEs) within the Data Validation Report of TOI-500 and announced the detection of a transiting signal with a period of $\mathrm{P}_{\mathrm{orb}} \approx 0.55$\,d, a depth of about $230$\,ppm, and a duration of T$_{14} \approx 1.0$\,hour. The transiting planet candidate passed all the validation tests from the Threshold Crossing Events (TCE), such as odd-even transit depth variation and ghost diagnostic tests, which helped to rule out an eclipsing binary scenario.


\subsubsection*{\textit{\textbf{LCOGT}} observations}

We observed TOI-500 continuously for $190$\,min in Sloan $r'$ band on 3 March 2019 and again for $156$\,min on 2 May 2019 in Pan-STARRS $z$-short band from the Las Cumbres Observatory Global Telescope (\lcogt, \cite{Brown:2013}) $1$\,m network nodes at South Africa Astronomical Observatory and Cerro Tololo Inter-American Observatory, respectively. The $4096\times4096$ \lcogt\ SINISTRO cameras have an image scale of $0\farcs389$ per pixel, resulting in a $26\arcmin\times26\arcmin$ field of view. The images were calibrated by the standard \lcogt\ {\tt BANZAI} pipeline \cite{McCully:2018}, and photometric data were extracted with {\tt AstroImageJ} \cite{Collins:2017}. Using the TOI-500\,b ephemeris from the \tess\ Sector $7$ SPOC data validation report to predict transit timing, each of our observations covered full transit duration windows, and the combined transits provided $75$\,min of phase coverage before and after transit. The SPOC pipeline transit depth of $231$\,ppm is generally too shallow to be reliably detected with ground-based observations, so we intentionally saturated TOI-500 to check for possible near eclipsing binaries (NEBs) that could be contaminating the \tess\ photometric aperture that generally extends $\sim$1$\arcmin$ from the target star. To account for possible contamination from the wings of neighboring star point spread functions (PSFs), we searched for NEBs out to $2\farcm5$ from the target star. We placed apertures according to Gaia DR2 positions and proper motion. If fully blended in the SPOC aperture, a neighboring star that is fainter than the target star by $9.1$ magnitudes in \tess-band could produce the SPOC-reported flux deficit at mid-transit (assuming a 100\% eclipse). To account for possible delta-magnitude differences between \tess-band and the follow-up bands, we included an extra $0.5$ magnitudes fainter (down to \tess-band magnitude $19$, i.e., $\Delta$mag\,$\approx$\,10). Our search ruled out NEBs in all $78$ neighboring stars that meet our search criteria (Figure~\ref{field}). All the searched stars had a minimum distance from TOI-500 of $12\arcsec$. We detected no additional source at smaller distances, excluding any blending induced on TOI-500 by a contamination source.

\subsubsection*{\textit{\textbf{Super-WASP}} observations}

The field of TOI-500 was observed with \textit{WASP-South} \cite{2006PASP..118.1407P} over four consecutive years from 2008 to 2012, in each year with observing spans of $170$ nights from October to March. Nearly $26$\,$000$ photometric data points were obtained, with a typical cadence on clear nights of $15$\,min. \textit{WASP-South} was then equipped with $200$\,mm, f/1.8 lenses backed by 2048\,$\times$\,2048 CCDs and observed with a $400-700$\,nm filter \cite{2006PASP..118.1407P}. TOI-500 is $4$\,magnitudes brighter than any other star in the $48\arcsec$ photometric extraction aperture. No sign of a transiting planet was found in \textit{Super-WASP} observations due to the high impact of the instrumental noise on the data, bound to mask the shallow transits of the planets.

We searched the data for a rotational modulation using a periodogram analysis \cite{2011PASP..123..547M}. The light curve from the 2008/2009 season shows a possible periodicity near $0.022$~d$^{-1}$, corresponding to a period of about $45$\,d, with an amplitude of $2$\,mmag and an estimated false-alarm probability of 1\,\%. This is not seen in the next two years. The $\sim$45\,d period is possibly seen again in 2011/2012, once more with an amplitude of 2\,mmag and a false-alarm probability near 1\,\% (Figure~\ref{fig:wasp}). Combining all 4 years of data again produces a peak at a period of $45\pm 5$\,d (where the error allows for the modulation being incoherent), with a $1$\%\ false-alarm probability and an amplitude of $1$\,mmag.

\subsection*{High-resolution imaging}

Sources that are not detected in seeing-limited photometry or by \gaia\ can lead to photometric contamination of the \tess\ light curve of TOI-500. Dilution of the transit depth can lead to an underestimated planet radius, or can make astrophysical false positives appear planetary in nature \cite{Ciardi2015}. We thus searched for nearby stellar companions using $4$ and $8$\,m class telescopes, providing robust limits on the presence of companions and the level of photometric dilution.

\subsubsection*{Lucky imaging with Gemini/Zorro}

On 16 March 2020, TOI-500 was observed using the Zorro speckle imager \cite{Scott2019}, mounted on the $8.1$\,m Gemini South telescope in Cerro Pachón, Chile. Zorro uses high speed electron-multiplying CCDs (EMCCDs) to simultaneously acquire data in two bands centered at $562$\,nm and $832$\,nm. The data were collected and reduced following the procedures described in \cite{Howell2011}. The resulting reconstructed image achieved a contrast of $\Delta\mathrm{mag}=7.4$ at a separation of $1\arcsec$ in the $832$\,nm band, without showing any contamination resource (see bottom panel of Figure~\ref{fig:zorro}).

\subsubsection*{High contrast imaging with SOAR/HRCam}

On 18 May 2019, TOI-500 was observed in $I$ band with a pixel scale of $0.01575\arcsec$\,pix$^{-1}$ using the HRCam imager, mounted on the $4.1$\,m Southern Astrophysical Research (SOAR) telescope at Cerro Tololo Inter-American Observatory, Chile. The data were acquired and reduced following the procedures described in \cite{Tokovinin2018} and \cite{Ziegler2020}. The resulting reconstructed image achieved a contrast of $\Delta\mathrm{mag}=7.2$ at a separation of $3\arcsec$ (see top panel of Figure~\ref{fig:soar}). The Zorro inner working angles of $17$ and $28$\,mas (at $562$\,nm and $832$\,nm respectively) yield spatial limits at the star of $\sim$1\,au (for d\,=\,47\,pc), near the orbital period semi-major axis of the outer planets. Any source within the speckle spatial limits cannot be a massive star because we would have detected it. A similar companion would have been disrupted long ago. Theoretical studies have shown that a close companion can truncate protoplanetary disks and newly forming planets \cite{Martin14, Jangcondell15} or disperse the disk before planets even begin to form \cite{Cieza09, Kraus12}. The obtained image contrast eliminates all other possible companions outside the inner working angle down to $\sim$M5V and out to a distance of $\sim$56\,au (at $1.2 \arcsec$).

\subsection*{\textit{\textbf{HARPS}} RV observations}
We observed TOI-500 with the High Accuracy Radial velocity Planet Searcher (\harps, \cite{Mayor2003}) spectrograph mounted at the ESO-$3.6$\,m telescope of La Silla Observatory, Chile. Between 22 March 2019 and 23 March 2020, we collected 197 high-resolution spectra with a resolving power of R=$\lambda/\Delta\lambda \approx 115\,000$, as part of the observing programs 1102.C-0923, 0103.C-0874, and 60.A-9709. We monitored the sky background using the second fibre of the instrument and set the exposure time to $900-2100$\,sec depending on sky conditions and constraints of the observing schedule. Given the relatively short orbital period of the transiting candidate, we adopted a multi-visit observing strategy, i.e., we acquired at least two spectra per night separated by at least one hour in most of the observing nights.

We reduced the data using the dedicated \harps\ Data Reduction Software (DRS) and computed the cross-correlation function (CCF) for each spectrum using a K5 numerical mask \cite{Baranne1996,Pepe2002a,Lovis2007}. We used the DRS to extract the full width at half maximum (FWHM) and the bisector inverse slope (BIS) of the CCF. We measured the Ca\,{\sc ii} H\,\&\,K lines activity indicator (S-index) using the code \texttt{TERRA} \cite{Anglada12}. We finally extracted differential RV measurements from the \harps\ spectra using the code \texttt{SERVAL} \cite{Zechmeister18}, which employs a template-matching algorithm specifically designed to derive precise radial velocities from high-resolution Echelle spectra of late K- and M-type dwarfs. The code provides also an additional activity indicator, namely, the differential line width (dLW) \cite{Zechmeister18}.

The \harps\ SERVAL RV measurements and their uncertainties are listed in Table~\ref{HARPS_data}, along with the FWHM, BIS, S-index, dLW, exposure time, and signal-to-noise ($\mathrm{S}/\mathrm{N}$) ratio per pixel at $550$\,nm. Time stamps are given in barycentric Julian date in barycentric dynamical time (BJD$_\mathrm{TDB}$).

\subsubsection*{Frequency analysis of the \textit{\textbf{HARPS}} RVs and activity indicators}

We computed the generalized Lomb-Scargle (GLS) \cite{Zechmeister09} periodogram of the \harps\ RVs and used the pre-whitening technique \cite{Hatzes2010,Gandolfi2017} to subsequently identify significant peaks and remove the corresponding periodic signals from the Doppler time series. We performed a least-squares sine-fit to the amplitude and phase at the dominant frequency found by the GLS periodogram and subtracted the fit from the \harps\ data. We iterated the process to identify the next most dominant frequency in the GLS periodogram of the RV residuals. We stopped the iteration once we reached the level of the noise and considered as significant only those peaks whose FAP is lower than $0.1$\,\%. Following the Monte Carlo bootstrap method \cite{Murdoch1993}, we estimated the FAP by computing the GLS periodograms of $10^6$ mock data sets obtained by randomly shuffling the RV measurements, while keeping the observation time-stamps fixed. We defined the FAP as the fraction of those periodograms whose highest power exceeds the power of the original observed data at any frequency.

Figure~\ref{periodogram_planet} displays the GLS periodograms of the \harps\ RV measurements and residuals. We found a very significant ($\mathrm{FAP} <0.0001$\,\%, no false positives out of $10^6$ trials, implying a $\mathrm{FAP}< 10^{-6}$) peak at $\mathrm{f}_\mathrm{d} =0.038$\,d$^{-1}$, corresponding to a period of $\mathrm{P}_\mathrm{d} = 26.3$\,d (Figure~\ref{periodogram_planet}, upper panel). The peak is surrounded by a series of equally-spaced secondary peaks separated by about $0.0035$\,d$^{-1}$, which are aliases of the dominant frequency at $0.038$\,d$^{-1}$, resulting from the window function (Figure~\ref{periodogram_planet}, bottom panel).

We performed a least-squares sine-fit to the amplitude and phase at $\mathrm{f}_\mathrm{d}$ and subtracted the best fit from the RV time series. The second panel of Figure~\ref{periodogram_planet} displays the periodogram of the residuals, following the subtraction of the signal at $\mathrm{f}_\mathrm{d}$. The most significant peak is found at $\mathrm{f}_\mathrm{c} = 0.151$\,d$^{-1}$ (FAP\,$<$\,0.1\,\%), which corresponds to a period of $\mathrm{P}_\mathrm{c} = 6.6$\,d. As in the previous case, the peak is surrounded by a series of equally-spaced aliases. We iterated the pre-whitening process and found two additional significant ($\mathrm{FAP}< 0.1$\,\%) peaks at $\mathrm{f}_\mathrm{e} = 0.016$\,d$^{-1}$ ($\mathrm{P}_\mathrm{e} = 60.7$\,d; third panel) and $\mathrm{f}_{\star} = 0.023$\,d$^{-1}$ ($\mathrm{P}_{\star} = 43.4$\,d; fourth panel). 

The periodograms in Figure~\ref{periodogram_planet} (right panels) show the presence of a Doppler signal at $\mathrm{f}_\mathrm{b} = 1.824$\,d$^{-1}$ ($\mathrm{P}_\mathrm{b} = 0.55$\,d) -- the transiting frequency detected in the \tess\ light curve -- whose power steadily increases once the other signals are gradually removed from the \harps\ time series. This peak is associated to the Doppler reflex motion induced by the USP planet, confirming the planetary nature of the transit signal detected in \tess\ data. Once the 4 signals at $\mathrm{f}_\mathrm{c}$, $\mathrm{f}_\mathrm{d}$, $\mathrm{f}_\mathrm{e}$, and $\mathrm{f}_{\star}$ are removed from the \harps\ time series (Figure~\ref{periodogram_planet}, fifth panel), the peak becomes significant ($\mathrm{FAP}\,< 0.1$\,\%).

The periodograms of the FWHM, dLW, and S-index shows a significant (FAP\,<\,0.1\,\%) excess of power at frequencies lower than the inverse of the baseline of our measurements. An inspection of the time series unveils the presence of a long term variation of the activity level (Figure~\ref{RV_data}), visible as an offset between the two observing seasons and likely associated to magnetic cycles. For each activity indicator, we accounted for the long term variation by subtracting the seasonal median values. Figure~\ref{activity indexes} displays the periodograms of the median-corrected activity indicators. The FWHM, dLW, and S-index show significant peaks (FAP\,<\,0.1\,\%) between 38 and 42 days, providing strong evidence that the signal at 43.4\,d seen in the HARPS RV residuals is due to stellar activity. As the same signal is also significantly detected in the \textit{Super-WASP} photometry (Fig.~\ref{fig:wasp}), we conclude that the stellar rotation period is 40-45 days. The signal detected in the HARPS RVs and activity indicators is associated to the presence of active regions appearing and disappearing on the visible stellar hemisphere as the star rotates about its axis. 

We note that the periodograms of the activity indicators do not show any significant peaks at $\mathrm{f}_\mathrm{b}$, $\mathrm{f}_\mathrm{c}$, $\mathrm{f}_\mathrm{d}$ and $\mathrm{f}_\mathrm{e}$, i.e., the frequencies detected in the \harps\ RVs, providing solid evidence that those signals are not associated to stellar activity, but they are rather induced by TOI-500\,b and three additional non-transiting planets with periods of $\sim 6.6$, $26.2$, and $61.5$\,d.

\subsubsection*{Fundamental stellar parameters}

We determined the stellar fundamental parameters of TOI-500 using the spectral analysis package Spectroscopy Made Easy (\texttt{SME}, version 5.2.2, \cite{Valenti96, Valenti17}). We performed the analysis of the co-added \harps\ spectrum -- which has an $\mathrm{S}/\mathrm{N}\sim 900$ in the continuum per pixel at $5500$\,\AA\ -- with the {\tt{MARCS}} model spectra (\url{https://marcs.astro.uu.se/}). \cite{MARCS} and the line lists from the Vienna atomic line database (VALD, \url{http://vald.astro.uu.se} \cite{Ryabchikova15}). The adopted methodology is the same as described in \cite[]{Fridlund17} and \cite{Persson18}. We measured the effective temperature $\mathrm{T}_\mathrm{eff}$ from the wings of the $\mathrm{H}_\alpha$ and $\mathrm{H}_\beta$ lines, and the surface gravity $\log \mathrm{g}_{\star}$  from the Ca and Mg\,b triplets around $6100$ and $5100$\,\AA, respectively. We derived the stellar projected rotational velocity $\mathrm{v}_{\star}\sin \mathrm{i}_{\star}$ and the iron relative abundance [Fe/H] from the narrow unblended iron lines between $6000$ and $6600$\,\AA. We fixed the micro- and macro-turbulent velocities using the values provided by the calibration equations of \cite{Gray84} ($\mathrm{v}_{\mathrm{mic}} = 0.5$\,km\,s$^{-1}$ and $\mathrm{v}_{\mathrm{mac}} = 1.5$\,km\,s$^{-1}$, respectively). We did not adopt the more recent equations from \cite{Bruntt10} and \cite{Doyle} as they are valid only for early K, G, and late F-type stars, while they are not calibrated for late K-type dwarf such as TOI-500. We checked our best-fitting model spectrum using the Na doublet at $5888$ and $5895$\,\AA. We found an effective temperature of $\mathrm{T}_\mathrm{eff}= 4440 \pm 100$\,K, a surface gravity of \mbox{$\log \mathrm{g}_{\star} = 4.50 \pm 0.06$\,(cgs)}, an iron and calcium abundance of $\mathrm{[Fe/H]} = 0.12 \pm 0.08$ and $\mathrm{[Ca/H]} = -0.01 \pm 0.10$, and a projected rotational velocity of \mbox{$\mathrm{v}_{\star}\sin \mathrm{i}_{\star} = 1.1 \pm 0.7$\,km\,s$^{-1}$}.

As a sanity check, we conducted an independent spectroscopic analysis of the co-added \harps\ spectrum and employed the code \texttt{SpecMatch-emp} \cite{Yee2017}, which utilizes hundreds of Keck/HIRES high-resolution template spectra of FGK stars whose effective temperature, iron content, and stellar radius have been accurately measured via interferometry,  asteroseismology, spectral synthesis, and spectrophotometry. We found an effective temperature of T$_\mathrm{eff}\,=\,4400\,\pm\,70$\,K, a stellar radius of $\mathrm{R}_{\star}\,=\,0.71\,\pm\,0.10~\mathrm{R}_{\odot}$, and an iron content of $\mathrm{[Fe/H]}\,=\,0.07\,\pm\,0.09$, with the effective temperature and iron abundance in excellent agreement with the parameter estimates with \texttt{SME}.

We measured the interstellar extinction along the line-of-sight to TOI-500 using the method described in \cite{Gandolfi2008}. Briefly, we built the spectral energy distribution (SED) of the star using the UBVRI \cite{Mermilliod1987} optical and JHKs \cite{cutri2003} near-infrared photometry and fitted the SED using the \texttt{BT-Settl-CIFIST} \cite{Baraffe2015} model spectrum with the same spectroscopic parameters  as the star. We adopted the extinction law of \cite{Cardelli1989} and assumed a total-to-selective extinction of $\mathrm{R}_\mathrm{v}=\mathrm{A}_\mathrm{v}/E(B-V) = 3.1$. We found that the interstellar reddening is negligible and consistent with zero ($\mathrm{A}_\mathrm{v} = 0.02 \pm 0.02$), as expected given proximity of the star ($\mathrm{d} \approx 47.39$\,pc; \cite{GaiaDR2}).
 
We combined the effective temperature and iron abundance determined  with \texttt{SME} with the \gaia\ DR2 parallax \cite{GaiaDR2} and the apparent $\mathrm{V}$-band magnitude\cite{Mermilliod1987} of $\mathrm{V} = 10.530$  to compute the stellar mass, radius, and age using the Bayesian web-tool {\tt{PARAM\,1.3}} \cite[\url{http://stev.oapd.inaf.it/cgi-bin/param_1.3}][]{daSilva06}. We added $0.06$\,mas to the nominal \textit{Gaia}’s parallax, to account for the systematic offset found by \cite{Stassun2018a} and \cite{Zinn2019}, and we assumed an uncertainty of $0.05$ on the apparent $\mathrm{V}$-band magnitude. We found a stellar mass of $\mathrm{M}_{\star}= 0.740\,\pm\,0.017~\mathrm{M}_{\odot}$, a stellar radius of $\mathrm{R}_{\star}\,=\,0.678\,\pm\,0.016~\mathrm{R}_{\odot}$, implying a surface gravity of $\log \mathrm{g}_{\star} = 4.618\,\pm\,0.017$ (cgs) in agreement within $2 \sigma$ with the spectroscopic value. We also found that the stellar radius agrees with the value derived using \texttt{SpecMatch-emp}, corroborating our analysis. We finally used the formalism as in \cite{Mamajek08} to estimate through gyrochronology the stellar age, which resulted to be 5.0\,$\pm$\,0.2~Gyr. This value agrees with the age of $4.7\,\pm\,4.0$~Gyr inferred with {\tt{PARAM\,1.3}}.

We also determined the local standard of rest (LSR) U, V and W space velocities of the parent star using the methods of \cite{2006MNRAS.367.1329R}, and, from these velocities, we computed the probability that TOI-500 belongs to the galactic thin disk, thick disk, or halo stellar population. Using the \gaia\ DR2 astrometric measurements of location, parallax, proper motion, and radial velocity, we derived the velocities reported in Table~\ref{stellar_values} and the following probabilities:
\begin{equation}
\mathrm{P}_\mathrm{thin} = 0.92963 \pm 0.00929 \\
\mathrm{P}_\mathrm{thick} = 0.07010 \pm 0.00002\\
\mathrm{P}_\mathrm{halo} = 0.0002767 \pm 0.0000008
\end{equation}
These values are in good agreement with kinematic membership probabilities computed independently by \cite{2020MNRAS.491.4365C}. The kinematic membership of TOI-500 in the galactic thin disk is consistent with our derived $\mathrm{[Fe/H]}$ and $\mathrm{[Ca/H]}$ values which are typical for galactic thin disk stars. The main results are summarized in Table \ref{stellar_values}.

\subsection*{Independent transit search}
\label{searching_transit}
To confirm the presence of the $0.55$-d transiting candidate announced by the \tess\ team and in order to search for additional candidates, we independently searched the \tess\ light curve for transit signals. We carried out our analysis using three different detrending algorithms and methods, as described in the paragraphs below.\\
\emph{Method 1}: We detrended the PDC-SAP light curve filtering out stellar activity and instrumental systematics with a Savitzky-Golay filter \cite{Savitzky1964} and we searched the time series for transit signals using the \texttt{DST} algorithm \cite{Cabrera2012}. While we confirmed the presence of the transiting candidate at $0.55$\,d, we did not identify any additional signal.\\
\emph{Method 2}. We also searched the light curve using the detection pipeline \texttt{EXOTRANS}. It combines a wavelet-based filter \texttt{VARLET} to remove discontinuities and stellar variations \cite{Grziwa16} and the advanced \texttt{BLS} \cite{Kovacs16}, which incorporates \texttt{PHALET}. The latter removes previously detected transits and searches the light curve for additional periodic events. \texttt{EXOTRANS} can detect multiple transits or transits masked by other strong periodic events (systematics, background binaries). We detected the transit signal at 0.55\,d. No significant additional transit signals were detected.\\ 
\emph{Method 3}. We detrended the \tess\ light curve with the code \texttt{wotan} \cite{Hippke19a}, which implements different detrending techniques. We chose to detrend the PDC-SAP light curve applying a cubic spline coupled to a sigma clipping algorithm, a well known methodology for removing activity trends \cite{Burt20, Wong20, Ikwut20}. We also applied the Transit Least Square method (TLS, \cite{Hippke19b}) to search for transits. 
\texttt{TLS} uses the transit model from Mandel \& Agol 2002 \cite{Mandel02} with the quadratic limb-darkening law. We fixed the limb darkening coefficients of TOI-500 to the values extracted from the \tess\ archive. We confirmed the USP transiting planet candidate and did not detect any additional transit signals.

The upper panel of Figure~\ref{Tess_data} shows the median-normalized PDC-SAP light curve of TOI-500 (black points) and the spline used to detrend the \tess\ data (red line) following \emph{Method 3}. The lower panel shows the corresponding detrended \tess\ light curve. The in-transit data points are highlighted with blue circles.

\subsection*{Joint analysis of the \textit{\textbf{TESS}} and \textit{\textbf{HARPS}} data}
We performed the joint analysis of the \tess\ transit photometry and \harps\ RV measurements using the code \texttt{pyaneti} \cite{Barragan19}, which generates posterior distributions of the fitted parameters using Markov chain Monte Carlo (MCMC) simulations coupled to a Bayesian framework. \texttt{pyaneti} uses the limb-darkened quadratic model from Mandel \& Agol 2002 \cite{Mandel02} for fitting the transit light curve. The code follows the $\mathrm{q}_1$ and $\mathrm{q}_2$ parametrization of the linear and quadratic limb darkening coefficients $\mathrm{u}_1$ and $\mathrm{u}_2$ as described in \cite{Kipping13}, and the parametrization of $\mathrm{e}$ and $\mathrm{\omega}$ proposed by \cite{Anderson11}. 

We used the PDC-SAP light curve, which was detrended following the procedure described in \emph{Method 3} in the previous section. We set Gaussian priors on $\mathrm{q}_1$ and $\mathrm{q}_2$ using the limb darkening coefficients derived by \cite{Claret17} for the \tess\ passband, imposing a conservative $1 \sigma$ uncertainty of $0.1$ on both the parameterized limb-darkening coefficients $\mathrm{q}_1$ and $\mathrm{q}_2$. A preliminary analysis showed that the transit light curve poorly constrains the scaled semi-major axis ($\mathrm{a}/\mathrm{R}_{\star}$) of planet b, owing to the shallowness of the transit signal.  We therefore constrained $\mathrm{a}/\mathrm{R}_{\star}$ using Kepler’s third law, the orbital period, and a Gaussian prior on the stellar density based on the derived stellar mass and radius. For the other transit parameters, we assumed uniform priors as reported in Table~\ref{table_prior}.

The RV model follows the results presented in the frequency analysis' section. Briefly, \texttt{pyaneti} accounts for the Doppler reflex motion of the 4 planets using Keplerian models. We modelled the RV stellar signal at the star’s rotation period as an additional coherent sine-like curve. We accounted for any variation not properly modelled by the coherent sine-curve, and/or any instrumental noise not included in the nominal RV uncertainties, by fitting for a Doppler jitter term. We adopted uniform priors for all the RV parameters, as summarized in Table~\ref{table_prior}.

We explored the parameter space with 500 chains, 500 iterations and a chains thin factor of 10 and tested for convergence with the Gelman-Rubin statistics. If the chains do not converge, \texttt{pyaneti} restarts new cycles of 5000 steps (500 iterations multiplied for the thin factor). We  produced the posterior distributions from the last set of $2.5\times 10^5$ samples, once the chains reached convergence. The inferred parameter estimates are the medians of the corresponding posterior distributions, while the associated uncertainties are defined as the $68$\,\% region of the distribution credible interval. The results are reported in the fourth column of Table~\ref{table_prior}.

Figure~\ref{planetb_phase_folded} shows the phase-folded transit light curve of TOI-500\,b along with the best-fitting model. The first row of Figure~\ref{Sinusoidal} displays the \harps\ \texttt{SERVAL} RV time series and the best-fitting Doppler model (stellar signal + 4 planets). The second, third, and fourth rows of Figure~\ref{Sinusoidal} shows the phase-folded \harps\ \texttt{SERVAL} Doppler measurements and the best-fitting models for TOI-500 b, c, d, e, and the star.

As a sanity check, we also estimated the Doppler reflex motion induced by the transiting planet using the floating chunk offset method described in \cite{Hatzes2011}. This technique is effective at measuring the mass of USP planets, while filtering out the long term RV variation induced by stellar activity and long period objects \cite{Gandolfi2017,Barragan2018}. Briefly, we divided the HARPS RVs into subsets of nightly measurements and analyzed only those radial velocities for which multiple measurements were acquired on the same night. The best-fitting orbital solution of TOI-500\,b was found using a Gaussian prior on the transit ephemeris -- as derived from the modelling of the transit light curve -- while allowing the RV semi-amplitude variation and nightly offsets to vary. We found a semi-amplitude of K$_\mathrm{b}\,=\,1.38\,\pm\,0.20$\,m\,s$^{-1}$, in very good agreement with the value reported in Table~\ref{table_prior}. 

\subsection*{Multi-dimensional Gaussian process analysis}

In order to study the influence of stellar activity in the \harps\ RV measurements, we proceeded to analyze our data-set using a multi-dimensional Gaussian Process (GP) approach \cite{Rajpaul} as implemented in \texttt{pyaneti} \cite{Barragan19,pyaneti2}. This approach has been useful to distinguish planet and activity induced RV signals using stellar activity indicators \cite{Barragan19b}. 

Given the evidence of multi-signals in the RVs, we first ran a multi-dimensional GP model of the time series of the FWHM and S-index activity indicators to characterize the scales of the star-induced signal. Following \cite{Rajpaul}, we assumed that the FWHM and S-index time series can be modelled as

\begin{equation}
    \begin{matrix}
    \Delta {\rm FWHM} & =  & \mathrm{F}_\mathrm{c}\, \mathrm{G}(\mathrm{t}), \\
    \var{\Delta {\rm S-index}} & = & \mathrm{S}_\mathrm{c}\, \mathrm{G}(\mathrm{t}), \\
\end{matrix}
\end{equation}

respectively. $\mathrm{S_c}$ and $\mathrm{F_c}$ are free parameters, which relate the time series to a GP-drawn function $\mathrm{G(t)}$ that describes the area covered by active regions on the stellar surface as a function of time. We created our co-variance matrix \cite{Rajpaul,pyaneti2} using the quasi-periodic kernel 

\begin{equation}
    \gamma(\mathrm{t}_\mathrm{i},\mathrm{t}_\mathrm{j}) = \exp 
    \left[
    - \frac{\sin^2[\pi(\mathrm{t}_\mathrm{i} - \mathrm{t}_\mathrm{j})/\mathrm{P}_{\rm GP}]}{2 \lambda_{\rm P}^2}
    - \frac{(\mathrm{t}_\mathrm{i} - \mathrm{t}_\mathrm{j})^2}{2\lambda_{\rm e}^2}
    \right],
    \label{eq:gamma}
\end{equation}

where $\mathrm{P}_{\rm GP}$ is the GP characteristic period, $\lambda_\mathrm{P}$ the inverse of the harmonic complexity, and $\lambda_\mathrm{e}$ is the long term evolution timescale. We created the residual vector for the GP regression by subtracting a constant offset to each activity indicator \cite{pyaneti2}.

We ran an MCMC sampling using \texttt{pyaneti}. We set wide Jeffreys priors on the multi-dimensional GP hyperparameters $\lambda_{\rm e} \in [50,250]$~d and $\lambda_{\rm P} \in [0.1,10]$, and a wide uniform prior on $\mathrm{P}_{\rm GP} \in [30,70]$~d. Figure~\ref{fig:gpstimeseries} shows the S-index and FWHM time series along with the GP model. The inferred values for the hyperparameters are  $\lambda_{\rm e} = 55.7^{+36.2}_{-21.2}$~d, $\lambda_{\rm P} = 1.5^{+1.0} _{-0.5}$, $\mathrm{P}_{\rm GP} = 43.1^{+2.8} _{-2.6}$~d. These results suggest that the signal has low harmonic complexity and that the rotation period of the star is close to $43$\,d, in agreement with the periodogram analysis of the RV measurements and activity indicators, and the results obtained with the \textit{Super-WASP} photometry. The low harmonic complexity of the process describing the stellar signal in the activity indicators suggests that the stellar signal in the RVs has a relatively low harmonic complexity too, and a quasi-sinusoidal behavior \cite[see][for more details]{pyaneti2}.

We proceeded to perform an analysis including light curve and RV time-series to characterize the planetary signals. For the transit modelling we followed the same approach as the one described in the previous section. We included the RV data in the multi-dimensional GP set-up, together with the FWHM and S-index activity indicators. We assumed that the RVs, FWHM, and S-index time series can be modelled as \cite{Rajpaul}

\begin{equation}
    \begin{matrix}
    \Delta {\rm RV} & = & \mathrm{V}_\mathrm{c}\, \mathrm{G}(\mathrm{t}) + \mathrm{V}_\mathrm{r}\, \dot{\mathrm{G}}(\mathrm{t}), \\
    \Delta {\rm FWHM} & = &  \mathrm{F}_\mathrm{c}\, \mathrm{G}(\mathrm{t}), \\
    \Delta \var{{\rm S-index}} & = &  \mathrm{S}_\mathrm{c}\, \mathrm{G}(\mathrm{t}), \\
\end{matrix}
\end{equation}

where $\dot{\mathrm{G}}(\mathrm{t})$ corresponds to the time derivative of the GP-drawn function $\mathrm{G(t)}$, while $\mathrm{V}_\mathrm{c}$ and $\mathrm{V}_\mathrm{r}$ are the amplitudes of $\mathrm{G(t)}$ and $\dot{\mathrm{G}}(\mathrm{t})$, respectively. The use of $\mathrm{\dot{G}(t)}$ in the modelling is needed to trace the active regions motion on the stellar surface \cite{Rajpaul,pyaneti2}. We created the co-variance matrix in a similar way as in the previous case, using the kernel in Eq.~(\ref{eq:gamma}). We also created the residual vector for the GP regression by subtracting 4 RV orbits accounting for Doppler signal induced by the 4 planets. For the innermost planet we set uniform priors based on the transit ephemeris (Table~\ref{table_prior}). For the other parameters of the Keplerian signals, we used wide uniform priors with period ranges based on the periodogram analysis. We note that the flexibility of this model was not able to provide a good fit for the time of inferior conjunction $\mathrm{T}_{0, \mathrm{e}}$ for TOI-500\,e. We therefore created a weakly informative Gaussian prior based on our periodogram analysis (Table~\ref{table_prior}). For the activity indicators we subtracted a constant offset \cite{pyaneti2}. We ran an MCMC sampling, with the same priors as before, except for the GP characteristic period, where we set $\mathrm{P}_{\rm GP} \in [35,50]$~d to speed up the convergence. Table~\ref{table_prior} shows the priors and sampled parameters used in this analysis.

Our modelling including the RVs, FWHM, and S-index recovers the Keplerian signals with periods of $0.55$, $6.64$, $26.23$, and $61.30$ days with Doppler semi-amplitudes of K$_\mathrm{b}$\,=\,\kb, K$_\mathrm{c}$\,=\,\kc, K$_\mathrm{d}$\,=\,\kd, and K$_\mathrm{e}$\,=\,\ke, respectively. These values, listed in the second column of Table~\ref{table_prior}, are consistent within \textbf{1\,$\sigma$} with those obtained using the sinusoidal model for the stellar signal presented in the previous section. The only exceptions are the orbital period and the RV semi-amplitude induced by the outer planet (TOI-500\,e), that differ by 2.2 and 1.7\,$\sigma$, respectively (here $\sigma$ is defined as the sum in quadrature of the respective uncertainties).

As discussed before, we note that the inferred process describing the stellar RV signal has a low harmonic complexity, a relatively low amplitude $\mathrm{V}_\mathrm{c} =$ \jAzero\,\ms, and a low amplitude regulating the GP derivative $\mathrm{V}_\mathrm{r} = $\,\jAuno\,m\,s$^{-1}$\,d$^{-1}$. The low harmonic complexity level behaviour is in agreement with the discussion made in \cite{pyaneti2}. Figure~\ref{fig:gpstimeseries} displays the time-series data sets together with the inferred models. Figure~\ref{fig:gpstimeseries} shows how the activity induced signal is well constrained by the activity indicators. Our analysis shows how in the low harmonic regime, stellar signals behave as sinusoidal curves. Since the two analyses provide consistent results, we decided to adopt the results obtained using the simplest sinusoidal model.

\subsection*{Stability analysis}

We carried out a set of dynamical simulations considering the gravitational interaction and the effect from general relativity (GR) to study the long-term stability of the system and check if some of the parameters, in particular those of the non-transiting planets, can be refined since they have no upper mass constraints and only lower mass limits from RVs. We took the stellar mass and radius reported in Table~\ref{stellar_values}, the planetary parameters listed in Table~\ref{table_prior}, and drew hundreds of samples from the parameters posteriors as initial parameters for the dynamical simulation. We used \texttt{rebound} \cite{Rein2012} with the standard IAS15 integrator \cite{Rein2015} to integrate over $10^8$ orbits for the inner planet, which correspond to a time span of $\sim$150\,kyr. The effect from GR was included via \texttt{reboundx} \cite{reboundx2020}. We could not explore a longer time span due to the highly multi-dimensional parameter space. We studied the parameter space for the non-transiting planets in more detail by drawing the true planetary masses from the reported minimum masses ($\mathrm{m}\sin \mathrm{i}$) listed in Table~\ref{table_prior}, allowing for inclinations between $0^{\circ}$ and $90^{\circ}$, where an inclination of $90^{\circ}$ means that the system in seen edge-on. We found that stable systems can exist for inclinations of the outer two planets TOI-500\,d and e between $40^{\circ}$ and $90^{\circ}$, and we could exclude orbits with inclinations $\mathrm{i}\,<\,30^{\circ}$ for planet c. Since the inner planet might have formed by inward migration via secular interaction if the outer planets form a typical Kepler multi-planet system (low eccentricities and small mutual inclinations), we studied the domain of small mutual inclinations ($\sim 4^{\circ}$) in more detail. In particular, we carried out dynamical simulations using the Stability of Planetary Orbital Configurations Klassifier (\texttt{SPOCK}, \cite{Tamayo2020a}), which can be used to study the stability of multi-planet systems with at least three planets and with a maximum mutual inclination of $\sim$11$^{\circ}$. Since this is the case for the task at hand, we employed \texttt{SPOCK}, which is much faster than \texttt{rebound}, allowing us to study a much longer time span. The speed-up in \texttt{SPOCK} is reached through the machine learning technique that is used to train stability classifications. First, numerical integration for the first $10^4$ orbits are carried out and then \texttt{SPOCK} predicts the stability of the system over $10^9$ orbits of the inner planet. We sampled the inclination of the transiting planet from the posterior solution in Table~\ref{table_prior} and calculated, depending on the sampled inclination, the inclination for the non-transiting planets taking into account the maximal allowed mutual inclination of 11$^{\circ}$. We took the values for the other parameters and for the USP planet from Table~\ref{stellar_values} and Table~\ref{table_prior}, and we drew $3\times10^4$ samples from the parameter posteriors as initial parameters for the simulation. We found that the system is stable for the whole parameter posterior space for $10^9$ orbits of the inner planet, which corresponds to a time span of $\sim$1.5\,Myr.

\subsubsection*{Low eccentricity migration process}
Once all planets have formed from the protoplanetary disk and the disk is dispersed, the system undergoes a phase of secular migration, during which the eccentricities are damped and the inner planet migrates towards the star, until it reaches its final orbit. The starting point for this migration would be a system of super-Earths and/or mini-Neptunes, from which the inner planet will become detached. A possible channel for the migration of USP systems accounts for an initial low eccentricity of the USP planet and its companions. To study this, we integrate the equations of motion for the secular evolution of the planets' eccentricities. We incorporate linear secular planet--planet interactions \cite{MD99}, which affect planets' eccentricities $e_j$ and longitudes of pericentre $\varpi_j$ through
\begin{equation}
    \dot{h}_j = \frac{1}{n_ja_j^2}
    \frac{\partial\mathcal{R}_j}{\partial k_j}, \quad
    \dot{k}_j = -\frac{1}{n_ja_j^2}
    \frac{\partial\mathcal{R}_j}{\partial h_j},
\end{equation}
where $h_j=e_j\sin\varpi_j$ and $k_j=e_j\cos\varpi_j$. Here, $n_j$ is each planet's mean motion and $a_j$ its semimajor axis. $\mathcal{R}_j$ is the standard secular disturbing function expanded to second order in eccentricities:
\begin{equation}
    \mathcal{R}_j = n_ja_j^2\left[\frac{1}{2}A_{jj}\left(h_j^2+k_j^2\right)
    +\sum_{k=1,k\ne j}^4
    A_{jk}\left(h_j h_k+k_j k_k\right)\right],
\end{equation}
where the matrix elements $A_{jj}, A_{jk}$ can be found in reference \cite{MD99}. We also incorporate general relativistic precession on all planets \cite{PuLai}:
\begin{equation}
    \dot{\varpi}_j = \frac{3GM_\star n_j}{c^2a_j},
\end{equation}
$c$ being the speed of light and $G$ the gravitational constant. On the innermost planet we also include terms for tidal precession \cite{PuLai}:
\begin{equation}
    \dot{\varpi}_\mathrm{b} = \frac{15k_{2,\mathrm{b}}n_\mathrm{b}}{2}
    \frac{M_\star}{m_\mathrm{b}}
    \left(\frac{R_\mathrm{b}}{a_\mathrm{b}}\right)^5,
\end{equation}
where $k_{2,\mathrm{b}}=1$ is the tidal Love number, and for decay of eccentricity and semimajor axis through tidal forces raised on the planet \cite{Jackson08}:
\begin{equation}
    \dot{e}_\mathrm{b} = -\frac{63\sqrt{GM_\star^3}R_\mathrm{b}^5}
    {4Q_\mathrm{b}m_\mathrm{b}a_\mathrm{b}^{13/2}}e_\mathrm{b},\quad
    \dot{a}_\mathrm{b} = -\frac{63\sqrt{GM_\star^3}R_\mathrm{b}^5}
    {2Q_\mathrm{b}m_\mathrm{b}a_\mathrm{b}^{13/2}}e_\mathrm{b}^2a_\mathrm{b},
\end{equation}
where $Q_\mathrm{b}=100$ is the (constant) tidal quality factor (for comparison, for Earth $Q/k \sim 1000$ \cite{Ray01}, for the Moon $Q/k \sim 1600$ \cite{Williams14}, for Mars $Q/k \sim 600$ \cite{Jacobson14} and for Io $Q/k \sim 70$ \cite{Lainey09}). These equations are converted to $\left(h_j,k_j\right)$ coordinates and integrated with the \texttt{scipy} Dormund--Price integrator \cite{2020SciPy-NMeth}.

We explored a range of initial semi-major axes for TOI-500\,b from 0.02 to 0.03 au, and initial eccentricities $e_\mathrm{b} = 0.05$, $e_\mathrm{c} = 0.05$, $e_\mathrm{d} \in \{0.05, 0.1\}$, $e_\mathrm{e} \in \{0.05, 0.1,0.15,0.2,0.25\}$. Initial longitudes of pericentre for the planets were set to 0, 90, 180 and 270 deg. The equations were integrated for a time of 5 Gyr, or until the semimajor axis of TOI-500\,b attained its current value.

We found that migration to the present location is possible within 5\,Gyr so long as the initial eccentricities of the planets are sufficiently high ($e_\mathrm{e}\gtrsim0.2$ when $e_\mathrm{b}=e_\mathrm{c}=e_\mathrm{d}=0.05$). When the initial $a_\mathrm{b} \gtrsim0.25$ au, the eccentricities of the two innermost planets can be excited high enough for their orbits to overlap. This renders the secular approximation described here invalid, although migration of TOI-500\,b to its current orbit may still be possible if collisions between the planets do not occur. With $a_\mathrm{b}\lesssim0.25$ au, orbit-crossing is avoided. We show one example of migration in Figure~\ref{fig:secular}, starting at $a_\mathrm{b} = 0.02$ au. Initial migration is rapid, with the planet migrating from $0.02$ to $0.013$ au in around 100 Myr, after which it slows. The present semimajor axis is attained after 2 Gyr, at which time migration is still proceeding, albeit very slowly. We also show in Figure~\ref{fig:secular} an analytical high-eccentricity migration track, where the planet is placed on a high-eccentricity orbit and circularises while maintaining its orbital angular momentum, emphasising the qualitatively different nature of the low-eccentricity pathway.

\subsection*{Atmospheric characterization of TOI-500\,b}

Ultra-short period planets present the intriguing possibility of atmospheric characterization to study secondary atmosphere creation \cite{Owen16}, extreme star-planet interactions \cite{Cauley17}, and to get clues as to the dynamical and migration history of the planet \cite{Millholland20}. In addition, given their extremely short period and frequent transits, they are attractive for atmospheric characterization from an observational perspective. Ultra-short period planets might have 3 different types of atmospheres: a Mercury type atmosphere \cite{Mura11}, a lava ocean atmosphere \cite{Briot10, Rouan11b, Barnes10}, or a silicate atmosphere \cite{Schaefer09, Ito15, Miguel19}. It would be very interesting to detect the atmosphere of TOI-500\,b because in all three cases the atmosphere would contain material from the crust, or even the interior of the planet. The detection of its atmosphere would thus open up the thrilling possibility to do mineralogy of an extrasolar planet. TOI-500\,b is among the top ten targets (currently ranked 8th) for hot terrestrial super-Earth planets (i.e., $\mathrm{R}_\mathrm{p} < 2$\,$\mathrm{R}_{\oplus}$ and $\mathrm{T}_{\mathrm{eq}} > 1000$\,K), joining targets GJ\,367\,b, 55\,Cnc\,e, HD\,219134\,b, K2-141\,b, GJ\,1252\,b, TOI-1807\,b, TOI-561\,b, TOI-1685\,b, and GJ\,9827\,b. Figure~\ref{fig:snrtoi} displays a relative atmospheric detection $\mathrm{S}/\mathrm{N}$ metric (normalized to TOI-500\,b) for all well-characterized transiting planets with $\mathrm{R}_\mathrm{p} < 2$\,$\mathrm{R}_{\oplus}$. The sample of exoplanets is taken from the NASA Exoplanet Archive (Available at \url{https://exoplanetarchive.ipac.caltech.edu/}). The atmospheric signal is calculated in a similar way in \cite{Niraula17} and is dominated by the atmospheric scale height, favoring hot, extended atmospheres, and the host star radius, favoring small, cool stars. The relative $\mathrm{S}/\mathrm{N}$ calculation scales with properties that make it favorable to detect and measure this signal. Our metric is similar to the transmission spectroscopy metric (TSM) in \cite{Kempton18}. The difference with our metric is that instead of calculating this per transit, we calculate it based on time, thus adding a $\mathrm{P}^{-0.5}$ term. Given the observational challenges of observing planets in transit with highly oversubscribed facilities, the frequency of transits is a very important constraint on obtaining atmospheric measurements of these exoplanets. We assume an effective scale height ($\mathrm{h}_{\mathrm{eff}} = 7\mathrm{H}$; \cite{Miller09}) using the equilibrium temperature, a Bond albedo of $\alpha = 0.3$, and an atmospheric mean molecular weight of $\mu = 20$. Because this is a relative assessment, and we are assuming identical properties for all the atmospheres in this sample, the precise value of these variables do not change the results. Silicate atmospheres of hot lava-ocean worlds should be detectable in lines of Na, $\mathrm{O}_2$, $\mathrm{O}$, and $\mathrm{SiO}$ \cite{Schaefer09}. Clearly, there are likely to be interesting variations in atmospheric properties among these exoplanets, which is precisely why it is important to observe a population of hot lava-ocean worlds, like TOI-500\,b. 

\section*{Code availability statement}
The numerical code used to test the low eccentricity migration pathway is available at \url{https://zenodo.org/} with the DOI:10.5281/zenodo.5877066.

\section*{Data availability statement}
\tess\ photometry is available at the Mikulski Archive for Space Telescopes (MAST) at \url{https://exo.mast.stsci.edu} under target name TOI-500.01. The raw \harps\ spectra can be retrieved from the ESO Science Archive Facility \url{http://archive.eso.org/cms.html} under ESO program IDs 1102.C-0923 (PI: Gandolfi), 0103.C-0442 (PI: Diaz), 0102.C-0338 and 0103.C-0548 (PI: Trifonov), 60.A-9700 and 60.A-9709 (ESO technical time). The ground-based photometry obtained with the LCO telescope, as well as the \soar\ and Gemini imaging data are available on the Exoplanet Follow-up Observing Program (ExoFOP) website \url{https://exofop.ipac.caltech.edu/tess/} under target name TOI-500.01. The raw Gemini data are available at \url{https://archive.gemini.edu/searchform} under Program ID  GS-2020A-Q-125. The archival \textit{WASP} data that support the findings of this study are available from the co-author Coel Hellier upon reasonable request. The archival \textit{SOAR} data that support the findings of this study are available from the co-author Carl Ziegler upon reasonable request. The extracted radial velocities and stellar activity indicators are listed in Table~\ref{HARPS_data}.

\section*{Acknowledgements}
This work was supported by the KESPRINT (\url{www.kesprint.science}) collaboration, an international consortium devoted to the characterization and research of exoplanets discovered with space-based missions. 

This paper includes data collected by the \textit{TESS} mission. Funding for the \textit{TESS} mission is provided by the NASA Explorer Program. We acknowledge the use of \textit{TESS} Alert data, which is currently in a beta test phase, from pipelines at the \textit{TESS} Science Office and at the \textit{TESS} Science Processing Operations Center. Resources supporting this work were provided by the NASA High-End Computing (HEC) Program through the NASA Advanced Supercomputing (NAS) Division at Ames Research Center for the production of the SPOC data products. This research has made use of the Exoplanet Follow-up Observation Program website, which is operated by the California Institute of Technology, under contract with the National Aeronautics and Space Administration under the Exoplanet Exploration Program. 

Based on observations made with the ESO-3.6\,m telescope at the European Southern Observatory (ESO), La Silla under ESO programs 1102.C-0923, 0102.C-0338, 0103.C-0442, 0103.C-0548, 60.A-9700, and 60.A-9709. We are very grateful to the ESO staff members for their precious support during the observations. We warmly thank Xavier Dumusque and Fran\c{c}ois Bouchy for coordinating the shared observations with \harps\ and Jaime Alvarado Montes, Xavier Delfosse, Guillaume Gaisn\'e, Melissa Hobson, and David Barrado Navascu\'es who helped collecting the \harps\ spectra.

This work has made use of data from the European Space Agency (ESA) mission {\it Gaia} (\url{https://www.cosmos.esa.int/gaia}), processed by the {\it Gaia} Data Processing and Analysis Consortium (DPAC, \url{https://www.cosmos.esa.int/web/gaia/dpac/consortium}). Funding for the DPAC has been provided by national institutions, in particular the institutions participating in the {\it Gaia} Multilateral Agreement.

This work makes use of observations from the LCOGT network. LCOGT telescope time was granted by NOIRLab through the Mid-Scale Innovations Program (MSIP). MSIP is funded by NSF. 

Some of the Observations in the paper made use of the High-Resolution Imaging instrument Zorro. Zorro was funded by the NASA Exoplanet Exploration Program and built at the NASA Ames Research Center by Steve B. Howell, Nic Scott, Elliott P. Horch, and Emmett Quigley. Data were reduced using a software pipeline originally written by Elliott Horch and Mark Everett. Zorro was mounted on the Gemini South telescope, and NIRI was mounted on the Gemini North telescope, of the international Gemini Observatory, a program of NSF’s OIR Lab, which is managed by the Association of Universities for Research in Astronomy (AURA) under a cooperative agreement with the National Science Foundation on behalf of the Gemini partnership: the National Science Foundation (United States), National Research Council (Canada), Agencia Nacional de Investigación y Desarrollo (Chile), Ministerio de Ciencia, Tecnología e Innovación (Argentina), Ministério da Ciência, Tecnologia, Inovações e Comunicações (Brazil), and Korea Astronomy and Space Science Institute (Republic of Korea). Data collected under program GN-2019A-LP-101.

Based in part on observations obtained at the Southern Astrophysical Research (SOAR) telescope, which is a joint project of the Minist\'{e}rio da Ci\^{e}ncia, Tecnologia e Inova\c{c}\~{o}es (MCTI/LNA) do Brasil, the US National Science Foundation’s NOIRLab, the University of North Carolina at Chapel Hill (UNC), and Michigan State University (MSU).

This research has made use of the NASA Exoplanet Archive, which is operated by the California Institute of Technology, under contract with the National Aeronautics and Space Administration under the Exoplanet Exploration Program.

LMS and DG gratefully acknowledge financial support from the CRT foundation under Grant No. 2018.2323 ``Gaseous or rocky? Unveiling the nature of small worlds''. EG acknowledges the generous support by the Th\"uringer Ministerium f\"ur Wirtschaft, Wissenschaft und Digitale Gesellschaft. IG, CMP, MF and AJM gratefully acknowledge the support of the Swedish National Space Agency (DNR 174/18, 65/19, 120/19C). JK gratefully acknowledges the support of the Swedish National Space Agency (DNR 2020-00104). SzCs, ME, KWFL, SG, and APH acknowledge support by DFG grants RA714/14-1 within the DFG Schwerpunkt SPP 1992, ``Exploring the Diversity of Extrasolar Planets''. MRD acknowledges the support by Comisi\'on Nacional de Investigaci\'on Cient\'ifica y Tecnol\'ogica (CONICYT)-PFCHA/Doctorado Nacional-21140646, Chile. TD acknowledges support from MIT's Kavli Institute as a Kavli postdoctoral fellow. T.T. further acknowledges support by the BNSF program "VIHREN-2021" project No.
\endgroup
\begingroup
\fontencoding{T2A}\selectfont КП-06-ДВ/5.
\endgroup
\begingroup
\fontencoding{T1} \selectfont
\section*{Author contributions statement}

LMSer performed the periodogram analysis and the joint analysis with \texttt{pyaneti}, wrote most of the text, and coordinated the contributions from the other co-authors. DGan performed the radial velocity analysis using the floating chunk offset method, wrote a significant part of the text, and is the principal investigator of the \harps\ large program, which enabled the discovery of 3 additional planets and the determinations of the planetary (minimum) masses. OBar performed the multi-dimensional Gaussian process analysis and wrote the relative section. JKor ran the stability analyses with \texttt{rebound} and \texttt{SPOCK} and wrote the relative section. AJMus and FDai described the most probable  formation/migration processes of the system. AJMus also ran the numerical simulation to test the low eccentricity migration pathway and wrote the relative section. MFri performed the spectral and chemical abundance analysis. KWFLam and SGrz searched the \tess\ light curve for transit signals. KACol performed the \lcogt\ observations and analyzed the data. JHLiv analyzed the \textit{GEMINI} and \soar\ imaging data. JAla, MRDia, FRod, and TTri contributed to the \harps\ RV follow up. WCoc computed the probability that TOI-500 belongs to different stellar populations. CHel analyzed the \textit{WASP-South} light curves. SBel contributed to the analysis of the radial velocity data. SRed explored the possibility to study the secondary atmosphere of TOI-550\,b, and wrote the relative section. SAlb, SzCsi, HJDee, MEsp, IGeo, EGof, EGue, APHat, RLuq, FMur, HLMOsb, EPal, CMPer AMSmi, VVEyl are members of the KESPRINT consortium and contributed to the \harps\ large program. CZie and AWMan performed the \soar\ imaging observations. ELNJen contributed to the \lcogt\ observations. SBHow performed the observations with \textit{GEMINI}/\textit{ZORRO}. JMJen, DLat, GRic, SSea, RVan, JNWin are the architects of the \tess\ mission. DACal, TDay, MFas, AWMan, PRow, ARud and JDTwi significantly contributed to the success of the \tess\ mission, which discovered the USP planet candidate. All the authors reviewed the manuscript. 

\section*{Competing interest statement}
We declare no competing interest.

\begin{table}
\centering
\caption{TOI-500 main identifiers, equatorial coordinates, proper motion, parallax, optical and infrared magnitudes, and fundamental parameters we refer to or we estimated within the present work. The acronyms listed in the third column refer to the Exoplanet Follow-up Observing Program (ExoFOP) database, \gaia\ Data Release 2 (DR2, \cite{GaiaDR2}), the \tess\ Input Catalog version 8 (TIC v8, \cite{Stassun2018}), the Two Micron All Sky Survey (2MASS, \cite{cutri2003}) catalog, and the Wide-field Infrared Survey Explorer (All{\it WISE}, \cite{cutri2014}) data release.}
\label{stellar_values}
\begin{tabular}{lrr}
\hline
Parameter & Value & Source \\
\hline
\multicolumn{3}{l}{\it Main identifiers}  \\
\noalign{\smallskip}
\multicolumn{2}{l}{TIC}{134200185} & ExoFOP \\
\multicolumn{2}{l}{CD}{-47 2804}  & ExoFOP \\
\multicolumn{2}{l}{HIP}{34269} & ExoFOP \\
\multicolumn{2}{l}{TYC}{8122-00785-1} & ExoFOP \\
\multicolumn{2}{l}{2MASS}{J07061396-4735137}  & ExoFOP \\
\multicolumn{2}{l}{\gaia\ DR2}{5509620021956148736}  & \gaia\ DR2 \\
\hline
\multicolumn{3}{l}{\it Equatorial coordinates, parallax, and proper motion}  \\
\noalign{\smallskip}
R.A. (J2000.0)	&    07$^\mathrm{h}$06$^\mathrm{m}$14.18$^\mathrm{s}$	& \gaia\ DR2 \\
Dec. (J2000.0)	& $-$47$\degr$35$\arcmin$16.14$\arcsec$	                & \gaia\ DR2 \\
$\pi$ (mas) 	& $21.0715\pm0.0209$                                    & \gaia\ DR2 \\ 
$\mu_\alpha$ (mas\,yr$^{-1}$) 	& $135.798 \pm 0.040$		& \gaia\ DR2 \\
$\mu_\delta$ (mas\,yr$^{-1}$) 	& $-146.251 \pm 0.037$		& \gaia\ DR2 \\
$\mathrm{U}$ $(\mathrm{km\,s^{-1}})$     & $37.70 \pm 0.05$          & This work      \\
$\mathrm{V}$ $(\mathrm{km\,s^{-1}})$     &$-$60.53$\pm$0.19          & This work     \\
$\mathrm{W}$ $(\mathrm{km\,s^{-1}})$     & $6.44 \pm 0.06$             & This work      \\
\noalign{\smallskip}

\hline
\multicolumn{3}{l}{\it Optical and near-infrared photometry} \\
\noalign{\smallskip}
\tess\                 & $9.402\pm0.006$     & TIC v8 \\
\noalign{\smallskip}
$\mathrm{B}$           & $11.668\pm0.050$   & TIC v8 \\
$\mathrm{V}$           & $10.540 \pm 0.030$   & TIC v8 \\
\noalign{\smallskip}
$\mathrm{J}$ 			&  $8.403\pm0.024$      & 2MASS \\
$\mathrm{H}$			&  $7.848\pm0.038$      & 2MASS \\
$\mathrm{K}$			&  $7.715\pm0.026$      & 2MASS \\
\noalign{\smallskip}
$\mathrm{W1}$			&  $7.630\pm0.030$      & All{\it WISE} \\
$\mathrm{W2}$			&  $7.736\pm0.020$      & All{\it WISE} \\
$\mathrm{W3}$           & $7.658\pm0.018$      & All{\it WISE} \\
$\mathrm{W4}$           & $7.617\pm0.124$      & All{\it WISE} \\
\hline
\multicolumn{3}{l}{\it Fundamental parameters}   \\
$\mathrm{v}_{\star} \sin \mathrm{i}_{\star}$\ (km\;s$^{-1}$)      & $1.1 \pm 0.7$       & This work \\
$\mathrm{T}_{\mathrm{eff}}$\ (K) & $4440 \pm 100$ & This work \\
$\log \mathrm{g}_{\star}$ (cgs) & $4.618 \pm 0.017$ & This work \\
$[\mathrm{Fe}/\mathrm{H}]$  & $0.12 \pm 0.08$ & This work \\
$[\mathrm{Ca}/\mathrm{H}]$ & $-0.01 \pm 0.10$ & This work \\
$\mathrm{M}_{\star}$ ($\mathrm{M_{\odot}}$) & $0.740 \pm 0.017$ & This work \\
$\mathrm{R}_{\star}$ ($\mathrm{R_\odot}$) & $0.678 \pm 0.016$  & This work \\
Age (Gyr) & $5.0 \pm 0.2$ & This work \\
Distance (pc) & $47.3924 \pm 0.0473$ & \gaia\ DR2 \\
$\mathrm{A_V}$ & 0.02\,$\pm$\,0.02 & This work \\
\hline
\end{tabular}
\end{table}

\begin{table*}
  \footnotesize
  \caption{TOI-500 system parameters as derived from the joint modelling of the \tess\ and \harps\ data. $\mathcal{U}$, $\mathcal{N}$ and $\mathcal{J}$ refer to uniform, Gaussian, and Jeffreys priors, respectively.  \label{table_prior}}  
  \centering
 \scalebox{0.9}{ \begin{tabular}{lccc}
  \hline
  \noalign{\smallskip}
  Parameter & Prior & Results with GP stellar model & Results with stellar sinusoidal model \\
  \noalign{\smallskip}
  \hline
  \noalign{\smallskip}
  \multicolumn{4}{l}{\emph{ \bf Model parameters for TOI-500 b}} \\
    Orbital period $\mathrm{P}_{\mathrm{b}}$ (days)  & $\mathcal{U}[0.5478 , 0.5485]$   & \Pb[] & \Pbbis[]\\
    Transit epoch $\mathrm{T}_{0, \mathrm{b}}$ (BJD$_\mathrm{TDB} - 2\,457\,000$)  & $\mathcal{U}[1468.3660, 1468.4140]$  & \Tzerob[] & \Tzerobbis[] \\
    $\sqrt{\mathrm{e_\mathrm{b}}} \sin \omega_{\star, \mathrm{b}}$ &  $\mathcal{U}[-1,1]$ & \esinb &  \esinbbis \\
    $\sqrt{\mathrm{e_\mathrm{b}}} \cos \omega_{\star, \mathrm{b}}$  &  $\mathcal{U}[-1,1]$ & \ecosb &  \ecosbbis \\
    Scaled planetary radius $\mathrm{R}_{\mathrm{b}}/\mathrm{R}_{\star}$ &  $\mathcal{U}[0,0.022]$ & \rrb[] & \rrbbis[] \\
    Impact parameter, $\mathrm{b_b}$ &  $\mathcal{U}[0,1]$  & \bb[] & \bbbis[] \\
    Radial velocity semi-amplitude variation $\mathrm{K_b}$ (ms$^{-1}$) &  $\mathcal{U}[0,4]$ & \kb[] & \kbbis[]\\
    \noalign{\smallskip}
    \multicolumn{4}{l}{\emph{ \bf Model parameters for TOI-500 c}} \\
    Orbital period $\mathrm{P}_{\mathrm{c}}$ (days)  &  $\mathcal{U}[6.5857,6.6857]$ & \Pc[] & \Pcbis[]\\
    Time of inferior conjunction $\mathrm{T}_{0, \mathrm{c}}$ (BJD$_\mathrm{TDB} - 2\,457\,000$)  & $\mathcal{U}[1559.7073 , 1564.1311]$  & \Tzeroc[] & \Tzerocbis[] \\
    $\sqrt{\mathrm{e_c}} \sin \omega_{\star, \mathrm{c}}$ &  $\mathcal{U}[-1,1]$ & \esinc &  \esincbis \\
    $\sqrt{\mathrm{e_c}} \cos \omega_{\star, \mathrm{c}}$  &  $\mathcal{U}[-1,1]$ & \ecosc &  \ecoscbis \\
    Radial velocity semi-amplitude variation $\mathrm{K_c}$ (ms$^{-1}$) &  $\mathcal{U}[0,4.5]$ & \kc[] & \kcbis[] \\
    \noalign{\smallskip}
    \multicolumn{4}{l}{\emph{ \bf Model parameters for TOI-500 d}} \\
    Orbital period $\mathrm{P}_{\mathrm{d}}$ (days)  &  $\mathcal{U}[25.4334,27.0334]$ & \Pd[] & \Pdbis[] \\
    Time of inferior conjunction $\mathrm{T}_{0, \mathrm{d}}$ (BJD$_\mathrm{TDB} - 2\,457\,000$)  & $\mathcal{U}[1578.4891, 1595.9780]$  & \Tzerod[] & \Tzerodbis[] \\
    $\sqrt{\mathrm{e_d}} \sin \omega_{\star, \mathrm{d}}$ &  $\mathcal{U}[-1,1]$ & \esind &  \esindbis \\
    $\sqrt{\mathrm{e_d}} \cos \omega_{\star, \mathrm{d}}$  &  $\mathcal{U}[-1,1]$ & \ecosd &  \ecosdbis \\
    Radial velocity semi-amplitude variation $\mathrm{K_d}$ (ms$^{-1}$) &  $\mathcal{U}[0,15]$ & \kd[] & \kdbis[] \\
    \noalign{\smallskip}
    \multicolumn{4}{l}{\emph{ \bf Model parameters for TOI-500 e}} \\
    Orbital period $\mathrm{P}_{\mathrm{e}}$ (days)  &  $\mathcal{U}[58.1620,64.1620]$ & \Pe[] & \Pebis[]\\
    Time of inferior conjunction $\mathrm{T}_{0, \mathrm{e}}$ (BJD$_\mathrm{TDB} - 2\,457\,000$)  & $\mathcal{U}[1834.4801, 1895.6421]$  & \ldots & \Tzeroebis[] \\
    Time of inferior conjunction $\mathrm{T}_{0, \mathrm{e}}$ (BJD$_\mathrm{TDB} - 2\,457\,000$)  & $\mathcal{N}[1865.8,3]$  & \Tzeroe[] & \ldots \\
    $\sqrt{\mathrm{e_e}} \sin \omega_{\star, \mathrm{e}}$ &  $\mathcal{U}[-1,1]$ & \esine &  \esinebis \\
    $\sqrt{\mathrm{e_e}} \cos \omega_{\star, \mathrm{e}}$ &  $\mathcal{U}[-1,1]$ & \ecose &  \ecosebis \\
    Radial velocity semi-amplitude variation $\mathrm{K_e}$ (ms$^{-1}$) &  $\mathcal{U}[0,10]$ & \ke[] & \kebis[] \\
    \noalign{\smallskip}
    \multicolumn{4}{l}{\emph{ \bf Model parameters of activity-induced RV signal}} \\
    Period $\mathrm{P}_{\star}$ (days)  &  $\mathcal{U}[41.6940,45.6940 ]$ & \ldots & \Pfbis[] \\
    Epoch $\mathrm{T}_{0, \star}$ (BJD - $2457000$)  & $\mathcal{U}[1562.1299, 1605.8239 ]$   & \ldots & \Tzerofbis[]\\
    Radial velocity semi-amplitude variation $\mathrm{K_\star}$ (ms$^{-1}$) &  $\mathcal{U}[0,4]$  & \ldots & \kfbis[]\\
    \multicolumn{3}{l}{\emph{ \bf Multi-dimensional GP parameters}} \\
    GP Period $P_{\rm GP}$ (days) &  $\mathcal{U}[35,50]$ & \jPGP[] & \ldots \\
    $\lambda_{\rm p}$ &  $\mathcal{J}[0.1,10]$ &  \jlambdap[] & \ldots \\
    $\lambda_{\rm e}$ (days) &  $\mathcal{J}[50,250]$ &  \jlambdae[] & \ldots \\
    $V_{\rm c}$ (\ms)  &  $\mathcal{U}[0,1]$ &  \jAzero & \ldots \\
    $V_{\rm r}$ (\ms\,d$^{-1}$) &  $\mathcal{U}[-1,1]$  & \jAuno & \ldots \\
    $F_{\rm c}$  (\kms)  & $\mathcal{U}[-1,1]$ & \jAdue & \ldots \\
    $S_{\rm c}$  &  $\mathcal{U}[-1,1]$ &  \jAquattro & \ldots \\
    \noalign{\smallskip}
    \multicolumn{4}{l}{\emph{ \bf Other system parameters}} \\
    Stellar density $\rho_\star$ ($\rm g\,cm^{-3}$) & $\mathcal{N}[3.34, 0.25]$ & \denstrbbis[] & \denstrb[] \\
    RV jitter term $\sigma_{\rm \harps}$ (ms$^{-1}$) & $\mathcal{U}[1.041 , 1.230]$ & \jRV[] & \jHARPSbis[]\\
    Parameterized limb darkening coefficient $\mathrm{q}_1$ & $\mathcal{N}[0.45, 0.10]$ & \qone & \qonebis \\
    Parameterized limb darkening coefficient $\mathrm{q}_2$ & $\mathcal{N}[0.38 , 0.10]$ & \qtwo & \qtwobis \\
    \noalign{\smallskip}
    \hline
    \noalign{\smallskip}
    \multicolumn{4}{l}{{Derived parameters for TOI-500 b}} \\
    Planet mass M$_\mathrm{b}$ ($\mathrm{M}_{\oplus}$)  & $\cdots$ & \mpb[] & \mpbbis[]\\
    Planet radius R$_\mathrm{b}$ ($\mathrm{R}_{\oplus}$)  & $\cdots$ & \rpb[] & \rpbbis[]\\
    Planet density $\rho_\mathrm{b}$ (${\rm g\,cm^{-3}}$) & $\cdots$ & \denpb[] & \denpbbis[] \\
    Scaled semi-major axis, $\mathrm{a_b}/\mathrm{R}_{\star}$ & $\cdots$ & \arb[] & \arb\\
    Semi-major axis $\mathrm{a_\mathrm{b}}$ (AU)  & $\cdots$ & \ab[] & \abbis[] \\
    Orbital inclination $\mathrm{i_\mathrm{b}}$ (deg)  & $\cdots$ & \ib[] & \ibbis[] \\
    $\mathrm{e_{\mathrm{b}}}$ & $\cdots$ & \eb & \ebbis \\
    $\omega_{\star\,\mathrm{b}}$ (deg) & $\cdots$ & \wbs[] & \wbbis[] \\
    Transit duration T$_{\mathrm{dur}, \mathrm{b}}$(hours) & $\cdots$ & \ttotb[] & \ttotbbis[] \\
    Transit depth T$_{\mathrm{depth}, \mathrm{b}}$(ppm) & $\cdots$ & \depthb[] & \Tdepthbis[] \\
    Equilibrium temperature $\mathrm{T}_{\rm eq, b}$ ($K$) & $\cdots$ & \Teqb[] & \Teqbbis[] \\
    Insolation $\mathrm{F}_{\rm b}$ ($\mathrm{F}_{\oplus}$)   & $\cdots$ & \insolationb[] & \insolationbbis[] \\
    \noalign{\smallskip}
    \multicolumn{4}{l}{\textbf{Derived parameters for the other planets}} \\
    $\mathrm{M}_\mathrm{c}\sin \mathrm{i}_{\mathrm{c}}$ ($\mathrm{M}_{\oplus}$)  & $\cdots$ & \mpc[] & \mpcbis[] \\
    $\mathrm{e}_\mathrm{c}$  & $\cdots$ & \ec[] & \ecbis[]  \\
    $\omega_{\star,\,\mathrm{c}} $ (deg)  &  $\cdots$ & \wc[] & \wcbis[]  \\
   $\mathrm{M}_\mathrm{d}\sin \mathrm{i}_{\mathrm{d}}$ ($\mathrm{M}_{\oplus}$)  & $\cdots$ & \mpd[] & \mpdbis[] \\
    $e_\mathrm{d}$  & $\cdots$ & \ed[] & \edbis[]  \\
    $\omega_{\star,\,\mathrm{d}} $ (deg)  & $\cdots$ & \wds[] & \wdbis[]  \\
    $\mathrm{M}_\mathrm{e}\sin \mathrm{i}_{\mathrm{e}}$ ($\mathrm{M}_{\oplus}$)  & $\cdots$ & \mpe[] & \mpebis[]\\
    $\mathrm{e}_\mathrm{e}$  & $\cdots$ & \ee[] & \eebis[]  \\
    $\omega_{\star,\,\mathrm{e}} $ (deg)  &  $\cdots$ & \we[] & \webis[]  \\
   \noalign{\smallskip}
    \multicolumn{4}{l}{\textbf{Other derived parameters}} \\
    Limb darkening $\mathrm{u}_1$ & $\cdots$ & \uone[] & \uonebis[]\\
    Limb darkening $\mathrm{u}_2$ & $\cdots$ & \utwo[] & \utwobis[]\\
   \noalign{\smallskip}
   \hline
   \end{tabular}}
\end{table*}

\newpage

\begin{figure}
    \centering
    \includegraphics[width = 17cm]{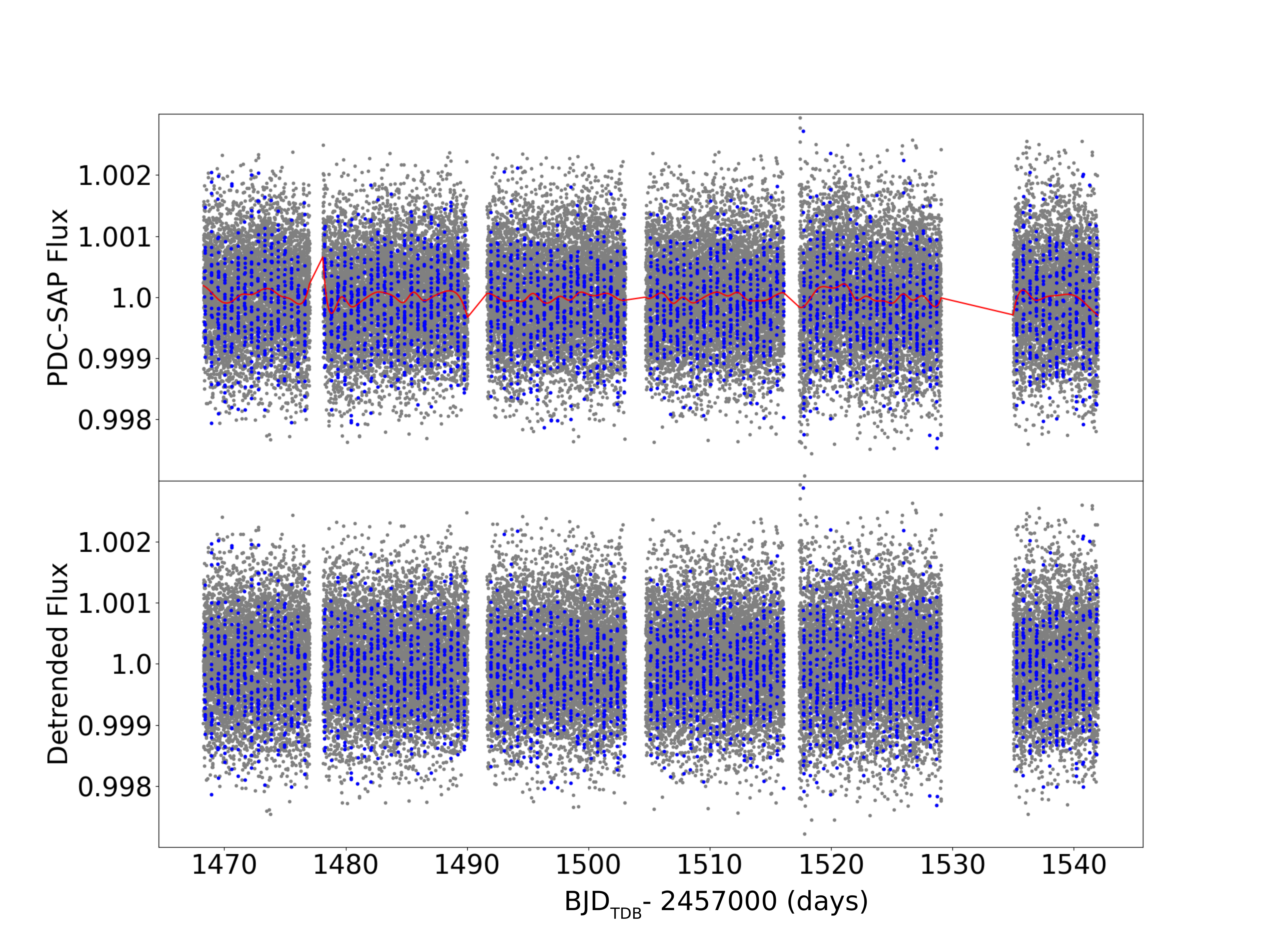}
    \caption{\emph{Upper panel}: median-normalized PDC-SAP \tess\ light curve of TOI-500 (grey points). The spline used to detrend the \tess\ data is overplotted with a red line. \emph{Lower panel}: detrended light curve, following the removal of outliers using a sigma clipping algorithm. The light curve was detrended following \textit{Method 3} as described in the ``Independent transit search'' Section. The data-points within the transits of TOI-500\,b are plotted with blue circles in both panels.}
    \label{Tess_data}
\end{figure}

\begin{figure}
    \centering
    \includegraphics[]{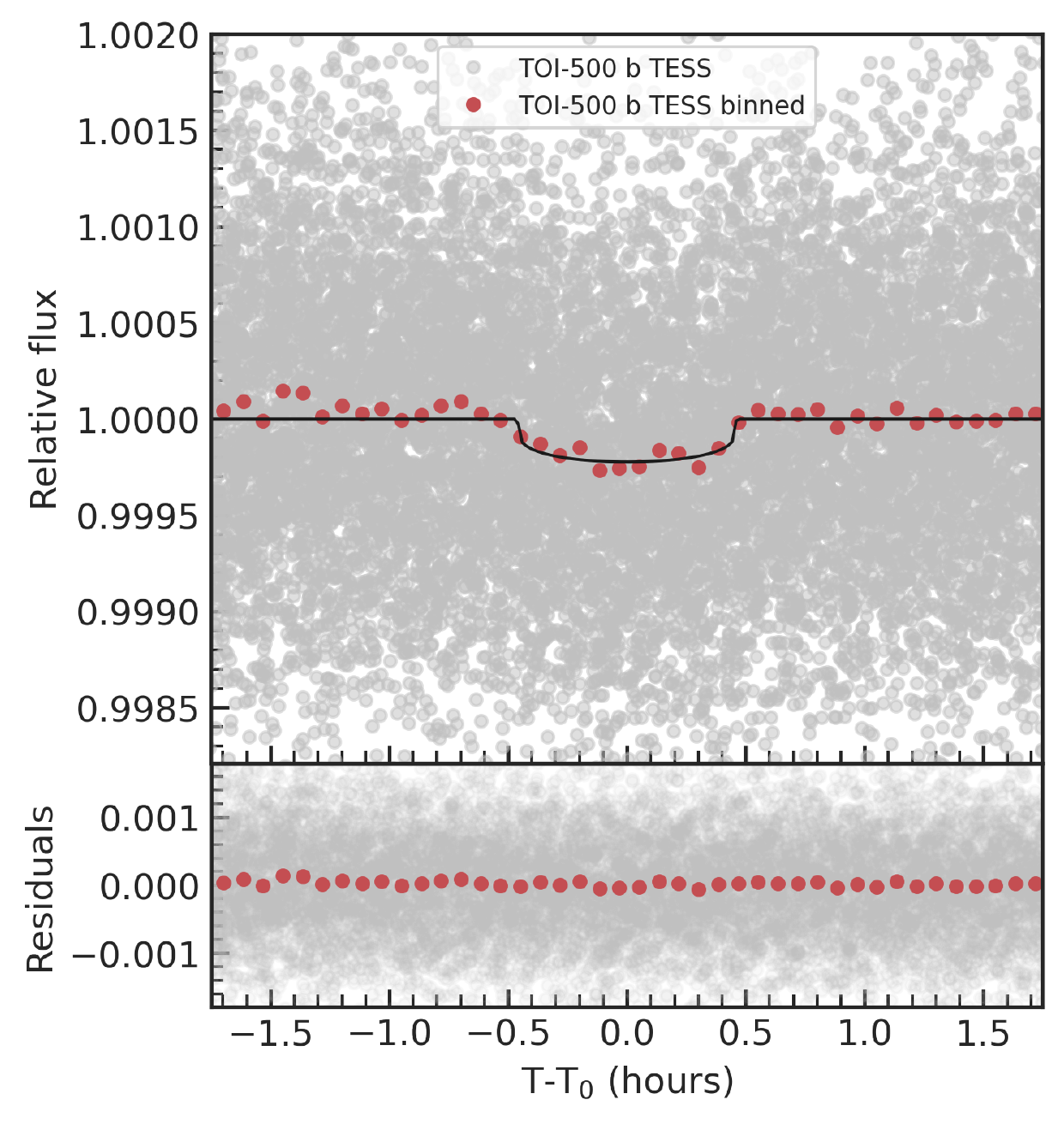}
    \caption{\emph{Upper panel:} Phase-folded \tess\ light curve of TOI-500\,b. \tess\ measurements are shown with light gray circles, along with the $5$\,min binned data (red circles), and the inferred transit model (solid black line). \emph{Lower panel}: residuals to the fit.}
    \label{planetb_phase_folded}
\end{figure}

\begin{figure}
    \centering
    \includegraphics[width = 15cm]{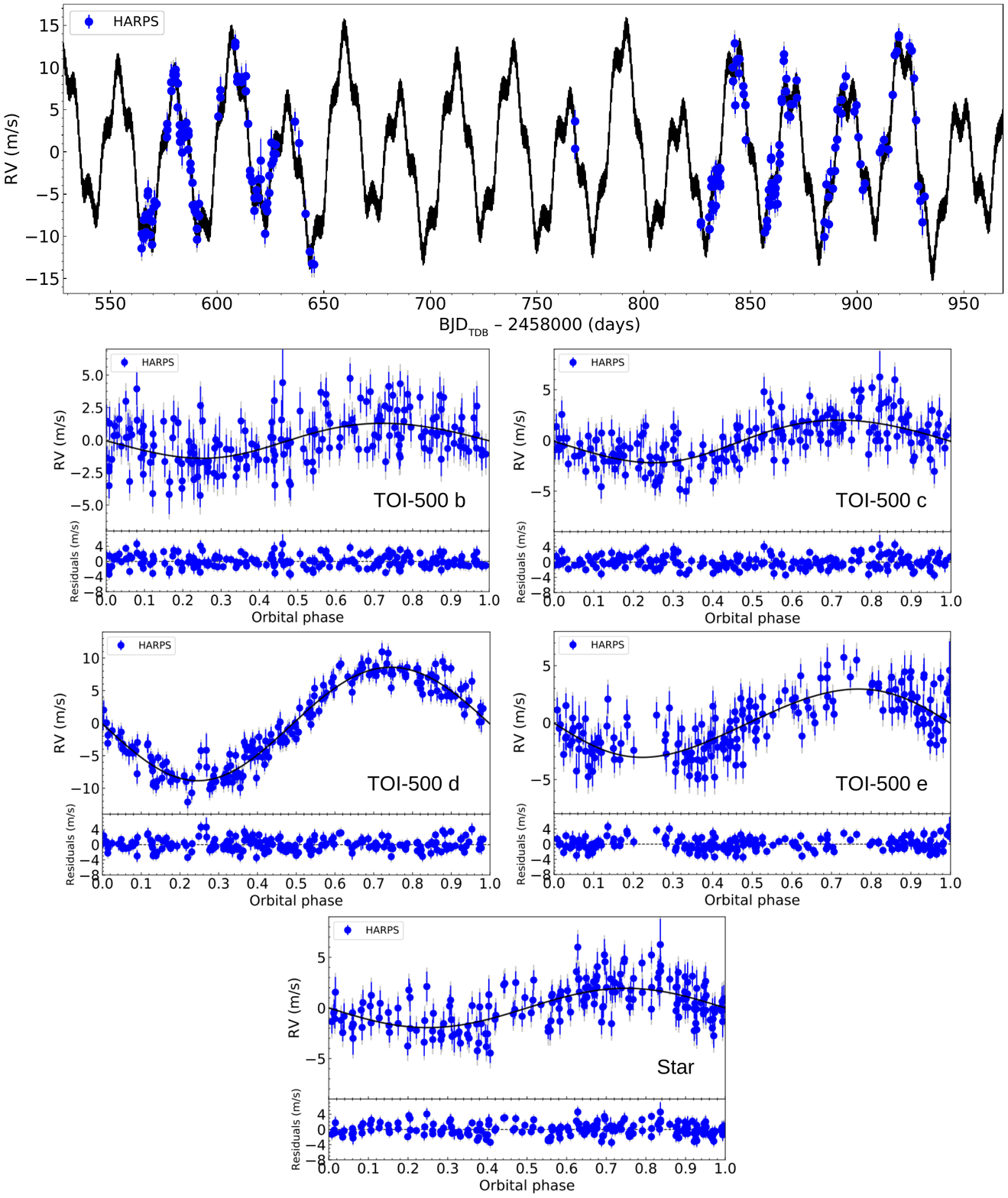}
    \caption{\emph{Upper row}: \harps\ \texttt{SERVAL} RVs (blue data points) and best-fitting 5-signal model (black line; 4 planets + stellar rotation). \emph{Second, third, and fourth rows:} Phase-folded RVs curve of TOI-500 b (\emph{first row, left panel}), TOI-500\,c (\emph{first row, right panel}), TOI-500\,d (\emph{second row, left panel}), TOI-500\,e (\emph{second row, right panel}), stellar rotation (\emph{third row}).} 
    \label{Sinusoidal}
\end{figure}

\begin{figure}
    \includegraphics[width = 17cm]{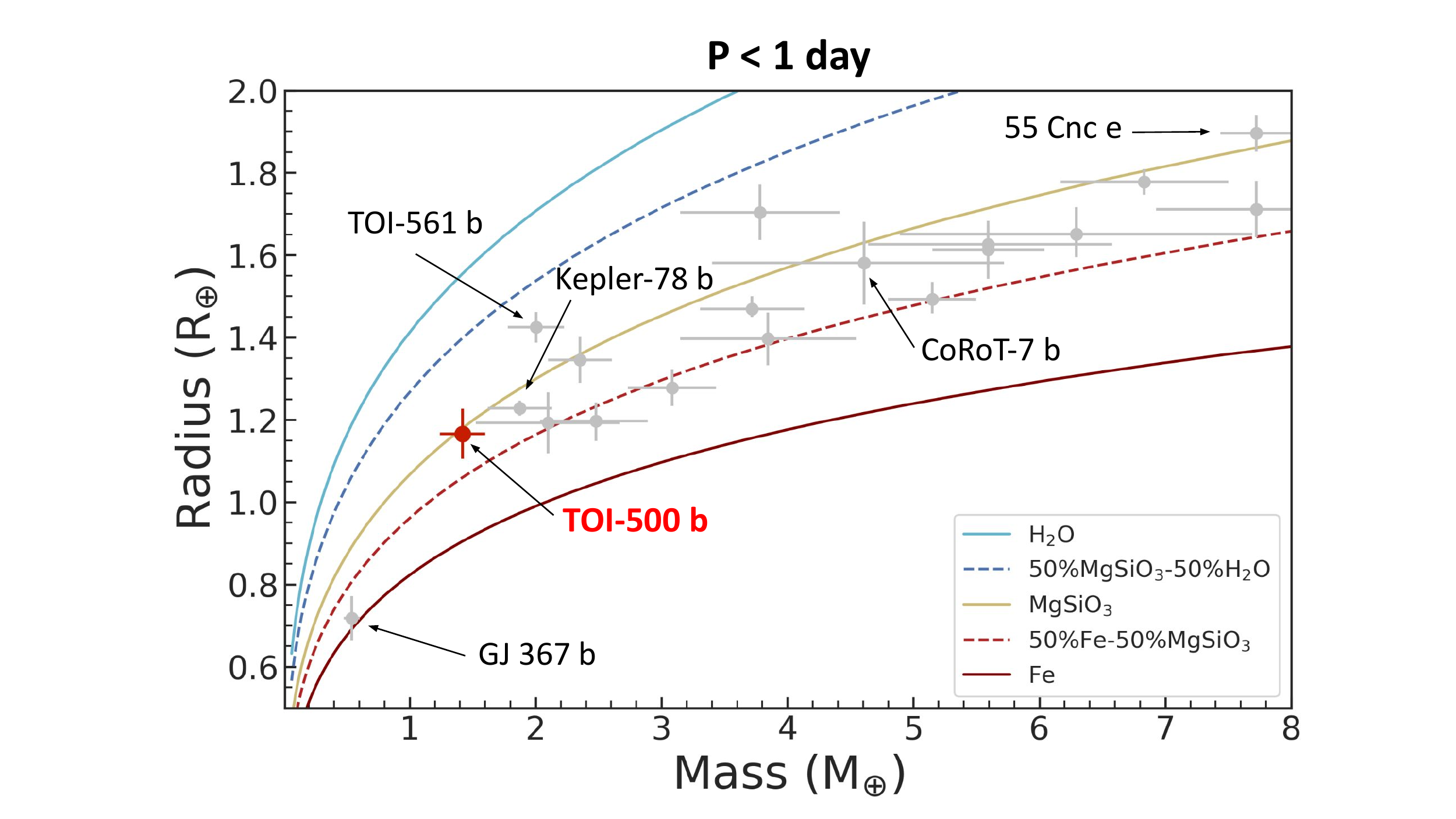}
    \caption{Mass-radius diagram for USP planets (P\,$<$\,1\,d) whose masses and radii are known with a precision better than $30$\,\%, as retrieved from the Transiting Extrasolar Planet Catalogue (TEPCat, \cite{Southworth11b}). The position of TOI-500\,b is highlighted with a red dot. The thick and dashed lines mark the bulk composition models from \cite{Zeng2016}.}
    \label{mass_radius}
\end{figure}

\begin{figure}
    \centering
    \includegraphics[width = 20cm]{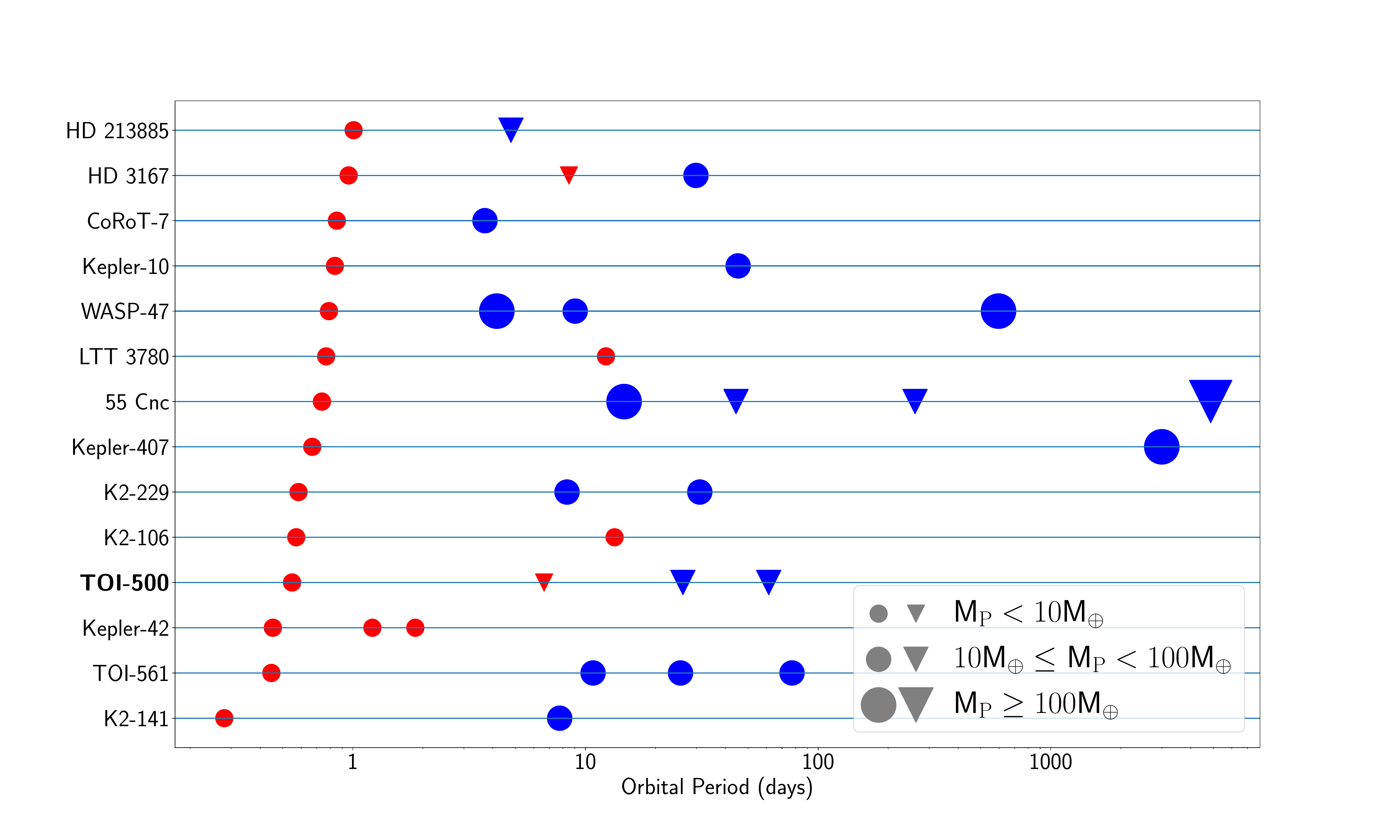}
    \caption{Multi-planetary systems hosting a USP planet whose mass and radius are known, as of December 2021. From top to bottom, systems are sorted by decreasing period of their USP planet. The x-axis displays the orbital period in logarithmic scale. The size of the symbols depends on the mass: we used the smallest symbols for planets with masses lower than $10~\mathrm{M}_{\oplus}$, the medium-size symbols for planets with masses between $10~\mathrm{M}_{\oplus}$ and $100~\mathrm{M}_{\oplus}$, the largest symbol for masses above $100~\mathrm{M}_{\oplus}$. We highlighted in red the Earth-like planets, while the gaseous planets are in blue. The planets with measured mass and radius are marked with dots, while those with only the radius or the minimum mass known are marked with triangles. For non-transiting planets we assumed that the planetary mass is the minimum mass. For transiting planets with no RV measurements, we used Equation~1 in \cite{Otegi20} to estimate the planetary mass. The systems were identified by cross-matching the The Extrasolar Planets Encyclopedia (\url{https://exoplanet.eu}) and TEPCat (\url{https://www.astro.keele.ac.uk/jkt/tepcat/}, \cite{Southworth}). Planetary parameters are extracted from \cite{Barragan18} and \cite{Malavolta18} for K2-141, \cite{Lacedelli20} for TOI-561, \cite{Mann17} and \cite{Muirhead12} for Kepler-42, \cite{Adams17}, \cite{Sinukoff17,Guenther17} and \cite{Livingston18} for K2-106, \cite{Santerne18} and \cite{Livingston18} for K2-229, \cite{Marcy14} for Kepler-407, \cite{Butler97, McArthur04, Marcy02, Fischer08, VonBraun11} and \cite{Winn11b} for 55 Cnc, \cite{Nowak20} and \cite{Cloutier20} for LTT 3780, \cite{Hellier12, SanchisOjeda2015, Becker15, Almenara16, NeveuVanMalle16} and \cite{Vanderburg17} for WASP-47, \cite{Fabrycky12} for Kepler-32, \cite{Batalha11, Fressin11, Santos15, Fogtmann14}, \cite{Weiss16} and \cite{Rajpaul17} for Kepler-10, \cite{Leger09, Queloz09, Hatzes2010, Hatzes2011, Haywood} \cite{Barros_1} and \cite{Faria} for CoRoT-7, \cite{Vanderburg16b} \cite{Gandolfi2017} and \cite{Christiansen17} for HD\,3167, \cite{Espinoza20} for HD\,213885 (aka TOI-141). TOI\,500 is highlighted in boldface.
    \label{USP_known}}
\end{figure}

\begin{figure}
   \hspace{-2.1cm}
   \includegraphics[width = 22cm]{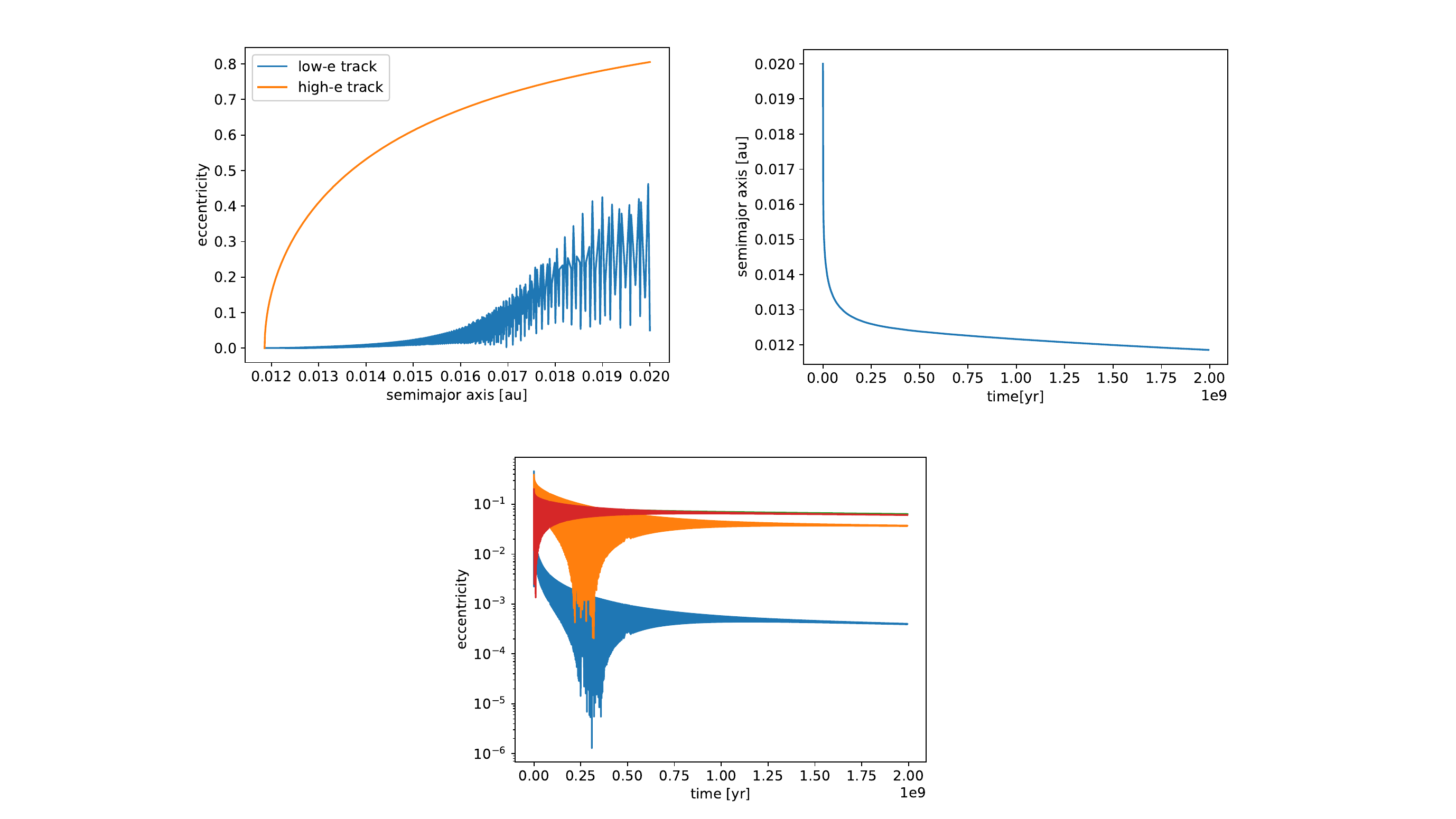}
     \caption{Example migration of TOI-500b from 0.02\,au to its current orbit. Top left panel: The blue curve shows the evolution of semimajor axis and eccentricity in a low-eccentricity pathway with initial eccentricities $e_\mathrm{b} = 0.05$, $e_\mathrm{c} = 0.05$,$e_\mathrm{d} = 0.05$, and $e_\mathrm{e} = 0.2$. The orange curve by contrast shows an analytical high-eccentricity migration track, where the planet rapidly circularises from a highly eccentric orbit while conserving its orbital angular momentum. Top right panel: the semimajor axis evolution of the USP in this system. Bottom panel: the eccentricity evolution of all planets in this system (USP in blue).}
    \label{fig:secular}
\end{figure}

\newpage
\printbibliography[]
\newpage
\newpage

\setcounter{figure}{0}
\setcounter{table}{0}

\renewcommand{\figurename}{Extended Data Figure}
\makeatletter 
\renewcommand{\thetable}{S\@arabic\c@table}
\makeatother

\begin{figure}
    \centering
    \includegraphics[width = 15cm]{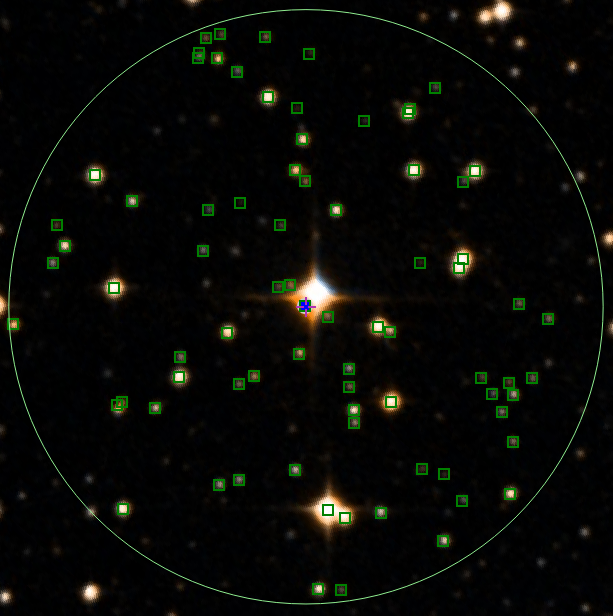}
    \caption{Field of view with the locations of the 78 \gaia\ DR2 stars checked for NEBs. The circle marks a $2.5\arcmin$ radius around TOI-500. The background image is from the digitized sky survey 2 (DSS2). The circles on each star represent the current Gaia DR2 position. The different timing between the DSS2 and DR2 databases is the reason for which some of the stars are shifted from the original position, as a consequence of their proper motion.}
    \label{field}
\end{figure}

\begin{figure}
    \centering
    \includegraphics[width=\columnwidth]{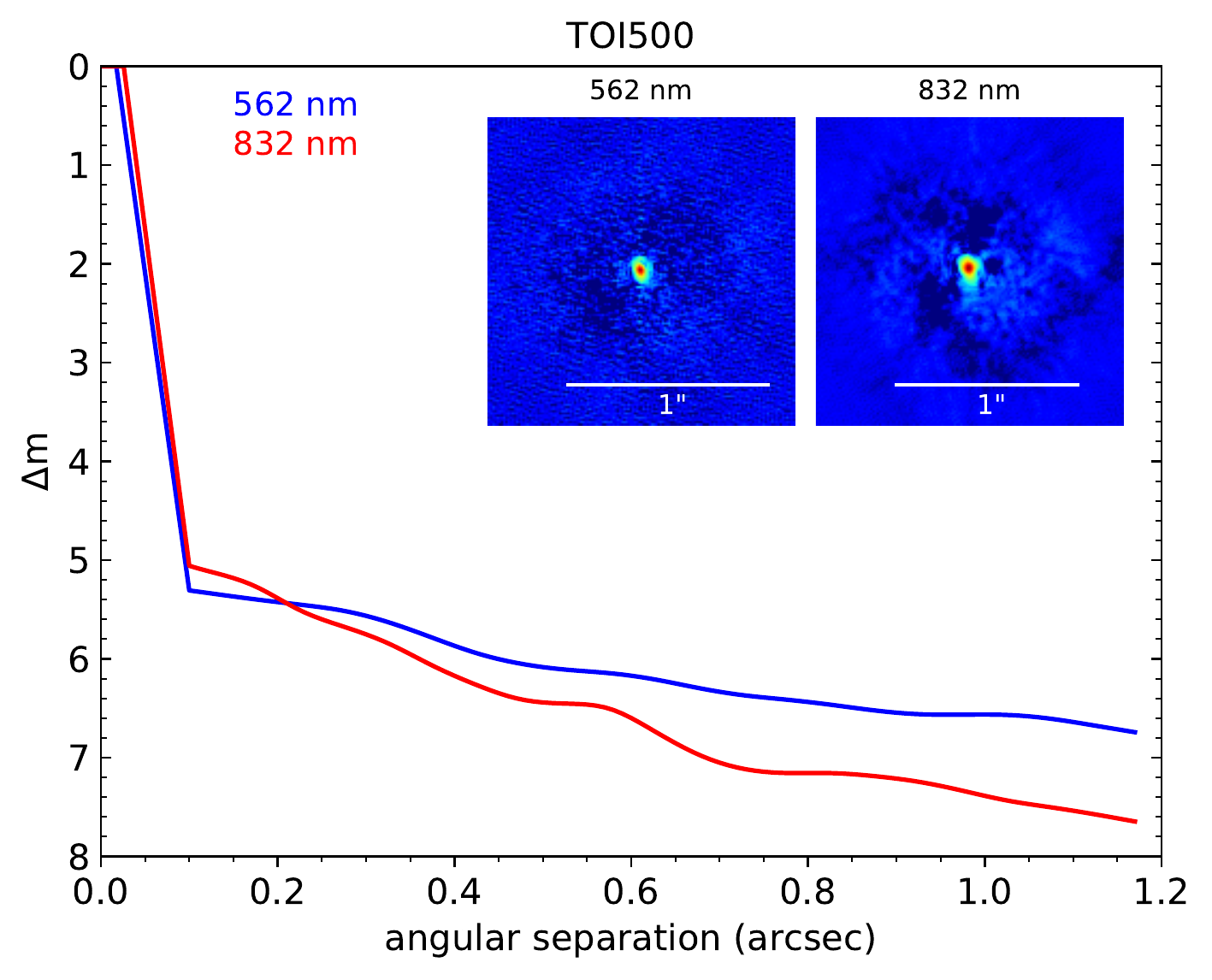}
    \caption{Gemini/Zorro 5-sigma contrast curves and $1.2\arcsec\times 1.2 \arcsec$ reconstructed images (inset).}
    \label{fig:zorro}
\end{figure}

\begin{figure}
    \centering
    \includegraphics[width=\columnwidth]{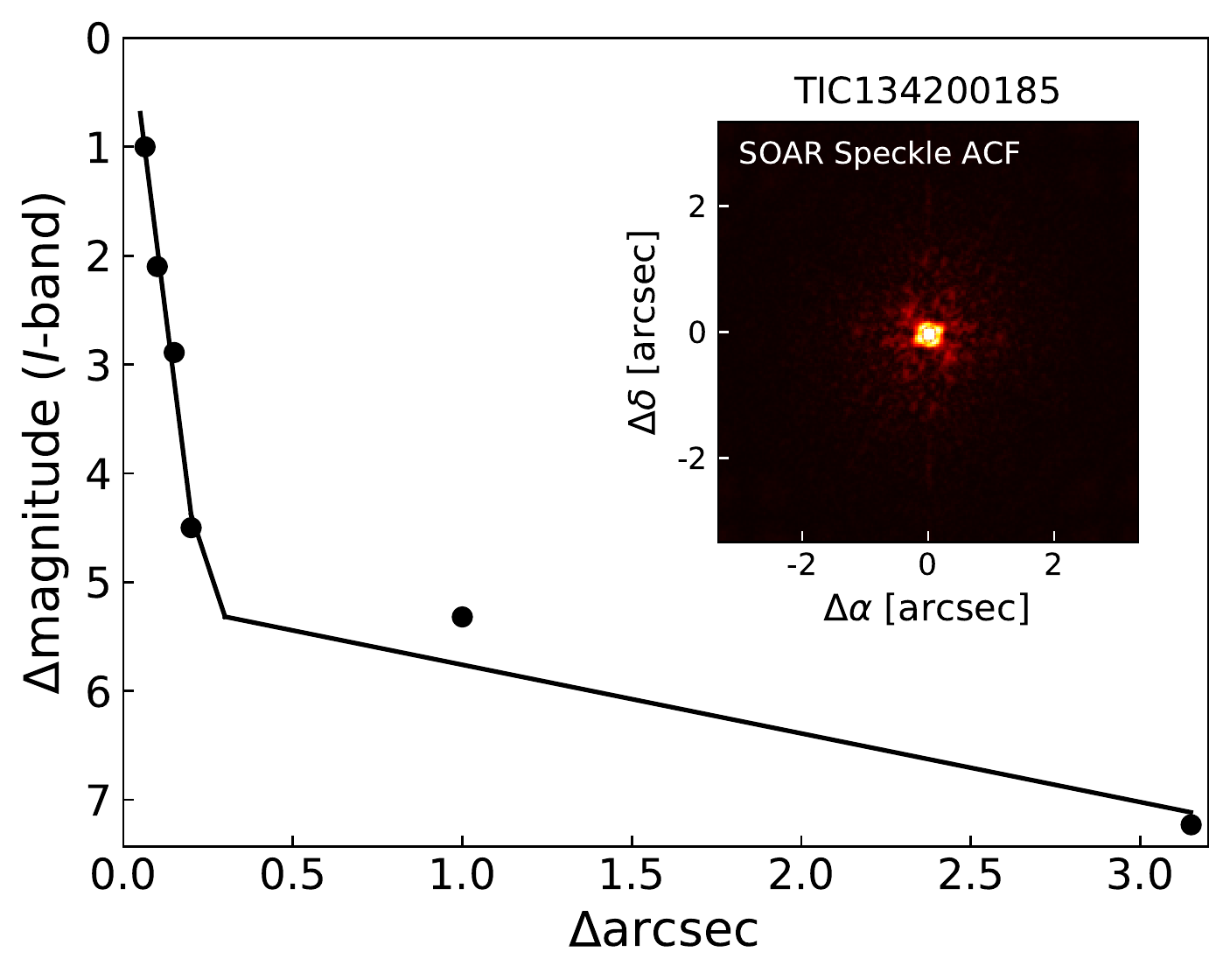}
    \caption{\soar\ contrast curve and $6\arcsec\times6\arcsec$ two dimensional auto-correlation function of \soar\ image (inset).}
    \label{fig:soar}
\end{figure}

\begin{figure}
    \hspace{-2.5cm}
    \centering
    \includegraphics[width = 21cm]{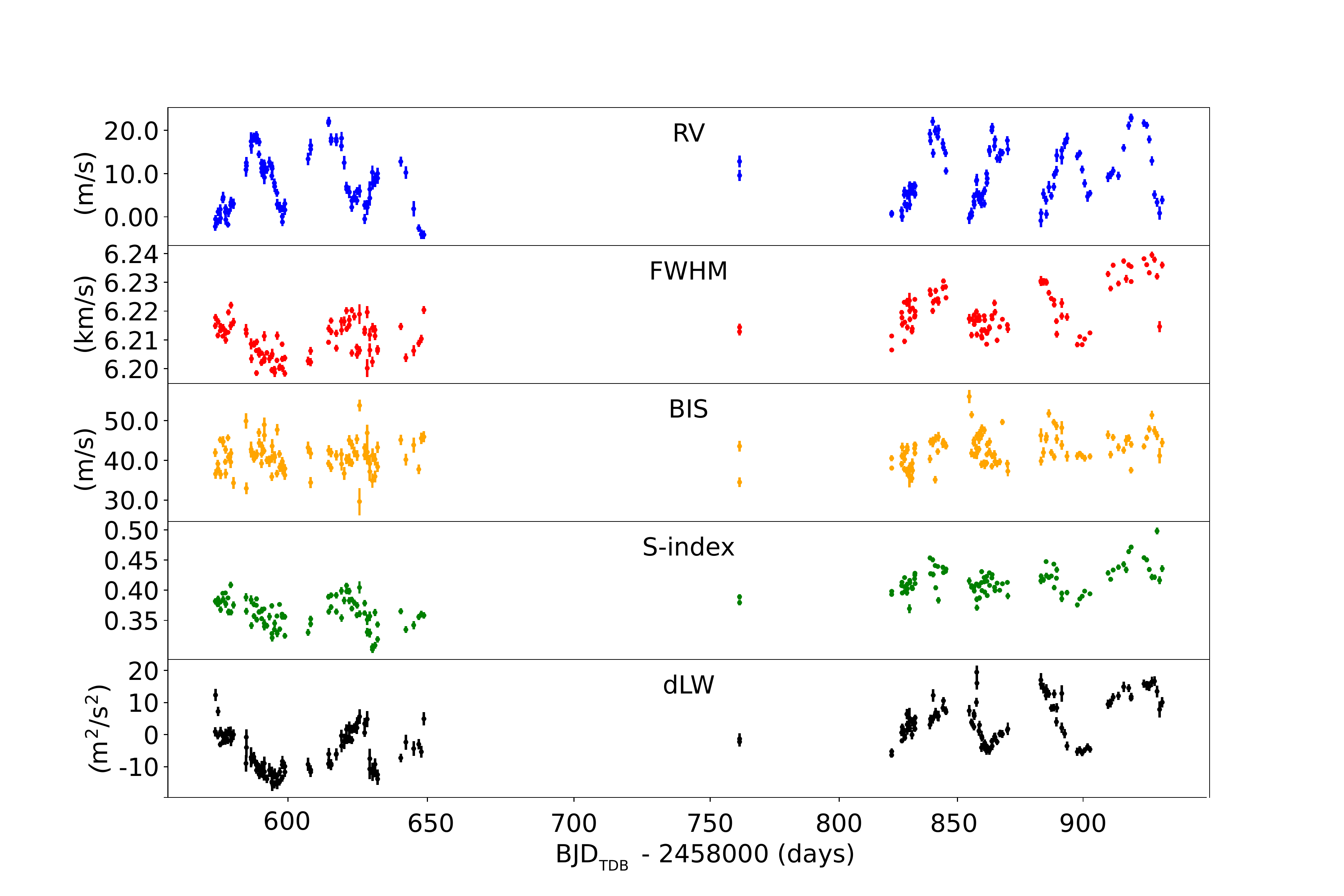}
    \caption{Time series of the \harps\ \texttt{SERVAL} RV measurements and activity indicators (FWHM, BIS, S-index, and dLW).}
    \label{RV_data}
\end{figure}

\begin{figure}
    \hspace{-5.5cm}
    \includegraphics[width = 29cm]{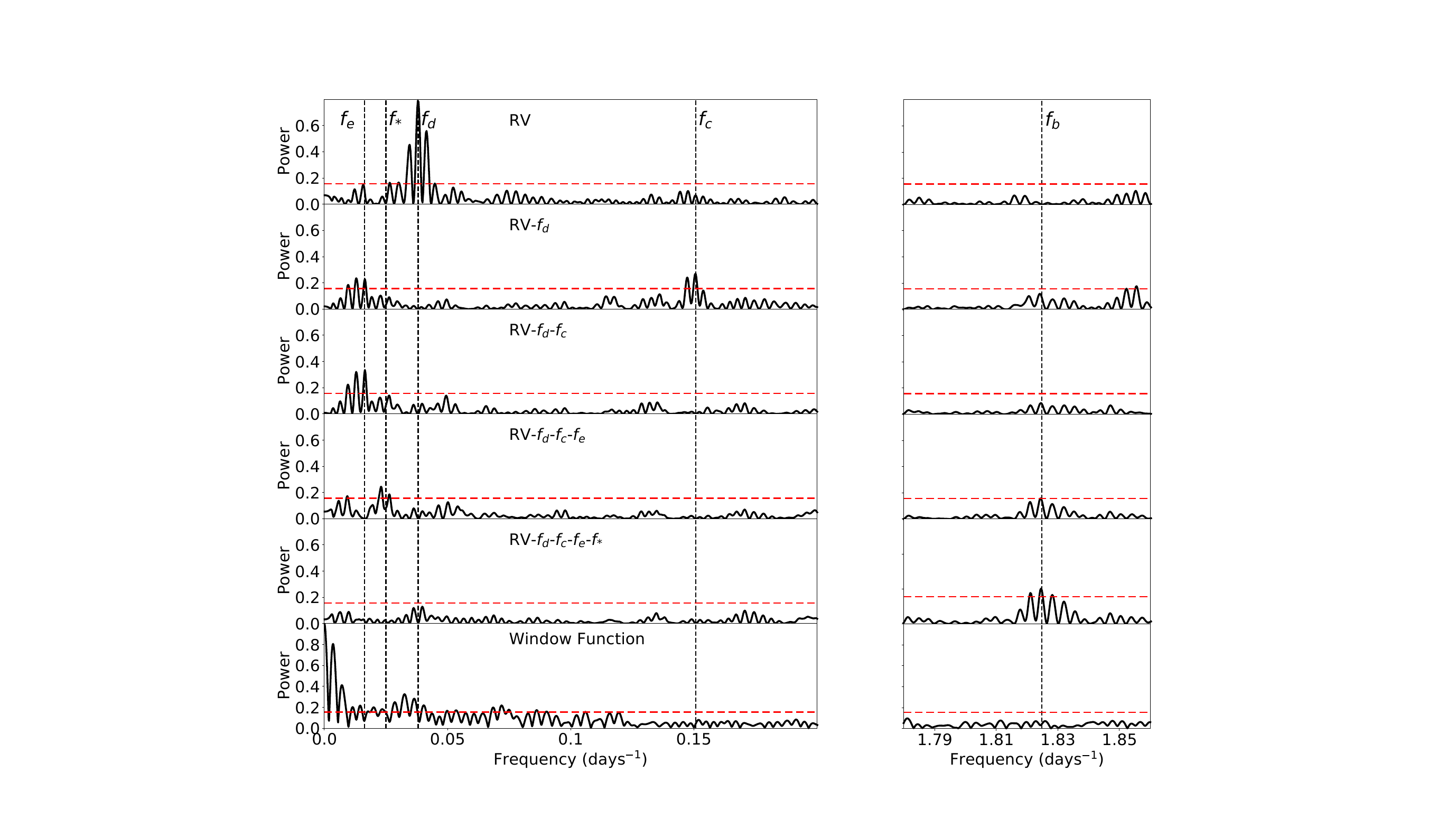}
    \caption{Generalized Lomb–Scargle periodograms of the \harps\ \texttt{SERVAL} RV measurements and residuals. The right and left columns cover two frequency ranges encompassing the Doppler signals of TOI-500\,c, d, e, and the stellar rotation frequency f$_\star$ (\emph{left panels}), and the orbital frequency of the USP planet TOI-500\,b (\emph{right panels}). \emph{From top to bottom}: RV data (\emph{upper panel}); RV residuals following the subtraction of the Doppler signal of TOI-500\,d (\emph{second panel}), TOI-500\,d and c (\emph{third panel}), TOI-500\,d, c, and e (\emph{fourth panel}), TOI-500\,d, c, and e plus the stellar signal at 43.4\,d (\emph{fifth panel}); window function (\emph{lower panel}). The red dashed horizontal lines mark the 0.1\,\% false alarm probability as derived using the bootstrap method. The vertical dashed lines mark the significant frequencies identified in the \harps\ data and discussed in the main text.}
    \label{periodogram_planet}
\end{figure}

\begin{figure}
    \hspace{-4.2cm}
    \includegraphics[width = 26cm]{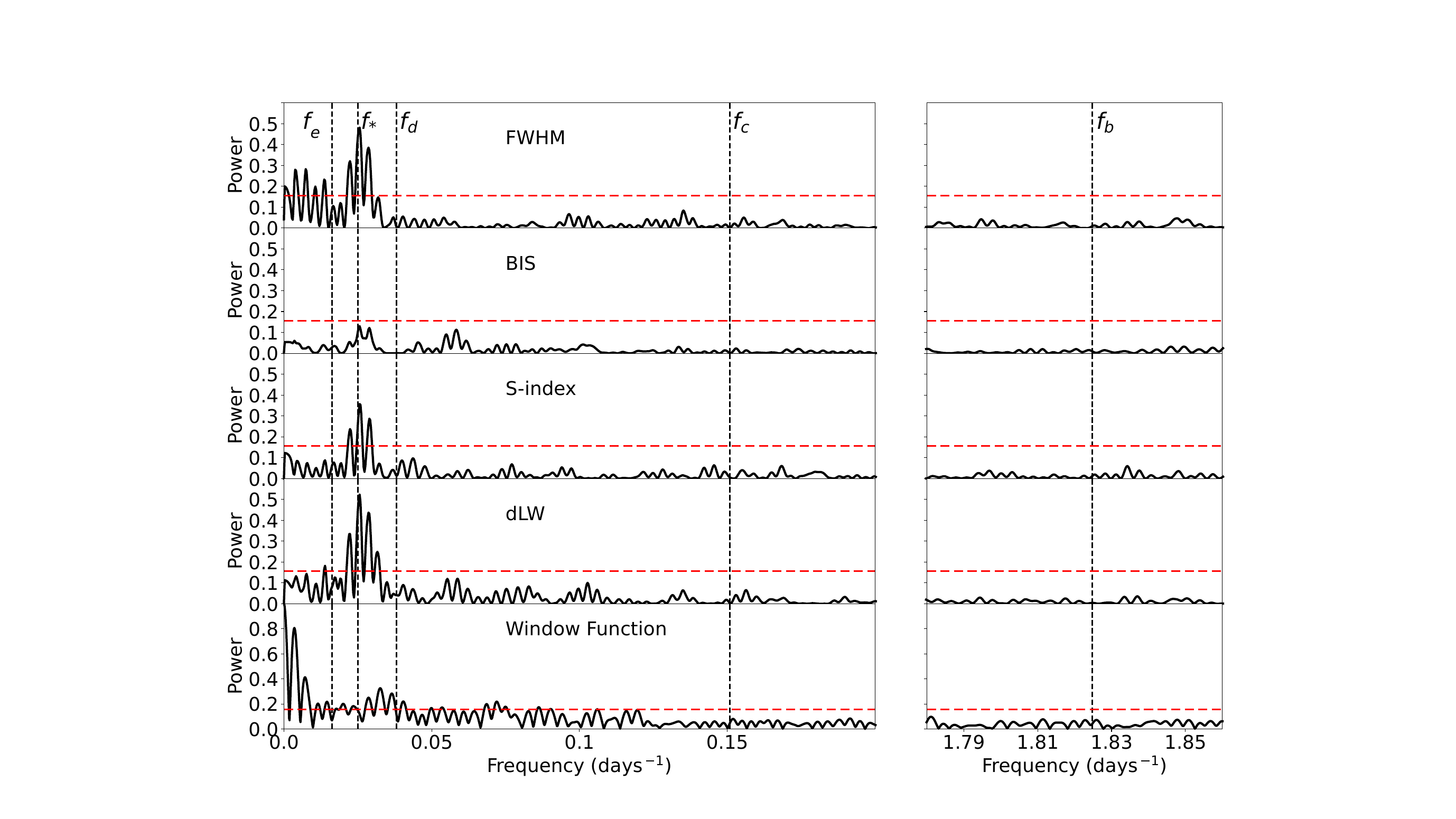}
    \caption{Generalized Lomb-Scargle periodograms of the activity indicators following the subtraction of the seasonal median values (see main text). The right and left columns cover two frequency ranges encompassing the Doppler signals induced by the 4 orbiting planets and stellar rotation. \emph{From top to bottom}: FWHM (\emph{upper panel}), BIS (\emph{second panel}), S-index (\emph{third panel}), dLW (\emph{fourth panel}), window function (\emph{lower panel}). The red dashed horizontal lines mark the 0.1\,\% false alarm probability as derived using the bootstrap method. The vertical dashed lines mark the significant frequencies identified in the \harps\ data and discussed in the main text.}
    \label{activity indexes}
\end{figure}

\begin{figure}
\centering
  \includegraphics[width = 17cm]{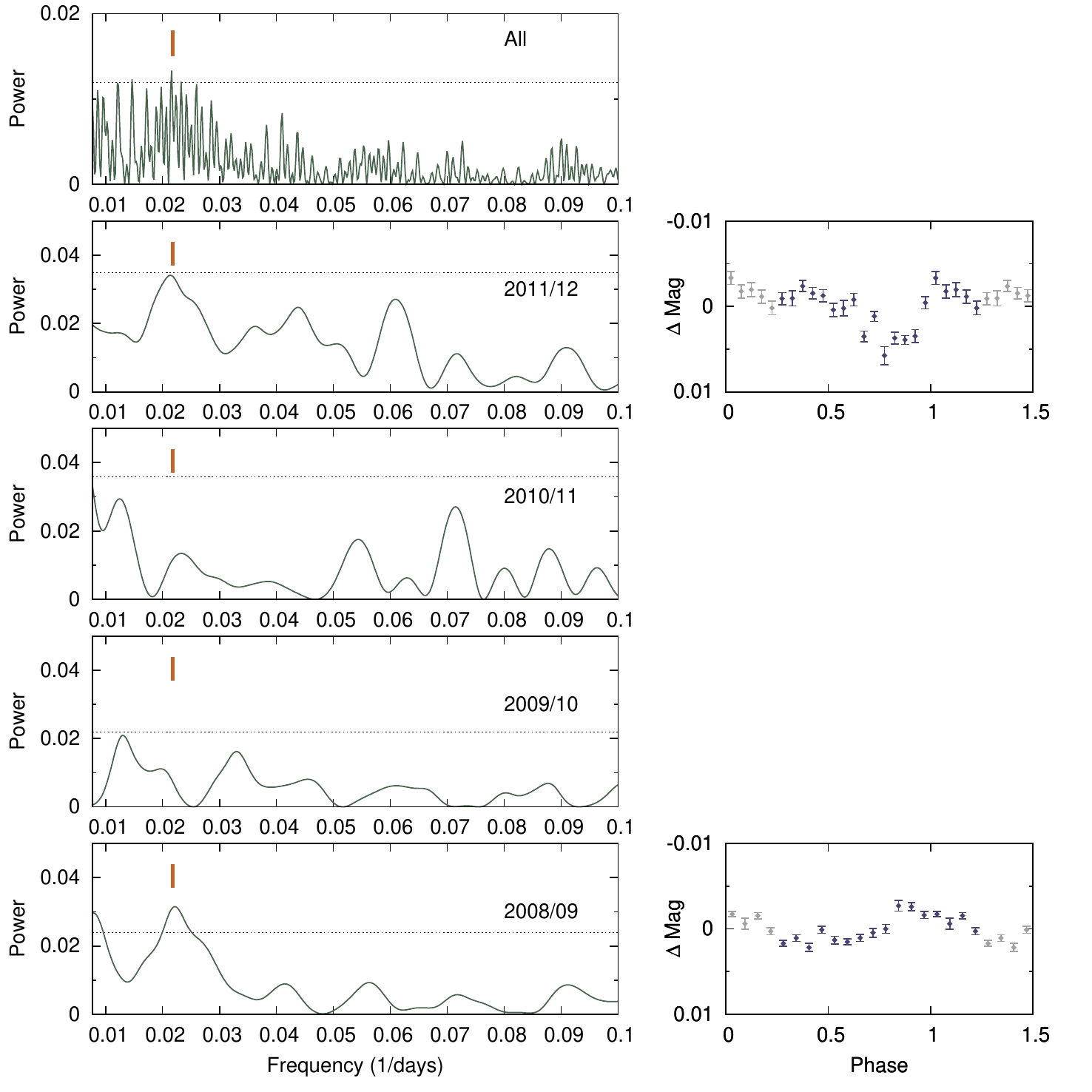}
  \caption{Lomb-Scargle periodograms of the \textit{WASP-South} light curves of TOI-500. \emph{Left panels, from bottom to top}: periodogram of the data acquired in 2008/2009, 2009/2010, 2010/2011, and 2011/2012. The \emph{upper left panel} displays the periodogram of the combined 4 years of data, showing a possible $0.022$\,d$^{-1}$ periodic signal, corresponding to a period of about 45\,d (\emph{upper panel}). This peak is marked in all the panels with a red thick line. The dotted horizontal lines show the $1$\%\ false alarm probability. The \emph{right panels} show the \textit{WASP-South} binned photometry folded at the $45$-d rotation period for the years 2008 and 2011, when the 45\,d signal is stronger. The displayed phases go from 0 to 1.5, in order to visualize better the periodicity of the photometric variability. The repeated data between phase 0 and 0.25, and phase 1.25 and 1.5 are shown with gray points.}
\label{fig:wasp}
\end{figure}

\begin{figure}
    \centering
    \includegraphics[width=\textwidth]{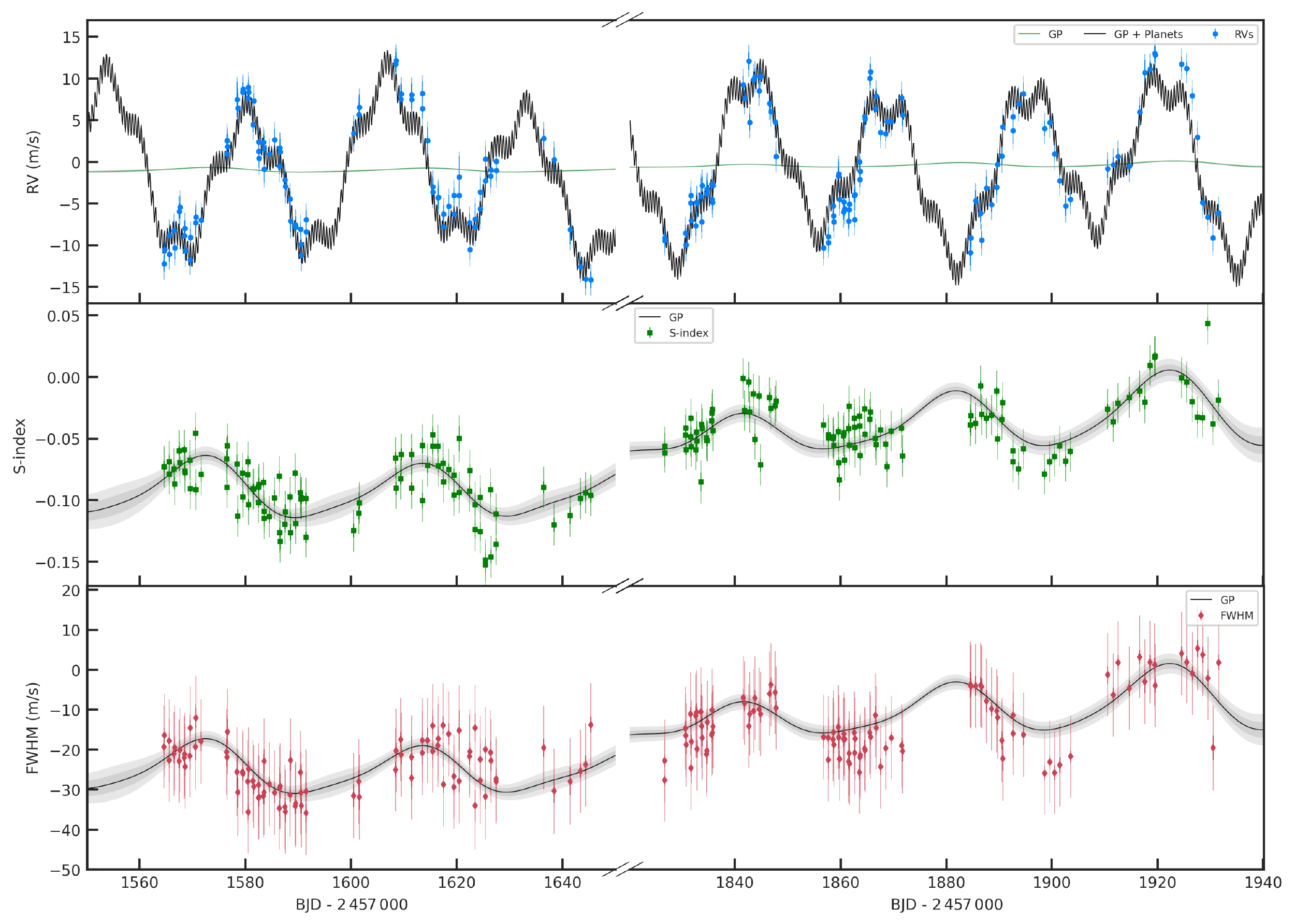}
    \caption{Median-subtracted \harps\ \emph{SERVAL} RVs (\emph{upper panel}), S-index (\emph{middle panel}), and FWHM of the cross-correlation function (\emph{lower panel}). \emph{Upper panel}: Median-subtracted HARPS \emph{SERVAL} RVs (blue data points), GP model (green line), and best-fitting (GP + 4 RV orbits) model (thick black line). \emph{Middle panel}: S-index (green data points) and GP model (thick black line). \emph{Lower panel}: FWHM (red data points) and GP model (thick black line). The solid lines mark the nominal error bars; the semitransparent lines account for the jitter terms. The shaded areas mark the 1 and 2\,$\sigma$ confidence interval of the GP models. Note the presence of a gap between  BJD$_\mathrm{TDB} - 2\,457\,000 = 1650$ and $1820$\,d}
    \label{fig:gpstimeseries}
\end{figure}

\begin{figure}
    \centering
   \hspace{-1cm}
  \includegraphics[width = 18cm]{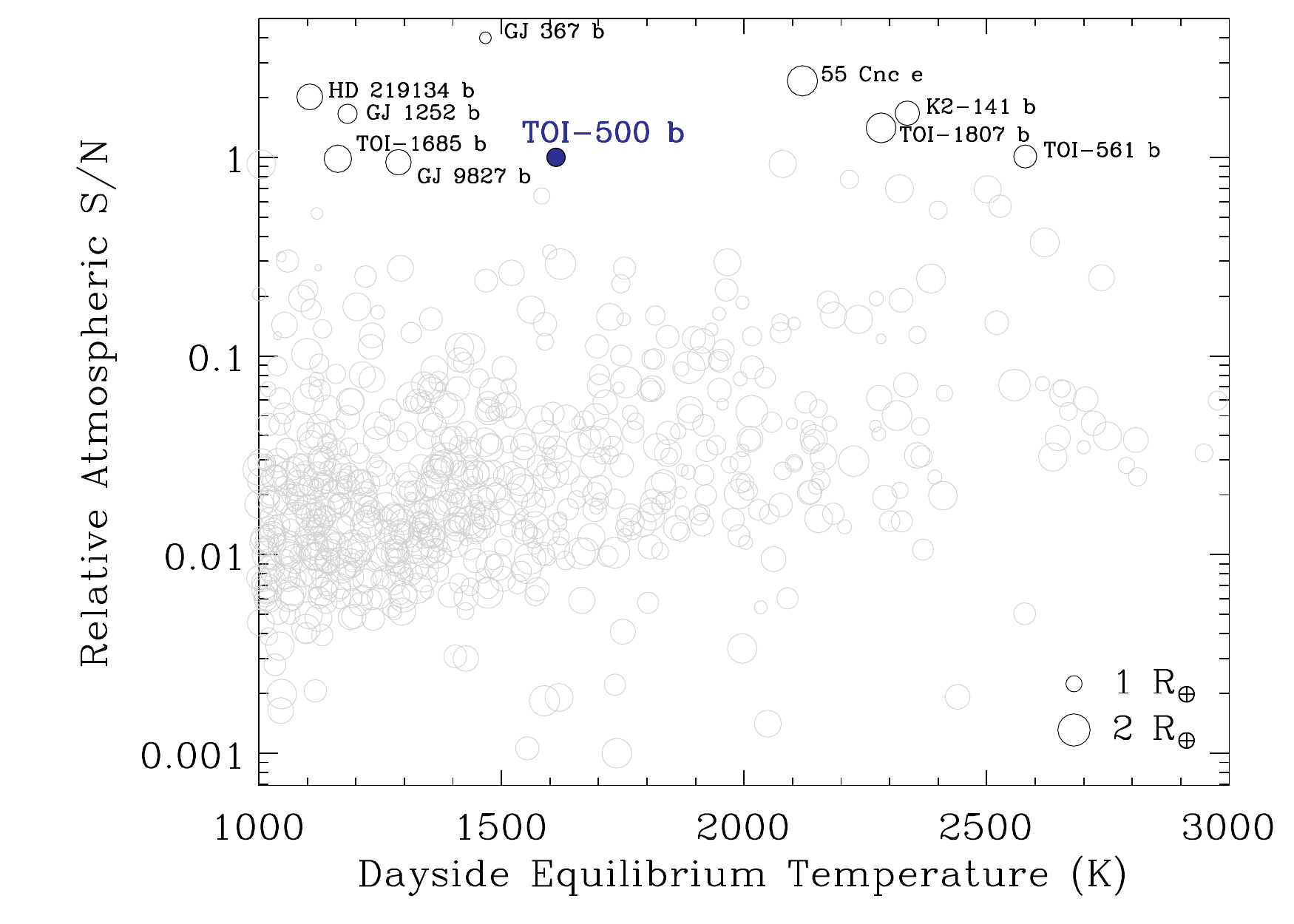}
    \caption{Scatter plot of exoplanets with atmospheric $\mathrm{S}/\mathrm{N}$ ratio as a function of the dayside predicted equilibrium temperature in Kelvin. TOI-500\,b is in a favorable position among the top 10 targets of interest for atmospheric characterization}
    \label{fig:snrtoi}
\end{figure}

\newpage
\begin{table*}
  \caption{List of favourable planets for atmospheric analysis. TOI-500\,b is ranked 8th. The relative atmospheric detection signal-to-noise (S/N) metric (normalized to TOI-500 b) is listed in the fourth column from the left. \label{relevant_targets}}
  \centering
  \begin{tabular}{rlccccc}
  \hline
  \noalign{\smallskip}
    Rank & Name & T$_\mathrm{eq}$ & S$/$N & R$_{\star}$ & P$_\mathrm{orb}$\\
         &      &    (K)          &       &(R$_{\odot}$)& (d) \\
  \noalign{\smallskip}
  \hline
  \noalign{\smallskip}
 1 & GJ\,367\,b    & 1467.0 & 3.996 & 0.46 & 0.3220 \\
 2 & 55\,Cnc\,e    & 2120.3 & 2.428 & 0.94 & 0.7365 \\
 3 & HD\,219134\,b & 1105.1 & 2.014 & 0.78 & 3.0929 \\
 4 & K2-141\,b     & 2336.5 & 1.668 & 0.68 & 0.2803 \\
 5 & GJ\,1252\,b   & 1182.9 & 1.656 & 0.39 & 0.5182 \\
 6 & TOI-1807\,b   & 2282.4 & 1.407 & 0.68 & 0.5494 \\
 7 & TOI-561\,b    & 2579.4 & 1.010 & 0.85 & 0.4466 \\
 8 & TOI-500\,b    & 1617.0 & 1.000 & 0.68 & 0.5482 \\
 9 & TOI-1685\,b   & 1163.0 & 0.984 & 0.49 & 0.6691 \\
10 & GJ\,9827\,b   & 1287.8 & 0.945 & 0.60 & 1.2090 \\
\noalign{\smallskip}
\hline
\end{tabular}
\end{table*}

\newpage
\newpage

\begin{longtable}{crrrrrrrrrr}
\caption{\label{HARPS_data} \harps\ \texttt{SERVAL} RV measurements and activity indicators of TOI-500. (a) Barycentric Julian dates are given in barycentric dynamical time; (b) S/N ratio per pixel at 550 nm.}\\
\hline
\noalign{\smallskip}
$\rm BJD_{TDB}^a$ & RV      & $\sigma_\mathrm{RV}$ &    BIS  & FWHM   & dLW & $\sigma_\mathrm{dLW}$ & $\var{{\rm S-index}}$ & $\sigma_\var{{\rm S-index}}$ & T$_\mathrm{exp}$ & S/N $^b$ \\
-2450000 (d)          & (\ms)  & (\ms)               &  (\ms) & (\ms)  & (m$^2$\,s$^{-2}$) & (m$^2$\,s$^{-2}$) & & & (s)              &   \\
\noalign{\smallskip}
\hline
\endfirsthead
\caption{Continued.}\\
\hline
\noalign{\smallskip}
$\rm BJD_{TDB}^a$ & RV      & $\sigma_\mathrm{RV}$ &    BIS  & FWHM   & dLW & $\sigma_\mathrm{dLW}$ & $\var{{\rm S-index}}$ & $\sigma_\var{{\rm S-index}}$ & T$_\mathrm{exp}$ & S/N $^b$ \\
-2450000 (d)          & (\ms)  & (\ms)               &  (\ms) & (\ms)  & (m$^2$\,s$^{-2}$) & (m$^2$\,s$^{-2}$) & & & (s)              &   \\
\noalign{\smallskip}
\hline
\noalign{\smallskip}
\endhead
\noalign{\smallskip}
\hline
\noalign{\smallskip}
\endfoot
\noalign{\smallskip}
\noalign{\smallskip}
8564.617845 &	-2.23 &    0.97 &   41.96 & 6214.88 &	 0.90 &    1.46 &   0.3819 &   0.0045 & 1800 &  82.9 \\
8564.695077 &	-0.53 &    0.91 &   36.59 & 6217.79 &	12.32 &    1.88 &   0.3814 &   0.0054 & 2200 &  67.8 \\
8565.605014 &	-1.07 &    0.94 &   39.16 & 6211.54 &	-0.37 &    1.09 &   0.3761 &   0.0043 & 1800 &  83.1 \\
8565.668295 &	 1.18 &    0.97 &   37.82 & 6216.40 &	 7.21 &    1.46 &   0.3856 &   0.0053 & 2100 &  75.4 \\
8566.525325 &	 1.74 &    1.19 &   45.19 & 6213.14 &	-3.06 &    0.91 &   0.3803 &   0.0034 & 1800 &  96.1 \\
8566.654902 &	-0.31 &    1.37 &   36.45 & 6214.65 &	 0.94 &    1.52 &   0.3678 &   0.0055 & 1800 &  69.7 \\
8567.511513 &	 4.05 &    0.87 &   44.69 & 6211.27 &	-2.27 &    0.95 &   0.3949 &   0.0034 & 1800 &  98.7 \\
8567.654371 &	 4.54 &    1.27 &   44.72 & 6214.11 &	-0.55 &    1.53 &   0.3849 &   0.0057 & 1800 &  68.2 \\
8568.511019 &	 1.15 &    0.75 &   39.63 & 6212.19 &	-2.22 &    0.96 &   0.3956 &   0.0033 & 1800 &  97.1 \\
8568.593482 &	 2.02 &    0.92 &   42.69 & 6213.03 &	-0.92 &    1.08 &   0.3783 &   0.0045 & 1800 &  80.1 \\
8568.666590 &	-0.65 &    1.18 &   36.68 & 6209.93 &	 0.24 &    1.58 &   0.3769 &   0.0062 & 1800 &  61.0 \\
8569.514237 &	-1.78 &    0.74 &   45.68 & 6212.61 &	-1.86 &    1.12 &   0.3873 &   0.0039 & 1800 &  85.8 \\
8569.671573 &	 0.95 &    1.02 &   40.86 & 6219.60 &	 0.95 &    1.49 &   0.3640 &   0.0054 & 2100 &  70.0 \\
8570.592087 &	 2.70 &    1.57 &   39.58 & 6214.88 &	 0.60 &    1.89 &   0.4088 &   0.0055 & 1800 &  55.1 \\
8570.666268 &	 3.36 &    1.29 &   41.81 & 6222.08 &	-1.90 &    1.66 &   0.3635 &   0.0050 & 2100 &  63.0 \\
8571.656132 &	 3.03 &    1.22 &   34.32 & 6216.22 &	 0.01 &    1.68 &   0.3755 &   0.0058 & 1800 &  60.9 \\
8576.500130 &	10.94 &    1.68 &   49.91 & 6213.56 &	-8.97 &    2.54 &   0.3883 &   0.0067 & 1800 &  41.2 \\
8576.618548 &	12.52 &    1.34 &   32.97 & 6212.25 &	-0.80 &    2.45 &   0.3652 &   0.0059 & 1800 &  48.4 \\
8576.676866 &	11.81 &    1.58 &   63.83 & 6218.66 &	-4.00 &    3.16 &   0.3988 &   0.0083 & 1800 &  37.4 \\
8578.493692 &	17.46 &    2.12 &   42.77 & 6208.61 &	-7.06 &    3.16 &   0.3841 &   0.0075 & 1800 &  35.9 \\
8578.605556 &	16.42 &    1.85 &   42.13 & 6203.45 &	-7.72 &    1.95 &   0.3416 &   0.0056 & 1800 &  55.8 \\
8579.512111 &	18.33 &    1.16 &   41.29 & 6208.62 &	-6.93 &    1.50 &   0.3768 &   0.0047 & 1800 &  61.7 \\
8579.612446 &	18.69 &    0.83 &   40.33 & 6208.17 &	-7.84 &    1.27 &   0.3572 &   0.0048 & 1800 &  73.2 \\
8580.501010 &	17.61 &    0.87 &   41.35 & 6206.24 &	-9.38 &    1.09 &   0.3751 &   0.0035 & 1800 &  96.6 \\
8580.573718 &	18.94 &    0.90 &   41.24 & 6198.49 &  -11.10 &    1.11 &   0.3858 &   0.0041 & 1800 &  79.5 \\
8580.640753 &	18.32 &    1.00 &   41.69 & 6209.33 &	-8.96 &    1.12 &   0.3511 &   0.0047 & 1800 &  82.9 \\
8581.515917 &	14.48 &    0.89 &   47.02 & 6206.36 &  -10.20 &    1.41 &   0.3640 &   0.0045 & 1800 &  65.8 \\
8581.613156 &	17.32 &    0.95 &   44.39 & 6204.95 &  -12.45 &    1.43 &   0.3636 &   0.0043 & 1800 &  81.2 \\
8582.519589 &	12.39 &    1.07 &   39.23 & 6202.06 &  -11.80 &    1.12 &   0.3653 &   0.0037 & 1800 &  75.7 \\
8582.570737 &	11.27 &    0.96 &   41.59 & 6205.35 &  -12.02 &    1.40 &   0.3528 &   0.0041 & 1800 &  75.2 \\
8582.652840 &	10.40 &    1.19 &   43.61 & 6202.30 &	-9.82 &    1.66 &   0.3674 &   0.0053 & 1800 &  71.6 \\
8583.486627 &	12.32 &    0.97 &   42.41 & 6202.56 &  -13.04 &    1.30 &   0.3691 &   0.0037 & 1800 &  83.8 \\
8583.575836 &	11.59 &    1.50 &   48.98 & 6211.31 &	-8.95 &    2.08 &   0.3454 &   0.0061 & 1800 &  55.5 \\
8583.638715 &	 9.13 &    1.54 &   46.31 & 6203.52 &  -11.54 &    1.99 &   0.3398 &   0.0063 & 1800 &  48.9 \\
8584.593727 &	10.93 &    0.98 &   40.04 & 6205.51 &  -13.79 &    1.36 &   0.3412 &   0.0049 & 1800 &  74.1 \\
8585.608930 &	12.65 &    1.29 &   39.77 & 6203.39 &  -11.05 &    2.18 &   0.3564 &   0.0063 & 1800 &  55.4 \\
8586.488934 &	 9.48 &    0.99 &   35.89 & 6199.49 &  -11.64 &    1.70 &   0.3742 &   0.0046 & 1800 &  64.1 \\
8586.539517 &	11.67 &    0.91 &   40.53 & 6204.55 &  -14.72 &    1.40 &   0.3282 &   0.0043 & 1800 &  72.4 \\
8586.632740 &	11.19 &    1.58 &   43.60 & 6205.06 &  -15.05 &    2.56 &   0.3212 &   0.0065 & 1800 &  48.8 \\
8587.500189 &	 7.83 &    1.07 &   41.25 & 6199.82 &  -15.51 &    1.30 &   0.3349 &   0.0043 & 1800 &  74.5 \\
8587.598973 &	 7.04 &    1.27 &   40.80 & 6198.63 &  -12.28 &    1.72 &   0.3453 &   0.0059 & 1800 &  55.6 \\
8588.491226 &	 5.54 &    0.91 &   36.67 & 6202.90 &  -12.88 &    1.10 &   0.3572 &   0.0038 & 1800 &  88.3 \\
8588.587229 &	 2.90 &    1.27 &   47.70 & 6211.48 &  -14.66 &    2.40 &   0.3284 &   0.0064 & 1200 &  53.5 \\
8589.488364 &	 2.42 &    1.02 &   41.62 & 6200.07 &  -14.74 &    1.13 &   0.3765 &   0.0034 & 1800 &  93.0 \\
8589.590912 &	 2.20 &    1.28 &   38.29 & 6200.69 &  -11.60 &    1.40 &   0.3356 &   0.0045 & 1800 &  74.1 \\
8590.491136 &	 1.91 &    0.85 &   39.21 & 6208.52 &  -13.50 &    1.11 &   0.3604 &   0.0034 & 1800 & 102.0 \\
8590.567485 &	 0.10 &    0.82 &   39.75 & 6203.37 &	-9.39 &    1.39 &   0.3557 &   0.0043 & 1800 &  91.9 \\
8590.615741 &	-1.18 &    1.01 &   37.67 & 6200.12 &	-8.32 &    1.75 &   0.3567 &   0.0054 & 1800 &  71.1 \\
8591.554693 &	 3.06 &    1.14 &   36.27 & 6198.32 &	-9.88 &    1.47 &   0.3561 &   0.0049 & 1800 &  80.0 \\
8591.575880 &	 1.60 &    1.19 &   37.93 & 6203.70 &  -11.64 &    1.53 &   0.3244 &   0.0046 & 1800 &  81.2 \\
8600.568282 &	13.40 &    1.45 &   43.23 & 6202.68 &	-9.25 &    2.03 &   0.3300 &   0.0056 & 1800 &  57.7 \\
8601.558423 &	15.65 &    1.46 &   34.43 & 6206.18 &  -11.67 &    1.56 &   0.3442 &   0.0051 & 1800 &  74.7 \\
8601.579818 &	16.51 &    1.56 &   41.74 & 6202.27 &  -11.01 &    1.54 &   0.3525 &   0.0054 & 1800 &  69.0 \\
8608.496002 &	21.73 &    0.96 &   39.24 & 6209.16 &	-9.08 &    1.55 &   0.3890 &   0.0047 & 1800 &  64.6 \\
8608.567149 &	22.14 &    0.96 &   42.52 & 6213.94 &	-6.05 &    1.86 &   0.3644 &   0.0052 & 1800 &  68.1 \\
8609.487316 &	17.50 &    0.98 &   38.13 & 6216.68 &	-9.09 &    1.49 &   0.3721 &   0.0046 & 1800 &  61.6 \\
8609.508461 &	18.14 &    1.16 &   41.97 & 6212.89 &	-9.52 &    1.57 &   0.3918 &   0.0049 & 1800 &  64.0 \\
8611.469679 &	18.04 &    1.28 &   41.33 & 6212.26 &	-6.19 &    1.91 &   0.3918 &   0.0053 & 1800 &  59.2 \\
8611.515874 &	17.48 &    1.16 &   41.14 & 6207.12 &	-5.99 &    1.44 &   0.3643 &   0.0050 & 1800 &  64.1 \\
8613.499052 &	16.39 &    1.18 &   41.53 & 6216.47 &	-0.33 &    1.83 &   0.3991 &   0.0064 & 1800 &  55.1 \\
8613.540859 &	18.18 &    1.46 &   39.09 & 6213.35 &	-3.50 &    1.96 &   0.3542 &   0.0062 & 1800 &  54.2 \\
8614.546470 &	12.53 &    1.60 &   36.71 & 6216.43 &	-2.10 &    2.35 &   0.3831 &   0.0064 & 1800 &  50.2 \\
8615.473966 &	 6.99 &    1.12 &   40.32 & 6220.15 &	 1.63 &    1.62 &   0.4077 &   0.0049 & 1800 &  64.0 \\
8615.515423 &	 6.38 &    1.04 &   40.32 & 6213.81 &	-1.50 &    1.45 &   0.3982 &   0.0050 & 1800 &  68.6 \\
8616.501331 &	 5.68 &    1.17 &   45.10 & 6215.14 &	 2.44 &    1.69 &   0.3985 &   0.0055 & 1800 &  67.5 \\
8616.536248 &	 5.75 &    1.45 &   40.15 & 6216.95 &	-0.86 &    1.80 &   0.3828 &   0.0060 & 1800 &  59.1 \\
8617.489895 &	 2.23 &    1.08 &   39.36 & 6220.28 &	 1.65 &    1.40 &   0.3846 &   0.0047 & 1800 &  67.8 \\
8617.531430 &	 3.73 &    1.07 &   43.93 & 6205.43 &	-1.70 &    1.23 &   0.3695 &   0.0050 & 1800 &  68.3 \\
8618.508507 &	 4.68 &    1.35 &   42.11 & 6218.11 &	 2.05 &    1.49 &   0.3794 &   0.0052 & 1800 &  62.8 \\
8619.500486 &	 3.75 &    1.05 &   45.34 & 6207.47 &	 2.13 &    1.93 &   0.3588 &   0.0051 & 1800 &  67.6 \\
8619.528213 &	 5.98 &    0.94 &   41.21 & 6204.84 &	 3.67 &    1.79 &   0.3749 &   0.0055 & 1800 &  64.5 \\
8620.488158 &	 8.18 &    2.55 &   29.63 & 6218.93 &	32.33 &    5.80 &   0.4046 &   0.0103 & 1800 &  26.8 \\
8620.513089 &	 5.96 &    1.47 &   53.83 & 6206.35 &	 5.61 &    2.26 &   0.3610 &   0.0058 & 1800 &  52.7 \\
8622.472536 &	 2.71 &    1.08 &   43.21 & 6212.53 &	 0.61 &    1.35 &   0.3620 &   0.0045 & 1800 &  77.8 \\
8622.512241 &	-0.50 &    1.19 &   41.34 & 6213.63 &	 3.41 &    1.76 &   0.3786 &   0.0053 & 1800 &  71.5 \\
8623.484304 &	 2.18 &    1.73 &   46.92 & 6219.64 &	 4.82 &    2.50 &   0.3306 &   0.0070 & 1800 &  43.4 \\
8623.502785 &	 3.13 &    1.96 &   42.09 & 6200.16 &	27.48 &    4.09 &   0.3512 &   0.0089 & 1800 &  35.4 \\
8624.478282 &	 4.34 &    1.78 &   39.18 & 6211.77 &  -10.70 &    3.15 &   0.3570 &   0.0081 & 1800 &  36.4 \\
8624.499478 &	 6.38 &    1.81 &   37.18 & 6206.45 &	-7.50 &    3.12 &   0.3291 &   0.0076 & 1800 &  37.5 \\
8625.483604 &	10.31 &    1.53 &   34.90 & 6202.39 &  -11.19 &    2.46 &   0.3058 &   0.0064 & 1800 &  44.1 \\
8625.503962 &	 7.74 &    1.42 &   41.30 & 6214.16 &  -11.92 &    2.50 &   0.3020 &   0.0063 & 1800 &  46.4 \\
8626.483767 &	 8.28 &    1.41 &   40.28 & 6213.48 &	-9.24 &    1.86 &   0.3632 &   0.0061 & 1800 &  50.1 \\
8626.504961 &	 8.99 &    1.25 &   35.96 & 6211.35 &  -11.47 &    1.89 &   0.3088 &   0.0055 & 1800 &  54.6 \\
8627.486767 &	10.03 &    1.35 &   43.28 & 6206.19 &  -12.49 &    1.62 &   0.3433 &   0.0052 & 1800 &  56.4 \\
8627.508376 &	 8.95 &    1.34 &   38.37 & 6206.87 &  -13.81 &    1.87 &   0.3188 &   0.0053 & 1800 &  54.1 \\
8636.481815 &	12.79 &    1.14 &   45.17 & 6214.65 &	-7.26 &    1.35 &   0.3651 &   0.0051 & 1800 &  74.5 \\
8638.460403 &	10.25 &    1.48 &   40.19 & 6203.79 &	-2.35 &    2.37 &   0.3347 &   0.0058 & 2400 &  51.6 \\
8641.484017 &	 1.86 &    1.77 &   43.87 & 6206.23 &	-4.36 &    2.27 &   0.3422 &   0.0067 & 2100 &  50.9 \\
8643.457885 &	-2.61 &    0.85 &   37.75 & 6208.79 &	-2.98 &    1.62 &   0.3559 &   0.0050 & 2100 &  71.5 \\
8644.458129 &	-4.03 &    1.14 &   45.56 & 6210.38 &	-5.35 &    1.85 &   0.3606 &   0.0056 & 2100 &  59.4 \\
8645.455945 &	-4.14 &    0.99 &   45.93 & 6220.40 &	 4.92 &    2.09 &   0.3583 &   0.0054 & 2100 &  60.5 \\
8767.792018 &	12.82 &    1.37 &   43.57 & 6212.82 &	-2.25 &    1.46 &   0.3891 &   0.0044 & 1800 &  62.6 \\
8767.812798 &	 9.58 &    1.25 &   34.51 & 6214.41 &	-1.32 &    1.78 &   0.3795 &   0.0044 & 1800 &  59.9 \\
8826.750232 &	 0.58 &    0.76 &   40.55 & 6211.35 &	-6.35 &    0.84 &   0.3931 &   0.0027 & 1800 &  98.3 \\
8826.771221 &	 0.88 &    0.69 &   38.06 & 6206.49 &	-5.22 &    0.97 &   0.3982 &   0.0028 & 1800 &  96.3 \\
8830.638609 &	 1.47 &    0.94 &   39.09 & 6219.53 &	 0.63 &    1.09 &   0.4076 &   0.0029 & 1800 &  97.0 \\
8830.743782 &	 0.04 &    1.03 &   41.10 & 6217.72 &	-1.93 &    0.78 &   0.4133 &   0.0024 & 1800 & 111.8 \\
8830.858651 &	 0.06 &    1.20 &   43.35 & 6215.38 &	 2.35 &    1.03 &   0.3955 &   0.0040 & 1200 &  85.6 \\
8831.631372 &	 5.09 &    0.86 &   37.83 & 6223.09 &	 1.12 &    0.96 &   0.4060 &   0.0032 & 1800 &  87.7 \\
8831.756621 &	 5.96 &    1.03 &   40.27 & 6209.52 &	-0.29 &    1.04 &   0.4211 &   0.0032 & 1800 &  81.3 \\
8831.833187 &	 2.96 &    0.91 &   41.78 & 6216.13 &	-1.03 &    0.90 &   0.3981 &   0.0032 & 1800 &  99.9 \\
8832.642369 &	 2.31 &    1.21 &   37.15 & 6222.68 &	 6.39 &    1.57 &   0.3957 &   0.0045 & 2400 &  60.0 \\
8832.761809 &	 5.18 &    1.12 &   43.35 & 6223.56 &	 3.14 &    1.23 &   0.4099 &   0.0040 & 1800 &  67.8 \\
8832.835968 &	 5.08 &    1.08 &   42.68 & 6214.36 &	 1.49 &    1.49 &   0.4024 &   0.0044 & 1800 &  69.9 \\
8833.638839 &	 5.85 &    2.17 &   35.89 & 6223.67 &	 5.12 &    3.25 &   0.3695 &   0.0074 & 2400 &  34.6 \\
8833.740462 &	 2.81 &    1.21 &   37.64 & 6220.11 &	 1.97 &    1.19 &   0.4163 &   0.0037 & 1800 &  73.5 \\
8833.832512 &	 7.13 &    1.10 &   38.47 & 6217.14 &	 4.72 &    1.03 &   0.4127 &   0.0044 & 2100 &  71.6 \\
8834.627722 &	 6.18 &    1.28 &   35.48 & 6212.98 &	-0.04 &    1.42 &   0.4051 &   0.0038 & 1800 &  73.4 \\
8834.729703 &	 6.73 &    1.17 &   39.26 & 6213.89 &	 3.96 &    1.38 &   0.4031 &   0.0042 & 1800 &  65.1 \\
8834.831917 &	 6.74 &    0.96 &   37.46 & 6220.99 &	 4.40 &    1.08 &   0.4028 &   0.0036 & 2100 &  86.4 \\
8835.635885 &	 7.31 &    0.91 &   43.03 & 6217.93 &	 1.97 &    0.92 &   0.4190 &   0.0030 & 1800 &  92.9 \\
8835.722447 &	 5.50 &    0.73 &   44.01 & 6224.08 &	 3.64 &    1.07 &   0.4253 &   0.0030 & 1800 &  83.5 \\
8835.778132 &	 5.13 &    0.89 &   41.83 & 6218.46 &	 1.79 &    0.97 &   0.4282 &   0.0030 & 1800 &  92.9 \\
8835.855598 &	 7.17 &    0.84 &   43.84 & 6219.91 &	 5.25 &    1.02 &   0.4111 &   0.0037 & 1800 &  93.0 \\
8841.624275 &	19.21 &    1.08 &   40.34 & 6227.22 &	 3.07 &    1.46 &   0.4536 &   0.0038 & 1800 &  67.3 \\
8841.819900 &	17.58 &    1.03 &   44.74 & 6225.84 &	 4.89 &    1.33 &   0.4275 &   0.0041 & 1800 &  78.0 \\
8842.657713 &	22.06 &    1.03 &   44.38 & 6220.11 &	 4.78 &    1.37 &   0.4506 &   0.0040 & 1800 &  63.8 \\
8842.862273 &	14.72 &    1.01 &   44.84 & 6223.12 &	12.23 &    1.96 &   0.4261 &   0.0049 & 1800 &  59.3 \\
8843.607790 &	19.85 &    1.03 &   35.13 & 6223.80 &	 6.22 &    1.27 &   0.4409 &   0.0037 & 1800 &  76.3 \\
8843.841628 &	20.19 &    0.97 &   45.56 & 6227.07 &	 6.94 &    1.37 &   0.4041 &   0.0044 & 1800 &  70.6 \\
8844.626024 &	18.51 &    0.85 &   42.24 & 6224.26 &	 5.60 &    1.32 &   0.4394 &   0.0034 & 1800 &  83.9 \\
8844.859809 &	20.22 &    1.10 &   45.93 & 6223.10 &	 5.76 &    1.71 &   0.3835 &   0.0052 & 1200 &  58.3 \\
8846.622475 &	16.98 &    1.16 &   44.26 & 6228.13 &	 8.31 &    1.17 &   0.4381 &   0.0041 & 1800 &  71.5 \\
8846.824397 &	16.04 &    0.88 &   44.73 & 6230.44 &	10.56 &    1.18 &   0.4293 &   0.0041 & 1800 &  84.6 \\
8847.632314 &	14.75 &    0.91 &   43.76 & 6228.44 &	 7.71 &    1.13 &   0.4312 &   0.0033 & 1800 &  83.4 \\
8847.820198 &	10.61 &    0.86 &   43.63 & 6224.63 &	 7.06 &    0.93 &   0.4351 &   0.0039 & 1800 &  92.8 \\
8856.804608 &	-0.32 &    1.35 &   56.09 & 6217.36 &	 7.45 &    1.76 &   0.4156 &   0.0060 & 1800 &  53.5 \\
8857.729722 &	 1.06 &    1.12 &   41.83 & 6211.64 &	 3.93 &    1.10 &   0.4049 &   0.0035 & 1800 &  85.6 \\
8857.750730 &	 0.30 &    0.90 &   51.49 & 6217.17 &	 3.73 &    1.17 &   0.4077 &   0.0038 & 1800 &  82.3 \\
8858.618012 &	 4.72 &    0.92 &   44.34 & 6216.65 &	 2.40 &    1.12 &   0.4048 &   0.0031 & 1800 &  88.9 \\
8858.663596 &	 3.54 &    0.96 &   45.08 & 6218.58 &	 6.50 &    1.34 &   0.4061 &   0.0036 & 1800 &  73.7 \\
8858.755827 &	 2.81 &    1.09 &   41.32 & 6215.40 &	 5.86 &    1.36 &   0.3990 &   0.0042 & 1800 &  69.9 \\
8859.605000 &	 8.31 &    1.13 &   43.07 & 6219.81 &	10.07 &    1.35 &   0.4104 &   0.0038 & 1800 &  71.9 \\
8859.718559 &	 8.53 &    1.44 &   41.73 & 6217.27 &	19.46 &    2.00 &   0.3851 &   0.0045 & 1800 &  55.0 \\
8859.804763 &	 5.50 &    1.14 &   46.14 & 6211.84 &	16.04 &    1.99 &   0.3709 &   0.0051 & 1800 &  63.2 \\
8860.578857 &	 3.99 &    0.96 &   42.81 & 6216.79 &	 1.18 &    1.08 &   0.4067 &   0.0034 & 1800 &  83.1 \\
8860.646806 &	 5.20 &    0.95 &   45.27 & 6218.11 &	 1.06 &    1.37 &   0.4092 &   0.0038 & 1800 &  67.6 \\
8860.800222 &	 4.33 &    0.82 &   46.93 & 6216.86 &	 3.01 &    1.27 &   0.3873 &   0.0044 & 1800 &  77.6 \\
8861.579038 &	 2.92 &    0.96 &   39.03 & 6211.23 &	-4.03 &    1.41 &   0.4129 &   0.0040 & 1800 &  68.2 \\
8861.633702 &	 4.23 &    1.32 &   46.34 & 6213.22 &	-0.49 &    1.66 &   0.4309 &   0.0044 & 1800 &  57.4 \\
8861.753368 &	 4.91 &    0.91 &   48.15 & 6210.64 &	-1.00 &    0.99 &   0.3994 &   0.0039 & 1800 &  82.4 \\
8862.596999 &	 5.94 &    0.76 &   39.55 & 6218.39 &	-3.64 &    1.00 &   0.4212 &   0.0028 & 1800 &  99.8 \\
8862.664555 &	 3.04 &    0.91 &   47.62 & 6213.40 &	-4.10 &    1.08 &   0.4139 &   0.0032 & 1800 &  82.2 \\
8862.817209 &	 6.06 &    0.98 &   38.79 & 6216.90 &	-2.59 &    1.09 &   0.3972 &   0.0042 & 1800 &  84.8 \\
8863.566881 &	10.02 &    0.94 &   39.34 & 6208.51 &	-5.15 &    1.11 &   0.4230 &   0.0035 & 1800 &  83.4 \\
8863.659962 &	 7.86 &    0.90 &   41.50 & 6212.23 &	-3.68 &    0.87 &   0.4149 &   0.0027 & 1800 & 101.1 \\
8863.793818 &	 8.87 &    0.88 &   44.01 & 6212.85 &	-4.02 &    1.00 &   0.3912 &   0.0039 & 1800 &  93.9 \\
8864.610798 &	15.46 &    1.10 &   42.22 & 6214.39 &	-4.90 &    1.40 &   0.4289 &   0.0035 & 1800 &  76.2 \\
8864.704636 &	15.13 &    1.27 &   44.70 & 6214.02 &	-4.74 &    1.07 &   0.4080 &   0.0035 & 1800 &  77.8 \\
8865.578149 &	20.00 &    0.81 &   38.53 & 6218.41 &	-2.04 &    1.05 &   0.4204 &   0.0034 & 1800 &  81.9 \\
8865.691720 &	20.76 &    0.97 &   41.14 & 6217.31 &	-3.57 &    1.23 &   0.4260 &   0.0038 & 1800 &  72.1 \\
8866.630389 &	16.35 &    1.19 &   41.49 & 6222.81 &	-0.63 &    1.21 &   0.4048 &   0.0040 & 1800 &  69.5 \\
8866.793359 &	17.86 &    0.94 &   39.71 & 6219.68 &	-0.82 &    1.48 &   0.3997 &   0.0046 & 1800 &  73.4 \\
8867.609423 &	13.54 &    1.02 &   39.20 & 6209.86 &	-2.26 &    1.02 &   0.4118 &   0.0037 & 1800 &  74.0 \\
8868.609708 &	13.35 &    0.89 &   39.65 & 6214.52 &	 0.27 &    1.06 &   0.4000 &   0.0032 & 1650 &  88.3 \\
8868.798870 &	14.84 &    1.01 &   34.38 & 6301.97 &	 0.40 &    1.12 &   0.3818 &   0.0041 & 1800 &  90.0 \\
8869.678127 &	14.85 &    0.81 &   49.66 & 6217.16 &	 0.18 &    1.22 &   0.4108 &   0.0032 & 1600 &  89.9 \\
8871.599314 &	17.68 &    1.00 &   39.20 & 6215.15 &	 1.44 &    1.08 &   0.4130 &   0.0035 & 1800 &  75.0 \\
8871.780380 &	15.59 &    1.31 &   37.28 & 6213.75 &	 1.79 &    1.93 &   0.3906 &   0.0055 & 1740 &  55.4 \\
8884.582627 &	-0.87 &    1.56 &   46.30 & 6230.48 &	17.00 &    2.15 &   0.4156 &   0.0057 & 1800 &  40.9 \\
8884.686466 &	 0.87 &    1.06 &   39.84 & 6229.99 &	15.71 &    1.68 &   0.4234 &   0.0047 & 1800 &  56.9 \\
8885.572076 &	 5.35 &    1.23 &   41.97 & 6230.12 &	14.48 &    1.70 &   0.4170 &   0.0049 & 1800 &  53.5 \\
8886.608284 &	 3.85 &    1.02 &   45.29 & 6230.35 &	12.16 &    1.45 &   0.4475 &   0.0037 & 1800 &  77.2 \\
8886.758254 &	 0.63 &    1.02 &   46.02 & 6229.89 &	14.02 &    1.74 &   0.4242 &   0.0046 & 1800 &  60.0 \\
8887.677809 &	 6.86 &    1.40 &   51.80 & 6226.36 &	12.66 &    1.22 &   0.4211 &   0.0041 & 1800 &  73.0 \\
8888.628038 &	 4.85 &    0.89 &   41.99 & 6224.37 &	 8.27 &    1.18 &   0.4236 &   0.0031 & 1800 &  89.4 \\
8889.605177 &	 6.95 &    0.90 &   49.63 & 6223.87 &	 8.38 &    1.17 &   0.4434 &   0.0035 & 1800 &  74.3 \\
8889.745868 &	 9.72 &    0.90 &   40.88 & 6222.20 &	12.70 &    1.35 &   0.4042 &   0.0045 & 1800 &  75.8 \\
8890.616595 &	10.66 &    1.14 &   48.67 & 6216.50 &	 3.99 &    1.45 &   0.4200 &   0.0039 & 1800 &  62.6 \\
8890.738064 &	14.21 &    1.51 &   45.37 & 6211.98 &	 8.32 &    1.48 &   0.4340 &   0.0049 & 1800 &  65.9 \\
8892.668702 &	15.37 &    1.00 &   43.87 & 6218.26 &	 2.11 &    1.61 &   0.3949 &   0.0043 & 1800 &  66.0 \\
8892.683853 &	13.74 &    1.62 &   48.22 & 6222.77 &	12.82 &    2.59 &   0.3858 &   0.0057 & 1800 &  37.3 \\
8893.776606 &	16.95 &    1.26 &   40.16 & 6306.54 &	 0.28 &    1.48 &   0.3802 &   0.0053 & 1800 &  65.7 \\
8894.702360 &	18.16 &    1.32 &   41.05 & 6217.90 &	-3.55 &    1.35 &   0.3962 &   0.0045 & 1800 &  65.9 \\
8898.713993 &	13.99 &    0.92 &   41.11 & 6208.31 &	-5.35 &    1.27 &   0.3758 &   0.0042 & 1800 &  77.9 \\
8899.670668 &	14.74 &    0.81 &   41.79 & 6211.12 &	-4.73 &    1.11 &   0.3858 &   0.0037 & 1800 &  89.2 \\
8900.589823 &	10.95 &    0.92 &   41.09 & 6208.33 &	-5.73 &    1.03 &   0.3901 &   0.0032 & 1800 &  87.8 \\
8901.545867 &	 7.75 &    0.96 &   40.56 & 6210.30 &	-4.82 &    0.91 &   0.3986 &   0.0029 & 1800 &  91.2 \\
8902.672747 &	 4.71 &    1.21 &   36.17 & 6267.48 &	-3.79 &    1.12 &   0.3864 &   0.0039 & 1800 &  84.3 \\
8903.606961 &	 5.47 &    0.75 &   41.01 & 6212.44 &	-4.56 &    1.01 &   0.3941 &   0.0031 & 1800 &  95.2 \\
8910.625898 &	 9.17 &    1.11 &   46.41 & 6232.90 &	 9.46 &    1.42 &   0.4285 &   0.0044 & 1800 &  72.3 \\
8911.635524 &	 9.65 &    0.93 &   41.47 & 6227.86 &	 9.85 &    1.41 &   0.4182 &   0.0039 & 1800 &  83.9 \\
8912.607148 &	10.63 &    0.98 &   45.80 & 6235.95 &	11.67 &    1.28 &   0.4336 &   0.0039 & 1800 &  83.9 \\
8914.660378 &	 9.48 &    0.96 &   43.32 & 6229.57 &	12.07 &    1.26 &   0.4381 &   0.0046 & 1800 &  79.5 \\
8916.670779 &	15.96 &    0.84 &   42.55 & 6237.40 &	14.90 &    1.61 &   0.4432 &   0.0050 & 1800 &  73.2 \\
8917.646069 &	20.69 &    1.48 &   45.03 & 6231.21 &	31.74 &    1.61 &   0.4340 &   0.0055 & 1800 &  63.5 \\
8918.622840 &	21.11 &    0.93 &   45.61 & 6236.04 &	14.52 &    1.23 &   0.4641 &   0.0040 & 1800 &  93.4 \\
8919.553301 &	23.03 &    0.82 &   37.54 & 6235.47 &	11.29 &    0.97 &   0.4711 &   0.0036 & 1800 &  84.6 \\
8919.574504 &	22.81 &    0.90 &   44.04 & 6230.27 &	11.82 &    1.25 &   0.4717 &   0.0036 & 1800 &  87.5 \\
8924.556320 &	21.71 &    0.88 &   43.51 & 6238.21 &	15.83 &    1.30 &   0.4540 &   0.0037 & 1800 &  85.1 \\
8925.598601 &	21.20 &    0.80 &   45.68 & 6236.12 &	15.29 &    1.45 &   0.4505 &   0.0043 & 1800 &  83.7 \\
8926.601623 &	17.91 &    0.90 &   47.89 & 6233.30 &	15.07 &    1.51 &   0.4344 &   0.0043 & 1800 &  85.4 \\
8927.613989 &	12.96 &    1.09 &   51.40 & 6239.52 &	16.43 &    1.60 &   0.4218 &   0.0052 & 1800 &  63.6 \\
8928.606783 &	 5.15 &    1.00 &   47.53 & 6237.91 &	16.65 &    1.54 &   0.4216 &   0.0046 & 1800 &  71.9 \\
8929.608598 &	 3.39 &    1.04 &   46.25 & 6232.04 &	13.47 &    1.85 &   0.4981 &   0.0055 & 1800 &  61.2 \\
8930.594169 &	 0.86 &    1.52 &   41.17 & 6214.62 &	 7.84 &    2.51 &   0.4166 &   0.0067 & 2100 &  40.8 \\
8931.628930 &	 3.91 &    0.99 &   44.49 & 6236.01 &	10.07 &    1.65 &   0.4359 &   0.0055 & 1800 &  63.6 \\
\end{longtable}

\endgroup

\end{document}